# Vortex breakdown and its topologies in turbulent flows within a typical swirl combustor geometry

*Nitesh Kumar Sahu* [a,*], *Anupam Dewan* [b] *and Mayank Kumar* [c,*]

a: Department of Fuel, Minerals & Metallurgical Engineering, Indian Institute of Technology (ISM) Dhanbad, Dhanbad, Jharkhand – 826004, India b: Department of Applied Mechanics, Indian Institute of Technology Delhi, Hauz Khas, New Delhi – 110016, India c: Department of Mechanical Engineering, Indian Institute of Technology Delhi, Hauz Khas, New Delhi – 110016, India

* Corresponding Author (1): Email - nitesh@iitism.ac.in, Tel: +91-32-6223-5158
* Corresponding Author (3): Email - kmayank@mech.iitd.ac.in, Tel: +91-11-2659-7392

## Abstract

We investigate vortex breakdown (VB) and its dominant topologies in turbulent, non-reacting flows within a canonical swirl combustor using large-eddy simulations (LES). A baseline configuration and operating conditions are first used to validate the LES solver, then five additional cases differing only by swirler vane-angle are simulated. The onset of VB is quantified using the generic swirl-number formulation, $SN_g$, by detecting an internal recirculation zone (IRZ) in the mean flow, excluding highly-intermittent VB cases. Analysis of the mean flow shows that $SN_g$ measured within $40$ mm downstream of swirler best represents the flow's swirl-strength compared with commonly used alternatives. A stable-VB first appears in the flow with $25^0$ vane-angle, $SN_g \approx 0.35$. $Q$-criterion iso-surfaces and velocity time-series at VC footprints show a single-helix VC to prevail across all investigated vane-angles; up to $60^0$, $SN_g = 0.98$. Weaker double-helix signatures also appear, resulting from quadratic self-interaction of single-helix tone for vane-angles$\leq 50^0$. They are quadratically independent in $60^0$ case, indicating association with a distinct helical hydrodynamic mode. Axisymmetric IRZ oscillations interact with helical-VC dynamics in $25^0$ and $60^0$ cases. VC precesses as stable limit-cycle oscillation via a marginally stable mode in $40^0 - 50^0$ cases, while it waxes and wanes strongly in $25^0$ and $60^0$ cases driven by stochastic forcing of a slightly stable mode. Alongside the coherent VC-strand, a weakly-coherent strand originates from swirler. Its precession frequency matches with the lowest coherent precession frequency, corroborating its precession-based origin. Overall, we establish critical-values, evaluation locations and a topology-map for predicting and interpreting VB-states in isothermal swirl combustor flows.

***Keywords***: Vortex breakdown; vortex core topology; vortex dynamics; precessing vortex core; swirl combustor

## *Acronyms*

| | |
|---|---|
| CB | Center-body |
| IRZ | Internal recirculation zone |
| ISL | Inner shear layer |
| OSL | Outer shear layer |
| $Re$ | Reynold number |
| $Re_{inlet}$ | Inlet Reynolds number |
| $SN_c$ | Conventional swirl number |
| $SN_g$ | Generic swirl number |
| VB | Vortex breakdown |
| VC | Vortex core |

## 1. Introduction

Swirl combustors are widely employed in combustion facilities, such as gas turbines [1,2] and pulverized coal-based reactors [3-5]. A typical swirl combustor design features an expansion plane connecting two tubes of different diameters and a swirl generator in the smaller diameter tube [1, 6-8]. These features are incorporated to induce vortex breakdown, essential for flame stabilization in the inherently turbulent flow of swirl combustors [9, 10]. Vortex breakdown is a phenomenon marked by the sudden emergence of a low-velocity recirculation zone in the core of a swirling flow [9]. Moreover, a vortex core in a swirling flow can break down into various topologies [9, 11-13], each of which uniquely influences the flow dynamics. The dynamics in turn, differently affect mixing [9, 14, 15], flame stability [16, 17], pollutant emissions [18], hot-spot formation [18-20] and instability characteristics [18-20] in swirl combustors. Therefore, comprehensive knowledge of the predominant vortex breakdown (VB) topologies and the parameter space governing both the onset of VB and the transitions among its various topologies in swirl combustors is crucial.

Since vortex breakdown was first observed over seven decades ago, numerous theories have been proposed to explain its onset mechanism and resulting flow topologies. A successful theory would identify the parameters that govern both the onset of VB and the transitions among its various topologies [9]. However, only a few recent theories, specifically those invoking a jump-like transition in vortex core dynamics about a critical state within a flow, have been able to partially explain the VB mechanism and identify the parameters governing it [9, 21-23]. Accordingly, experimental observations and numerical simulations are generally relied upon for identifying the parameters and their critical thresholds governing the onset and topology of VB [11, 13, 21]. An ensemble of flow-based and geometry-based parameters is expected to be of relevance [11-13, 21]. Combustion is known to modulate VB characteristics by altering flow properties [8, 20]. However, in the present study, we examine the effect of relevant parameters decoupled from combustion. Accordingly, we focus only on non-reacting, isothermal flow inside swirl combustors. The influence of combustion parameters such as heat release rate, temperature of flow, etc., on VB characteristics in swirl combustors will be explored in our forthcoming work.

Based on the previous studies [11-13, 21], swirl strength emerges as a relevant flow parameter. Earlier researchers [11-13, 24, 25] used various scaling-based definitions—such as the circulation number, $\Omega$, and the inverse of Rossby number, $Ro^{-1}$, shown in $\text{Eq.}\,(1)$—to quantify swirl strength of isothermal flows. However, evaluation of these quantities in flows without solid-body rotation require additional assumptions and supplementary analysis [21, 24, 25]. Building on the conservation laws, Chigier and coworkers [26–28] introduced a non-dimensional number naming it swirl number, shown in $\mathrm{Eq.}\,(2)$, which can be evaluated directly from local flow-field data. Their goal was to develop a relevant parameter that remains constant in the absence of viscous dissipation and otherwise exhibits a net decay along the flow, with weak variation rates, in a region of fixed-cross-section away from any flow transition. Owing to these traits, swirl number has become the standard metric for quantifying swirl intensity [29]. Its most generic formulation for an isothermal flow is given in $\mathrm{Eq.}\,(3)$ [29]. In practice, Chigier *et al.* [26] also proposed a simplified version, $SN_p$, $[\mathrm{Eq.}\,(4)]$ which neglects the mean of the fluctuating terms—trading some generality for ease of use, as we will discuss later.

$$\Omega = \frac{\Gamma}{<U_a>_{A-avg}\, d_c} \qquad\qquad Ro = \frac{2<U_a>_{A-avg}}{\omega_c d_c} \tag{1}$$

$$SN = \frac{2*time-mean\ of\ the\ axial\ flux\ of\ tangential\ momentum}{d_c*time-mean\ of\ the\ axial\ flux\ of\ axial\ momentum} \tag{2}$$

$$SN_g = \frac{2\iint_{A_c} \rho(<U_a><U_\theta>+<u'_a u'_\theta>) r dA}{d_c \iint_{A_c} \{(<p>-p_{ref})+\rho(<U_a><U_a>+<u'_a u'_a>)\} dA} \qquad U_\theta = \frac{(\vec{r}\times\vec{U}).\hat{e}_a}{|\vec{r}|=r} \tag{3}$$

$$SN_p = \frac{2\iint_{A_c} \rho(<U_a><U_\theta>) r dA}{d_c \iint_{A_c} \{(<p>-p_{ref})+\rho<U_a><U_a>\} dA} \tag{4}$$

here $\Gamma$ is the flow circulation, $< U_a >_{A-avg}$ is the area-averaged time-mean axial-velocity component, $\omega_c$ is a characteristic rotation rate and $d_c$ is a characteristic diameter, $< U_a >$ *and* $< U_\theta >$ are the time-mean axial & azimuthal/ tangential components of velocity, respectively, $< u'_a u'_\theta >$ is their cross-correlation and $< u'_a u'_a >$ is the variance of axial-velocity fluctuations, $< p >$ is the time-mean local static pressure, $p_{ref}$ is a reference pressure. All formulations assume statistically stationary flow. $\hat{e}_a$ is unit-vector along flow.

The foregoing overview on swirl numbers is to provide a consistent basis for the subsequent discussions in the present work. Sarpkaya [11] as well as Fahler and Leibovich [12] experimentally investigated the impact of Reynolds number ($Re$) and circulation number ($\Omega$) on the onset and topology of VB inside a diverging tube with the cone angle of $3^0$. They observed that the interplay of these parameters governs the onset and topology of VB in the investigated geometry. VB occurred only when $Re$ exceeded $3700$ in the flow with the lowest investigated circulation number, $\Omega = 1.07$, whereas VB was observed for all the investigated Reynolds numbers above $700$ in the flow with $\Omega = 3.00$. A total of six VB topologies/modes were observed in the diverging tube [12]. Only two of these VB modes, namely spiral VB (single-helix mode) and bubble VB, were predominant in flows with $Re \geq 4000$. Bubble VB was the only mode observed at higher values of $\Omega$ in these flows. The bubble topology became unstable below a certain threshold value of $\Omega$ where it intermittently transitioned into either a single-helix or double-helix configuration. The VB topology stabilized exclusively as the single-helix configuration below a further lower threshold value of $\Omega$, i.e., only spiral VB occurred at lower values of $\Omega$ in flows with $Re \geq 4000$. These threshold values of $\Omega$ decreased with increase in $Re$. Syred and Beer [21] reparametrized Sarpkaya's results [11] by replacing $\Omega$ with a highly simplified formulation of swirl number expressed here as $SN_c$ and defined in $\text{Eq.}\,(5)$. As expected, the results show that bubble VB predominated at higher values of $SN_c$, while spiral VB predominated at lower values of $SN_c$ in flows with $Re \geq 4000$.

$$SN_c = \frac{2\int_o^{R_o} \rho < U_a > < U_\theta > 2\pi r^2 dr}{d_c \int_o^{R_o} \{\rho < U_a > < U_a >\} 2\pi r dr} \tag{5}$$

Escudier and Zehnder [13] investigated the impact of divergence/cone angle on VB characteristics by systematically testing various short diverging tubes with cone angles up to $25^0$ within a confined facility. Their results demonstrated that larger divergence angles reduce the critical swirl number and $Re$ for the onset of VB. Furthermore, only the bubble and spiral VB modes predominated in flows with $Re \geq 4000$ across all the investigated divergence angles. Bubble VB exclusively appeared at higher values of $\Omega$, while spiral VB was the only mode visible at lower values of $\Omega$ under these conditions.

Based on the aforementioned studies [11-13], it is likely that the predominant VB modes within a typical swirl combustor configuration, described earlier, are also independent of $Re_{inlet}$ at least for $Re_{inlet} \geq 10^4$ ($Re_{inlet}$ was used in [11-13]). This range of $Re_{inlet}$ is typical for practical reactor operations [3-6] and is therefore the focus of the present work. Accordingly, it seems reasonable to identify the prevailing VB topologies in an isothermal swirl combustor by investigating it over a wide range of operating conditions that influence VB at any $Re_{inlet}$ above $10^4$. Swirl strength was identified as one such parameter in the above studies. Furthermore, in a typical swirl combustor geometry, the expansion ratio, $ER$ [Eq. (6)], could also be a parameter influencing the VB topology.

$$ER = \frac{hydraulic\ dia\ of\ tube\ after\ expansion\ plane}{hydraulic\ dia\ of\ tube\ before\ expansion\ plane} \tag{6}$$

A few numerical studies [30, 31] have investigated predominant VB modes in swirl combustor geometries at inlet Reynolds numbers up to 3000. However, no attempt has yet been made to resolve the dominant VB topologies in swirl combustors for $Re_{inlet} \geq 10^4$. Recently, Vanierschot *et al.* [32] investigated VB topology in a swirling flow using tomographic PIV velocity-field measurements, albeit at only a single value each of $SN_c$ and $Re_{inlet}$ (operating conditions are such that the effect of outlet contraction on VB characteristics is negligible [33]). $Re_{inlet}$ was approximately $1.4 \times 10^4$, $SN_c$ was $0.36$ and $ER$ of the investigated setup was just above $20.0$ [32]. $SN_c$ was measured at the expansion plane immediately downstream of the swirler. The VC structure at each measured instant was revealed using the iso-surface of a suitable value of the $Q$-criterion. These iso-surfaces exhibited the double-helix VC structure across the investigated time period. When the instantaneous velocity field data were phase-averaged, to isolate the large-scale flow structures, the reproduced iso-surface of the $Q$-criterion again exhibited a double-helix VC. Accordingly, Vanierschot *et al.* [32] reported the formation of a stable double-helix VB topology. Lu *et al.* [34] numerically investigated flow fields with $SN_c = 0.30$ and $0.50$, $Re_{inlet} \approx 1.25 \times 10^5$ within a swirl combustor having $ER \approx 1.5$. $SN_c$ was calculated just downstream of the swirler. The VB appeared intermittently at $SN_c = 0.30$, while a stable VB was observed at $SN_c = 0.50$. The instantaneous fields of velocity, pressure and iso-surfaces of vorticity presented in the article clearly depict the formation of a stable double-helix VB in the latter case, which is of practical interest. However, the double-helix VB was rarely observed in earlier discussed confined swirling flow investigations with $Re_{inlet} \geq 4000$ [11-13], and when it appeared it was unstable or intermittent [11-13]. Notably, the experiments in [32] and simulations in [34] did not adopt any systematic approach to stabilize this rare VB mode, making it an intriguing subject for further investigation.

One possible reason for this unique occurrence might be the difference in the profile of the center-body (CB) of swirlers used in [32] and [34] compared to the works discussed earlier [11-13]; (CB with a bluff-body profile creates a separation bubble at its tip that eases flame ignition by providing a hot spot [36]). Vanierschot *et al.* [32] and Lu *et al.* [34] used center-bodies with bluff-body profiles, whereas earlier discussed studies [11-13] employed either streamlined center-bodies or omitted them entirely. Vortex shedding from a bluff body evolves into a coherent helical structure in its wake under conditions suitable to excite the wavemaker of a global helical instability [35]. The interaction of this response with a confined swirling flow may have induced the formation of the stable double-helix VB observed in [32, 34]. To isolate the primary cause, we simulate an isothermal swirl combustor with CB having a bluff-body profiled and operating at $Re_{inlet} \geq 10^4$, where if a double-helix is observed we investigate the mechanism of its formation.

Taamallah *et al.* [20] simulated both reacting and non-reacting flows with $SN_c \approx 0.70$ and $Re_{inlet} = 2.0 \times 10^4$ in a swirl combustor with $ER = 2.0$ and a swirler with a sharp-edged CB using FVM-filtered large-eddy simulation (LES) approach. The sharp-edged CB is effectively a bluff-body. As mentioned earlier, the focus of the present study is exclusively on non-reacting flows. Taamallah *et al.* [20] reported the formation of a nearly stable single-helix VB that occasionally transitioned into a double-helix configuration. This dominant topology along with its transition behavior has not been observed in other swirl combustors of interest [32, 34]. However, Taamallah *et al.* [20] did not provide sufficient evidence to substantiate the formation of the claimed VB topologies. Their result showing 2-D slice of vortex core (VC) on a streamwise plane, using lambda-2/$\lambda_2$-criterion [38] at several instances, reveals only a single patch. Still, Taamallah *et al.* [20] concluded the occasional presence of a double-helix configuration without any additional justification. Furthermore, they employed first-order accurate spatial discretization schemes for a few quantities along with a non-zero sub-grid closure in their model, which is a combination known to introduce significant numerical diffusion into discretized governing-equations [37]. Hence, an appropriate LES along with a proper presentation of the results, 2-D and 3-D presentation of VC like in [32, 34, 38], is necessary for the setup in [20] to correctly identify the corresponding VB topology. Notably, the geometries studied in [20, 32, 34] effectively correspond to the typical swirl combustor configuration, which is specified in the very beginning of this section.

Furthermore, the possibility of a spiral VB topology forming in [20] rather than the double-helix VB observed in [32] and [34] cannot be dismissed since the value of $SN_c$ among these studies differ significantly. Other VB topologies may also form in swirl combustors with $Re_{inlet} \geq 10^4$ at swirl strengths different from those considered in [20, 32, 34]. Examining flow fields across a wide range of swirler vane-angles in the setup of [20], with other variables held constant, will resolve the dependence of VB topologies on swirl strength. As discussed earlier, the investigation would not only resolve the VB topologies that predominate in the swirl combustor flow in [20] but also in any typical swirl combustor operating at $Re_{inlet} \geq 10^4$. Recall that a typical swirl combustor geometry corresponds to a combustor having two tubes connected by an expansion plane with a swirler placed in the smaller diameter tube. Although $ER$ could also influence these dominant topologies, we will argue in the *Results and Discussion* section that its effect on VB topology can be delineated based on the available literature and the additional investigations proposed above. Further, $ER \geq 1.5$ is common in reactor geometries from lab-scale to large-scale facilities [3-5, 20, 32, 34] and therefore we exclusively focus on this range in the present work.

Next, we examine the parameters governing the onset of VB in swirl combustors. As stated earlier, Sarpkaya [11] as well as Syred and Beer [21] demonstrated that the interplay of swirl strength and $Re$ governed the onset of VB in diverging tubes. However, the onset of VB in swirling jets issuing from straight nozzles with $Re_{nozzle} \geq 1200$ depended only upon swirl strength [21, 39]. Hallets and Toews [40] confirmed the applicability of this behaviour to swirl combustors for a given $ER$ when $Re_{inlet} \geq 10^4$. They also found the critical value of swirl strength to depend on the inlet velocity profile. However, we do not focus on this aspect in the present work for brevity. Furthermore, the critical value was found to be almost independent of $ER$ in swirl combustors with $ER \geq 1.5$ [40]. Notably, this observation is of importance since our focus in the present work is only on reactor geometries with $ER \geq 1.5$, as mentioned earlier. Hallets and Toews [40] measured the swirl strength in the experiments using a simplified formulation of $SN$, $SN_s$ $[\text{Eq.}\,(7)]$, at a location upstream of the expansion plane.

$$SN_S = \frac{2\int_o^{Ro} 2\pi\rho < U_a >< U_\theta > r^2 dr}{d_{inlet}\,\{\rho_{inlet}(\pi r_{inlet}^2) U_{a,mean-inlet}^2\}} \quad (7)$$

here $U_{a,mean-inlet}$ is the mass flux weighted mean axial-velocity.

Further, we anticipate that the critical value of any other swirl number formulation would also exhibit a similar degree of independence from $ER$ and $Re_{inlet}$ for swirl combustors with $ER \geq 1.5$ operating at $Re_{inlet} \geq 10^4$. Both Terhaar *et al*. [41] and Vanierschot *et al.* [32] employed $SN_c$ $[\mathrm{Eq.}\,(5)]$ to quantify swirl strength in their swirl combustors operating under conditions relevant to the present work. However, a comparison of their findings reveals substantially different critical values of $SN_c$. Terhaar *et al.* [41] in an experimental-cum-numerical study measured the critical value of $SN_c$ well above the expansion plane to be $0.47$ for a non-reacting flow with $Re_{inlet} = 5.0 \times 10^4$ in a swirl combustor with $ER \approx 3.0$. In contrast, Vanierschot *et al.* [32] observed a well-developed internal recirculation zone in a swirling flow with $Re_{inlet} = 1.4 \times 10^4$ within a combustor with $ER \approx 20.0$ even at $SN_c = 0.36$ measured just after the expansion plane (operating under conditions where effect of outlet contraction on VB is not substantial).

According to Vignat *et al*. [29], the discrepancy mentioned above arose because $SN_c$ varies significantly within swirl combustors with a dramatic variation across the expansion plane. Further, the plane of measurement of $SN_c$ in the studies [32] and [41] is arbitrary with no correspondence. The investigation of an isothermal swirl combustor with $ER = 100$ operating at $Re_{inlet} = 1.1 \times 10^4$ in [29] indicates a similar behavior for other simplified $SN$ formulations. They argued that a useful comparison of flow fields based on $SN$ requires a $SN$ formulation whose value does not vary significantly along the flow, or at least within geometrically similar sections, across the flows being compared used to quantify swirl strength.

Vignat *et al*. [29] observed that $SN_g$, $\mathrm{Eq.}(3)$, remained nearly constant in two distinct regions of a typical swirl combustor: (i) the inlet tube and (ii) a region located downstream and sufficiently away from the expansion plane, with a different value in each region. This behavior is expected because $SN_g$, $\mathrm{Eq.}(3)$, is based on the conceptual basis of the $SN$ proposition without any simplification, discussed earlier. In contrast, simplified swirl number formulations did not exhibit such behavior. $SN_c$, $\mathrm{Eq.}(5)$, was shown in [29] to undergo large non-monotonic spatial variations even in the above two regions, with a strong dependence on local geometrical changes and other flow parameters. Consequently, the values of $SN_c$ measured at a corresponding location in two swirl-combustors may differ significantly even if the corresponding values of $SN_g$ are nearly the same. $SN_p$ $[\mathrm{Eq.}(4)]$, another simplified form of $SN$ performs worse. It was discussed in [29] that since $SN_p$ is obtained by omitting the turbulent fluctuation terms it could often yield negative swirl strengths, a clear physical inconsistency. Therefore, Vignat *et al.* [29] discarded $SN_p$ and it is likewise excluded from the present investigation. Based on a similar argument, Vignat *et al.* [29] reported the generic formulation, $SN_g$ $[\mathrm{Eq.}(3)]$, as the most appropriate measure of swirl strength in isothermal flows.

However, the work in [29] did not resolve the critical values of $SN_g$ for swirl-combustors. Nor did they explore the dependence of its critical value on parameters like $ER$ and $Re_{inlet}$, as addressed by the work in [40] for $SN_s$. The suitability of the $SN_s$ formulation is not explicitly discussed in [29], which is notable given that its critical value for the onset of VB was reported to exhibit only weak dependence on other key parameters in [40]. Further, Vignat *et al.* [29] did not specify how far the second suitable region should be from the expansion plane to measure $SN_g$, region of almost constant $SN_g$, so that its value could appropriately separate out flows with different swirl strength and nor did they clarify its possible dependence on other key parameters. They also did not express any preference between the two suitable regions, inlet tube vs region downstream of expansion plane, where the measured value of $SN_g$ should be used to characterize a flow's swirling strength. We shed some light on this aspect in the present work.

A key observation that supports the use of $SN_g$ as the most appropriate formulation is its property of remaining nearly constant in regions of the flow with constant cross-section and away from flow transitions. Even though $SN_g$ varies slowly in these regions its value is not constant. Hence, resolving a value of $SN_g$ to characterize a flow's swirl strength in long tubes would require further standardization. Our present investigation addresses this aspect, to some extent, and the issues outlined in the preceding paragraph.

Finally, to address the gaps and dilemmas identified in the literature presented above, a strategy based on numerical investigations is adopted in the present work. The isothermal swirl combustor flow in Taamallah *et al.* [20] is simulated by solving appropriately discretized Favre-filtered governing equations using a segregated, pressure-based compressible solver of the OpenFOAM-v10 open-source CFD software. The predicted velocity field for this base case is validated against the corresponding experimental data [20]. Then the validated solver is employed to simulate five additional cases each with a different swirler vane-angle while keeping all the other flow and geometry parameters the same as in the base case [20]. The results are analysed to (1) quantify the critical swirl strength using the generic formulation, $SN_g$ [Eq. (3)] [29], (2) verify the generality of its appropriateness over $SN_c$, Eq. (5), and $SN_s$, Eq. (7), and (3) address a group of related queries discussed two paragraphs earlier. Further, the 3-D structures of VCs in the simulated cases are examined to assess the dependence of VB topologies on swirl strength and to identify the dominant topologies. If we observe a stable double-helix VC, we will investigate a possible mechanism for its formation that also yields a reason for its absence in confined swirling flows with no CB or a perfectly streamlined CB in [11-13]. The critical value of $SN_g$ and the resolved VB topology characteristics may be applicable to any relevant swirl combustor flows. Notably, relevant swirl combustor flows in the present work correspond to swirl combustors with a sudden expansion such that $ER \geq 1.5$, a negligible outlet contraction and operating at $Re_{inlet} \geq 10^4$.

## 2. Geometry and flow conditions

According to the present investigation strategy, we develop a computational model to re-investigate the flow field in [20]. We adopt the computational domain and isothermal flow conditions in [20] as our base case. The domain is a truncated section of a lab-scale swirl combustor facility in the reacting gas dynamics lab at MIT [6, 7, 20]. Figures 1a and 1b show the 3-D geometry and a cross-section of the computational domain. It comprises a 38 mm Ø × 120 mm long tube and a 76 mm Ø × 225 mm long tube connected via an expansion plane. An axial swirler with a sharp-edged center-body and eight vanes is mounted midway in the smaller tube. Figure 1c details the swirler geometry and highlights the key flow structures that typically form inside swirl combustors. The swirl vanes are inclined at $45^0$ to the flow direction, not shown in figure 1c. The thickness and height of swirl vanes are not mentioned in [20] and therefore we estimated them through the scaling of a detailed swirler image in an earlier publication [42], yielding vane thickness and height of 1 mm and 24 mm, respectively.

Air at atmospheric pressure and temperature of $300$ K enters axially into the computational domain through the smaller tube with a mean speed of $8.2$ m/s, yielding $Re_{inlet} = 2 \times 10^4$ based on $\nu_{air@300\,K} = 1.56 \times 10^{-5}$ m$^2$/s. Air then sweeps over the swirler to acquire swirl and subsequently expands into the larger diameter tube. The axial entry is to mimic the effect of choke plate positioned upstream of the computational domain in the experimental setup [20].

A comprehensive model developed in the present work to simulate the above base flow field is validated against the corresponding experimental data in [7, 20]. Using the validated model, we further investigate the flow fields corresponding to various other vane-angles keeping intact other features of swirl combustor geometry and flow conditions.

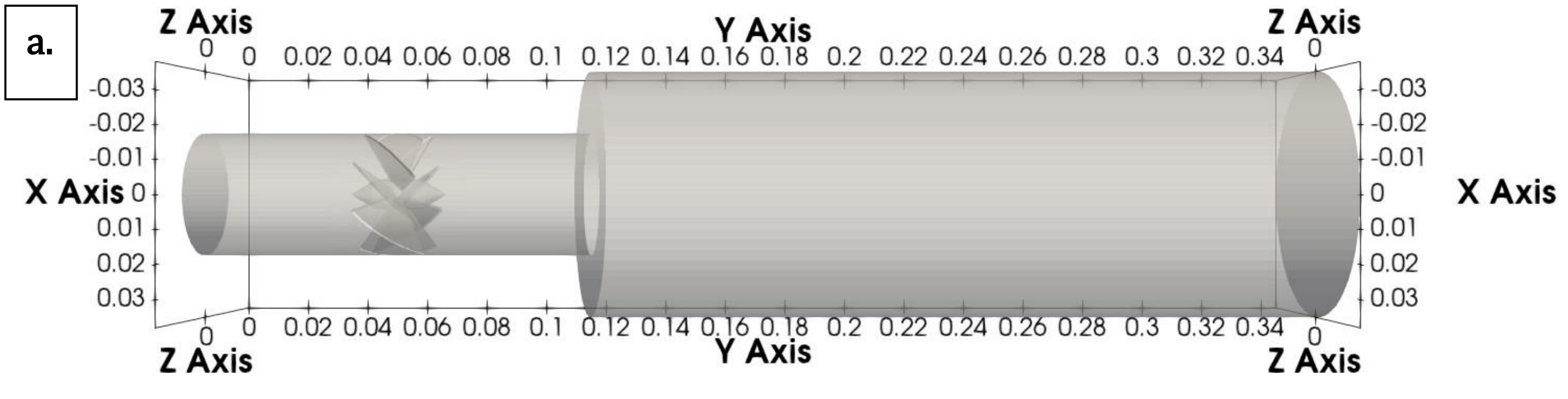

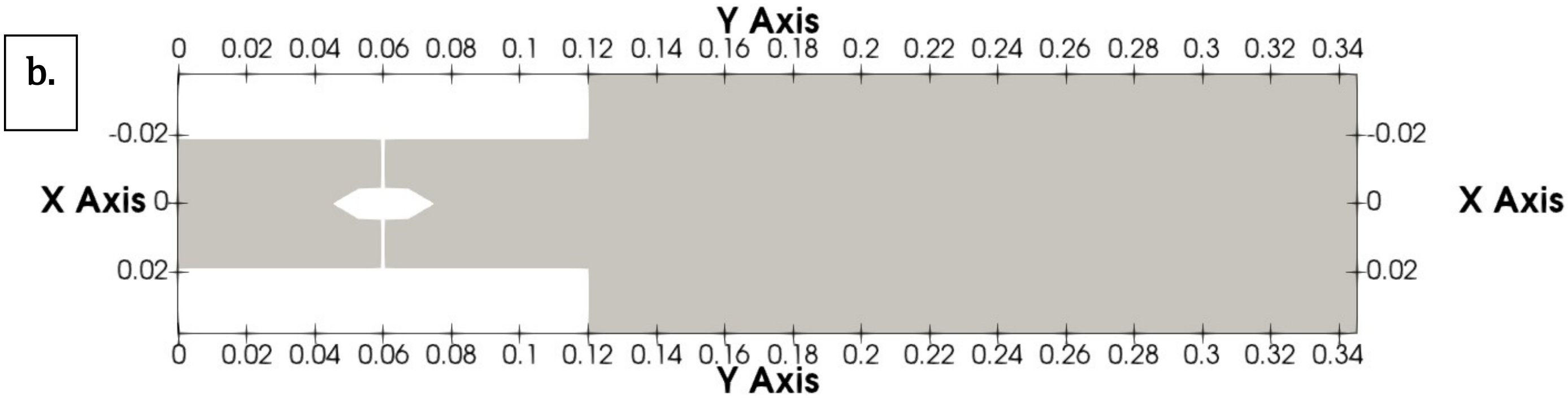

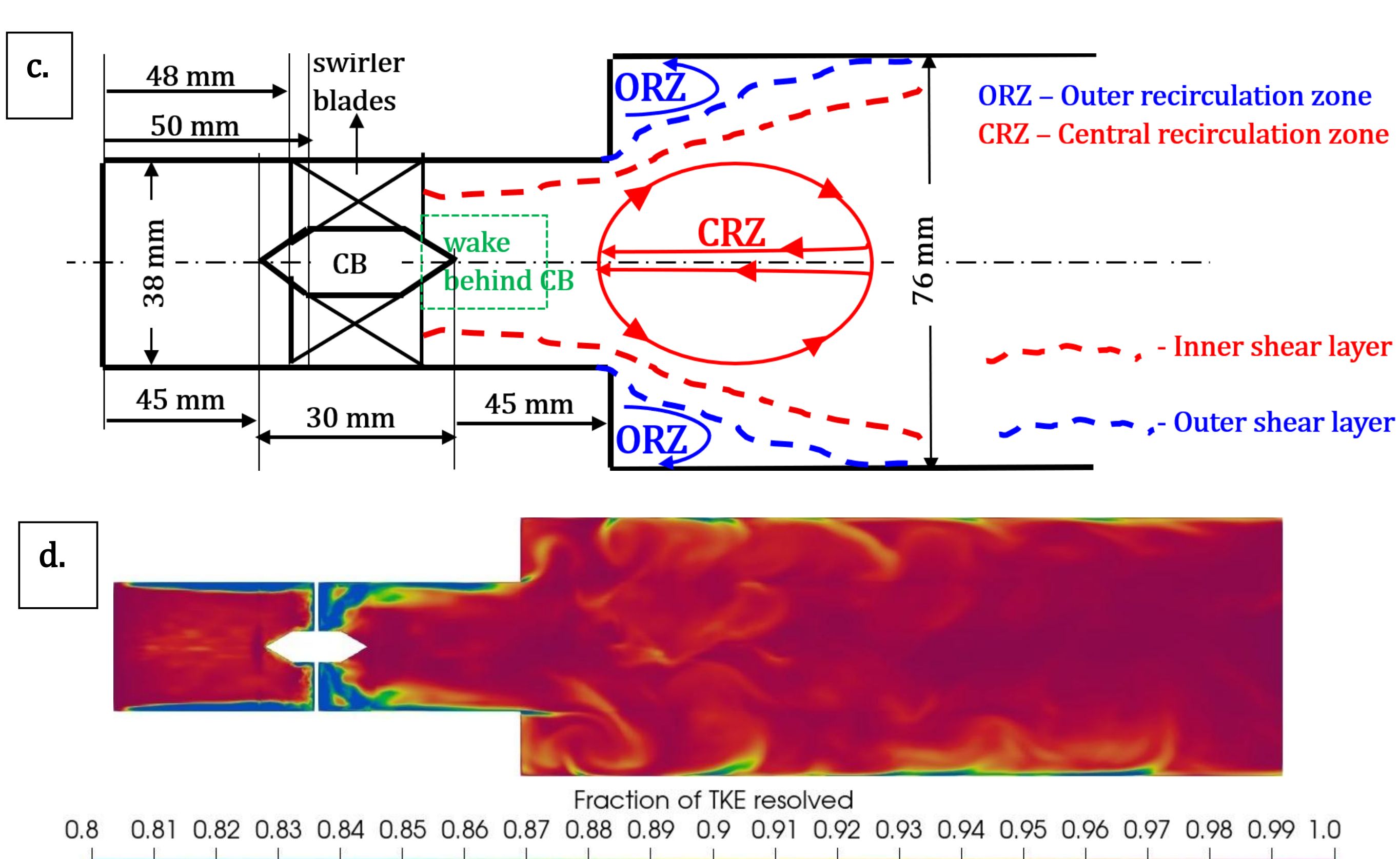

**Figure 1. (a)** Computational geometry of the base case with an expansion plane and a swirler body consisting of a center-body (CB) & eight swirler vanes each inclined at $45^0$ from the flow direction. It is truncated portion of a larger experimental set-up used in [7, 20]. **(b)** A streamwise plane passing from the centreline of the computational domain. Dimensions in **(a)** and **(b)** are in m. **(c)** Schematic of the typical flow structures that form in such geometrical configurations. **(d)** Contours of the fraction of resolved turbulent kinetic energy (TKE).

# 3. Mathematical formulation and numerical solver

## 3.1. Governing equations and boundary conditions

The compressible flow governing equations are solved using the finite-volume method (FVM) for simulations in the present work. Although the inlet Mach number here is low ($Ma = \frac{8.2}{332} \approx 0.025$), we retain the compressible formulation to capture the vortico-acoustic interaction, which is known to influence vortex structures even at such low Mach numbers [34]. $Re_{inlet} = 2 \times 10^4$ across all the cases simulated in the present work. A wide spectrum of length and time scales persist in each of these turbulent flows, where resolving nearly all the scales would incur a prohibitive cost [43]. Therefore, we adopt the LES approach that resolves energetic eddies. Technically, LES resolves scales up to a fraction of the inertial sub-range in turbulence spectrum and the effect of smaller scales on resolved scales is typically modelled [43]. This is achieved in LES by solving spatially filtered governing equations. Spatial filtering removes length-scales smaller than the filter width from flow variables and the corresponding temporal scales, owing to spatio-temporal coupling in turbulence [44]. However, it introduces some residual or sub-filtered quantities that either are modeled or neglected for closure, depending upon the inherent dissipation of the numerical schemes used in the solution of LES equations [37, 43]. We use here OpenFOAM-v10, an open-source CFD software, in which FVM discretization implicitly performs filtering operation with a filter width equivalent to the size of corresponding grid element [37]. Interestingly, the truncation error from lower-order accurate discretization schemes replaces the need for modeling of sub-filtered/sub-grid quantities [37, 43]. However, a second-order accurate or better numerical discretization schemes may be required to enhance numerical accuracy with which sub-grid modeling improves LES predictions [37]. To clarify, a version of *modified* governing equations, i.e., *PDEs* satisfied by the numerical solutions [43], retaining only the leading order truncation errors are given below [6, 45].

$$\frac{\partial \bar{\rho}}{\partial t} + \frac{\partial (\bar{\rho}\, \widetilde{u_i})}{\partial x_i} = 0 \tag{8}$$

$$\frac{\partial \bar{\rho} \widetilde{u_i}}{\partial t} + \frac{\partial \left(\bar{\rho} \widetilde{u_i} \widetilde{u_j}\right)}{\partial x_j} = -\frac{\partial \bar{p}}{\partial x_i} + \frac{\partial}{\partial x_j}\left(\overline{\tau_{ij}} - \tau_{ij}^{\mathrm{D}} - \tau_{ij}^{\mathrm{SGS}}\right) \tag{9}$$

$$\frac{\partial \bar{\rho} \tilde{h}}{\partial t} + \frac{\partial \left(\bar{\rho} \widetilde{u_i} \tilde{h}\right)}{\partial x_i} + \frac{\partial \bar{\rho} \tilde{K}}{\partial t} + \frac{\partial (\bar{\rho} \widetilde{u_i} \tilde{K})}{\partial x_i} - \frac{\partial \bar{p}}{\partial t} = \frac{\partial}{\partial x_i}\left(-\overline{q_i} - H_i^{\mathrm{D}} - h_i^{\mathrm{SGS}}\right) \tag{10}$$

These equations result from the Taylor series expansion of the FVM-discretized governing equations [43], where the time-derivatives in temporal-truncation errors are converted to spatial-derivatives. $\overline{\phi(\boldsymbol{x})}$ represents filtered version of a scalar variable $\phi(\boldsymbol{x})$, $\boldsymbol{x}$ is a location vector. The tilde symbol ($\sim$) denotes density-weighted filtering, also called Favre-filtering, $\widetilde{\phi(\boldsymbol{x})} = \frac{\overline{\rho(\boldsymbol{x})\phi(\boldsymbol{x})}}{\overline{\rho(\boldsymbol{x})}}$. $K = \frac{u_i u_i}{2}$ denotes specific kinetic energy and standard notations are used for other flow variables and fluid properties. $\tau_{ij}^{\mathrm{SGS}}$ and $h_i^{\mathrm{SGS}}$ are the sub-grid terms arising from filtering the convective fluxes of $u$ and $h$, respectively. In the energy equation for sub-sonic flows in OpenFOAM-v10 [34, 45], the analogous sub-grid term of $K$ and the viscous-work term are neglected, likely because they approximately balance each other in such flows [34, 45]. The derivatives of $\tau_{ij}^{\mathrm{D}}$ and $H_i^{\mathrm{D}}$ represent the concerned leading order truncation errors. The filtered fluxes in an isothermal flow can be reasonably modeled as [34, 45]

$$\overline{\tau_{ij}} = 2\bar{\mu}(\widetilde{S_{ij}} - \frac{1}{3}\delta_{ij}\widetilde{S_{kk}}) \qquad \overline{q_i} = -\bar{\rho}\bar{\alpha}\frac{\partial \tilde{h}}{\partial x_i} \tag{11}$$

here $\alpha$ and $k$ of air are treated as a constant in the isothermal flow being simulated and $S_{ij} = \frac{1}{2}(\frac{\partial u_i}{\partial x_j} + \frac{\partial u_j}{\partial x_i})$ is the strain rate tensor. We use the classic k-equation transport model approach in OpenFOAM-v10 for closure of other sub-grid quantities. Here $\tau_{ij}^{\mathrm{SGS}}$ and $h_i^{\mathrm{SGS}}$ are modelled using eddy-viscosity/eddy-diffusivity hypothesis.

$$\tau_{ij}^{\mathrm{SGS}} = \frac{2}{3}\bar{\rho}k_{\mathrm{SGS}}\delta_{ij} - 2\bar{\rho}\nu_{\mathrm{SGS}}(\widetilde{S_{ij}} - \frac{1}{3}\delta_{ij}\widetilde{S_{kk}}) \qquad h_i^{\mathrm{SGS}} = -\bar{\rho}\frac{\nu_{\mathrm{SGS}}}{Pr_{\mathrm{SGS}}}\frac{\partial \tilde{h}}{\partial x_i} \tag{12}$$

$$\tilde{K} = \frac{\widetilde{u_l}\widetilde{u_l}}{2} + k_{\mathrm{SGS}} \tag{13}$$

with sub-grid Prandtl number, $Pr_{\mathrm{SGS}} = 1.0$ [46]. $k_{\mathrm{SGS}} = \frac{1}{2}(\widetilde{u_l u_l} - \widetilde{u_l}\widetilde{u_l})$ is sub-grid kinetic energy and $\nu_{\mathrm{SGS}}$ is sub-grid kinematic viscosity. $\nu_{\mathrm{SGS}}$ is modeled as

$$\nu_{\mathrm{SGS}} = C_K\Delta\sqrt{k_{\mathrm{SGS}}} \tag{14}$$

where, coefficient $C_K = 0.094$ and $\Delta$ is filtered width for which cube-root-of-cell-volume, $\sqrt[3]{V}$, formulation with Van-Driest damping is used since this damping improves predictions in the vicinity of wall [47].

$$\Delta = min(\frac{\kappa y}{C_S}D, \sqrt[3]{V}) \qquad D = 1 - e^{-\frac{y^+}{A^+}} \tag{15}$$

here $D$ is damping factor. $A^+ = 26, \kappa = 0.41, C_S = 0.158$ are model constants [47] and $y$ is the distance from wall. Finally, the transport equation of $k^{\mathrm{SGS}}$ for closure is

$$\frac{\partial \bar{\rho}k_{\mathrm{SGS}}}{\partial t} + \frac{\partial[\bar{\rho}\widetilde{u_i}k_{\mathrm{SGS}}]}{\partial x_i} = \frac{\partial}{\partial x_i}(\bar{\rho}(\nu + \nu_{\mathrm{SGS}})\frac{\partial k_{\mathrm{SGS}}}{\partial x_i}) + -\bar{\rho}\tau_{ij}^{\mathrm{SGS}}:\widetilde{S_{ij}} - \bar{\rho}C_e\frac{{k_{\mathrm{SGS}}}^{3/2}}{\Delta} \tag{16}$$

where $-\bar{\rho}\tau_{ij}^{\mathrm{SGS}}:\widetilde{S_{ij}} = \bar{\rho}[\nu_{\mathrm{SGS}}\{(\frac{\partial \widetilde{u_i}}{\partial x_j})^2 + \frac{\partial \widetilde{u_i}}{\partial x_j}\frac{\partial \widetilde{u_j}}{\partial x_i} - \frac{2}{3}\left(\frac{\partial \widetilde{u_l}}{\partial x_l}\right)^2\} - \frac{2}{3}k_{\mathrm{SGS}}\frac{\partial \widetilde{u_l}}{\partial x_l}]$ models the production of $k^{\mathrm{SGS}}$. The last term in Eq. (16) models the destruction of $k_{\mathrm{SGS}}$ with coefficient $C_e = 1.048$ [47]. Readers are suggested to visit OpenFOAM-v10 documentation [48] and related resources for more details about the model.

The main boundary conditions (b. cs.) for the base-case simulation, consistent with the experimental flow conditions reported in [7, 20], are summarized in table 1.

**Table 1.** Boundary conditions considered in the present study [7, 20].

| $\boldsymbol{B.Cs.}$ / **Patch** | $\bar{p}$ | $\tilde{u}$ | $\tilde{T}$ | $k^{\mathrm{SGS}}$ |
|---|---|---|---|---|
| Inlet | Zero gradient | Turbulent inlet; $<\tilde{u}>_{noise-mean}=$ 8.2 m/s; turbulence intensity = 5% | Fixed value; 300 K | Fixed value; $10^{-5}$ $m^2/s^2$ |
| Outlet | Wave transmissive | Zero gradient | Fixed value; 300 K | Zero gradient |
| Wall (every) | Zero gradient | No slip; 0 m/s | Fixed value; 300 K | Fixed value; 0 $m^2/s^2$ |

In the experimental setup [7, 20], a choke plate was placed upstream of the domain that is computationally investigated in the present work to stabilize the mass flow rate entering the domain against acoustic fluctuations. We apply a velocity b. c. along with a zero-gradient pressure b. c. at the inlet to reproduce this effect numerically. Specifically, *turbulentInlet fvPatch* in OpenFOAM-v10 [49] is selected as the inlet velocity b.c. since it imposes desirable spatio-temporal fluctuations apart from the mean velocity to replicate turbulent characteristics of flow entering the domain. The spatio-temporal fluctuations are generated using a random number generator, which we set to yield a turbulent intensity ($T.I.$) of $5\%$ in the present work. This value of $T.I.$ is based on the empirical correlation $T.I. = \frac{0.16}{Re_{inlet}^{1/8}}$ [50] since turbulent characteristics of flow entering the domain are not reported in related earlier publications [6-8, 20, 42]. Wave transmissive b. c. is employed for pressure at the outlet, following Poinsot and Lele [51], to minimize spurious reflection of wave into the computational domain. At the inlet, a quite small value is assigned to $k^{\mathrm{SGS}}$ rather than zero to prevent the simulations from crashing [52]. The value of $k^{\mathrm{SGS}}$ at the wall is set to zero since $y^+ < 3.5$ in the mesh [53] across all the cases simulated in the present work. More details about the mesh are mentioned in the next section.

## 3.2. Numerical solver

The physical behavior of each term in the governing equations (GEs), $\mathrm{Eqs.}\ (8) - (10)$ and $(16)$ in the present work, categorizes the numerical schemes appropriate for their FVM-discretization. A set of appropriate schemes are used for the collocated FVM-discretization in the present work, where each scheme is at least second order accurate. The use of higher order, sixth- or seventh-order, reconstructions is restricted from the employment of tetrahedral grid elements around the swirler to fully capture geometrical details of the swirler in the present computational mesh [54]. The temporal terms are discretized using the *pure Crank-Nicolson* scheme. The convective terms are discretized using the *van-Leer* interpolation scheme, owing to its TVD behavior that minimizes numerical dispersion arising from spatial-truncation errors. Gradients at cell nodes are discretized using the *cell-limited least-squares* scheme. The *cell-limited* aspect ensures bounded interpolation of any variable to neighboring faces. The gradient of enthalpy at cell-faces in the FV integration of effective heat flux term, $\text{Eqs.}\,(11)$ and $(12)$, is discretized using a *central-difference* scheme. The scheme is augmented by a non-orthogonal correction, the *snGrad* scheme, in the regions of computational mesh with skewed elements. The mesh in the present case has a skewness below $0.8$, and therefore only a limited correction is applied. The strain rate at cell-faces in the FV integration of effective stress tensor, $\text{Eqs.}\,(11)$ and $(12)$, is interpolated to adjacent cell nodes using a *linear* interpolation scheme. These strain rates are then discretized using the predefined gradient scheme, here it is the *cell-limited least-squares* scheme. Mass flux and fluid properties are interpolated to cell nodes using a *cubic* scheme and the *Rhie-chow* correction is applied to avoid pressure-velocity decoupling on collocated grid arrangement used in OpenFOAM-v10.

The resulting algebraic equations are solved using the rhoPimpleFoam solver, which is an integration of the PIMPLE algorithm with the iterative linear solvers appropriate for the discretized GEs. PIMPLE is a pressure-based, transient, segregated algorithm suitable for compressible flows with $Ma \leq 1$ [55]. It combines the outer corrector loop structure of the SIMPLE algorithm [55] with inner PISO-style pressure–velocity corrector loops, so that the pressure–velocity coupling can be iterated within each time step [55]. In the present work, five outer SIMPLE-like corrector iterations (PIMPLE outer loop passes) are used; within each outer corrector, the energy and turbulence equations are solved, and the momentum equation is advanced (momentum predictor), with thermodynamic variables (and density) updated accordingly. Then, two inner PISO-like corrector iterations (PISO loop passes) are performed, each consisting of a pressure equation solve followed by flux correction and velocity (momentum) correction. A non-orthogonality correction loop is additionally applied within the pressure solution. Please refer to OpenFOAM-v10 documentation [48] and [55] for more details about the solver. All the equations are solved with full relaxation (relaxation factor = $1.0$) in the present work.

Usually, the above comprehensive solver, the solver along with numerical schemes, remains stable even if the maximum Courant number, $Co_{max}$, exceeds $1.0$ [55]. However, enforcing $Co_{max} < 1.0$ enhances temporal accuracy and boundedness of the solution, especially for boundedness with second and higher-order temporal schemes [56]. The timestep is accordingly constrained in the present work and more about it is discussed in the next section.

# 4. Computational grid, grid resolution and solver validation

Next, we describe the spatial mesh and timestep used to discretize the GEs in the present simulations. We use hexahedral mesh elements throughout the investigated computational domains, except in their swirlers—where tetrahedral elements are used to capture the geometric details. The mesh size and timestep are constrained by the LES requirement of resolving at least $80\%$ of turbulent kinetic energy (TKE) locally [43]

$$Fraction\ of\ TKE_{resolved} = \frac{TKE_{resolved}}{TKE_{resolved} + k_{SGS}} \geq 0.80 \quad (17)$$

here $TKE_{resolved} = \frac{\langle \tilde{u}_i \tilde{u}_i \rangle - \langle \tilde{u}_i \rangle \langle \tilde{u}_i \rangle}{2}$ and $\langle \cdot \rangle$ denotes temporal mean.

Pope's criterion [43] for LES in isotropic turbulence stipulates resolving eddies larger than one-sixth of the integral length scale $l_0$. It is commonly applied for $R_\lambda \lesssim 1000$, where $R_\lambda$ is $Re$ based on the Taylor micro-scale. $R_\lambda \approx \sqrt{2Re_{inlet}}$ from [43] yields $R_\lambda \approx 200$. Considering $l_0 \approx r_{inlet} = 19$ mm in the present work yields a cut-off eddy size $(l_0/6) \approx 3.2$ mm. To represent eddies of size $O(l_0/6)$ at least 2 cells are required across this scale, i.e., a mesh size $\Delta \lesssim (l_0/6)/2 \approx 1.6$mm**,** consistent with [43]. Accordingly, we design a mesh with the maximum cell edge length limited to 1.2 mm and with additional refinements around the regions highlighted in figure 1c to resolve the shear layers and the wakes downstream of the swirler's center-body. This baseline mesh ($M1$) for each investigated case is expected to satisfy Pope's LES resolution criterion ($Fraction\ of\ TKE_{resolved} \geq 0.80$). The selection of maximum edge length as 1.2 mm instead of 1.6 mm is a conservative approach as the entire flow field is not expected to be isotropic. $M1$ has refinement near the wall with a first node height of 0.04 mm, resulting in $y^+ < 3.0$ in the base case and below 3.5 across all the investigated cases. Typically, $y^+ \leq 5.0$ with lower computational requirement is reasonable in LES for engineering applications [29, 53]. We do not attempt to resolve the region around swirler vanes owing to prohibitive computational costs. The smallest mesh element, of size 0.034 mm, appears in the swirler on $M1$. $M1$ for each swirl-combustor case investigated in the present work contains approximately 1.7 million mesh elements. The overall metrics of $M1$ computational grids are minimum orthogonality = 0.18, maximum aspect ratio $\approx 30$ and maximum skewness = 0.8.

Time-step, $\Delta t$, selection is guided by the cut-off eddy time scale, which we estimate using the correlations in [43] to be $\mathcal{O}(1\text{ms})$. However, finer structures may appear in the shear layer and the wake of the swirler's CB, motivating a smaller time step on the order of $0.05$ ms. Monitoring $Co_{max}$ shows that $\Delta t = 0.05$ ms yields $Co_{max} > 1.0$ and reducing $\Delta t$ to $2.50$ $\mu$s brought $Co_{max}$ below $0.9$ in all simulations performed in the present work.

The scaled residuals of all variables dropped below $10^{-6}$ every timestep, before or in the fifth outer-corrector loop across all the simulations in the present work. The convergence at each timestep is ensured by monitoring the variation of pressure and axial-velocity with iterations at two points within the computational domain. The flow-through time is approximately $0.120$ s in the investigated cases. Mean flow fields are obtained by averaging over ten flow-through times, discarding the first three flow-through times.

Figure 1d shows a contour of the fraction of TKE resolved for $M1$-based simulation in the base case, confirming more than $80\%$ TKE resolution in the regions of interest; analogous contours for the remaining vane-angle cases are provided in subsequent sections. To further assess grid sensitivity in the present implicit-filter LES, we monitor three quantities of interest (QoIs) representing the mean-flow VB state and the dominant coherent dynamics: (i) the CRZ length, $L_{CRZ}$, defined as the axial extent of the contiguous region on the combustor centerline where the time-mean axial velocity is negative, (ii) the precession frequency, $f_p$, since precession was observed experimentally in the base case in [7, 20] and (iii) the maximum resolved turbulent kinetic energy, $TKE_{resolved,max}$, (a second-order turbulence statistic) which typically occurs in the shear layer around CRZ.

The precession frequency, $f_p$, is obtained via Fast Fourier Transform (FFT) of the time series of x-velocity fluctuations; details are provided in the *Validation* section of the present paper. The refinement modifies the effective numerical filter in implicit LES; therefore, strict invariance of the full turbulence field is not expected. However, once the mesh satisfies the intended LES resolution level ($Fraction\ of\ TKE_{resolved} \geq 0.80$), these large-scale QoIs are expected to exhibit weak sensitivity to further refinement.

Accordingly, a grid-sensitivity study is performed for the base case using three systematically refined meshes: the baseline mesh ($M1$, 1.7 million elements), designed to satisfy $Fraction\ of\ TKE_{resolved} \geq 0.80$, and two refined meshes obtained by applying a uniform refinement factor of 1.11 in each spatial direction, thereby yielding $M2$ with 2.4 million elements and $M3$ with 3.3 million elements. The resulting values of $L_{CRZ}$, $f_p$, and $TKE_{resolved,max}$ are listed in table 2, showing insignificant variation with refinement, consistent with the weak sensitivity of QoIs once the large-scale structures are well-resolved. A time-step sensitivity check is additionally performed on $M1$ by halving the time step from 2.50 $\mu$s to 1.25 $\mu$s and the corresponding QoIs are listed in table 3, again indicating negligible sensitivity to time-step refinement.

Owing to computational limitations, the above grid/time-step sensitivity study is performed only for the base case. The remaining cases differ only in swirler vane-angles (swirler geometry), while the combustor geometry, operating conditions, and target LES resolution criteria are unchanged. Hence, $\Delta t = 2.5\ \mu$s and the mesh sizing with refinement strategy used to construct $M1$ are retained for the full parametric set, and analogous contours of the fraction of resolved TKE (as in figure 1(d)) are presented for each case in the subsequent sections.

**Table 2.** Grid sensitivity of key QoIs in the base case over meshes with refinement ratio of 1.11.

| Mesh element count | $L_{CRZ}$ (m) | $f_p$ (Hz) | $TKE_{resolved,max}$ (m$^2$/s$^2$) |
|---|---|---|---|
| $M1$ (1.7 million) | 0.2250 | 130.0 | 29.045 |
| $M2$ (2.4 million) | 0.2254 | 127.5 | 29.324 |
| $M3$ (3.3 million) | 0.2254 | 127.5 | 29.419 |

**Table 3.** Time-step sensitivity of key QoIs in the base case using mesh $M1$ (1.7 million).

| Time step | $L_{CRZ}$ (m) | $f_p$ (Hz) | $TKE_{resolved,max}$ (m$^2$/s$^2$) |
|---|---|---|---|
| $\Delta t = 2.5\ \mu s$ | 0.2250 | 130.0 | 29.045 |
| $\Delta t = 1.25\ \mu s$ | 0.2253 | 130.0 | 29.191 |

## 5. Validation

To validate the developed comprehensive computational model, we first compare its time-averaged predictions of the normalized axial-, radial- and tangential-velocity components ($U_{axial}$, $U_{radial}$ and $U_\theta$) on the $M1$ mesh for the base case with the corresponding experimental data reported in [7, 20]. Figure 2 presents the radial distributions of these velocity components at four streamwise locations, where $y$ denotes the distance downstream from the expansion plane. The present solver reproduces these radial-profiles with reasonable accuracy. Notably, it correctly captures the location of the upstream stagnation point of the IRZ in the mean flow-field, i.e., it accurately predicts the mean axial-velocity on the centerline at $y/R = 0$. This value was ill-predicted in simulation results of [20]. The experiments in [20] also report the presence of a precessing vortex core (PVC) with a frequency of $120$ Hz [7] in the investigated isothermal flow, which is adopted as the base case in the present work. To determine the corresponding precession frequency in our simulated base flow, we perform a Fast Fourier Transform (FFT) on the time-series of the x-velocity fluctuations. The velocity fluctuations are recorded for more than three flow-through times, sampled every 250$^{th}$ timestep, at a point on the domain axis located 5 mm upstream of the expansion plane. The resulting frequency spectrum in figure 3 exhibits a single sharp peak at $130$ Hz, yielding a Strouhal number $St = \frac{f \times d_{inlet}}{U_{inlet}} = 0.60$. This value is an acceptable deviation from the experimental result, since certain geometrical and flow details are not reported in [7, 20]. The rationale for choosing this probe location will be evident in the *Results and Discussion* section of the present work.

Overall, the developed computational model reproduces both the mean flow field and its dominant unsteady dynamics with good fidelity in the base case. Having established this level of accuracy, we proceed with our computational investigation using the validated solver.

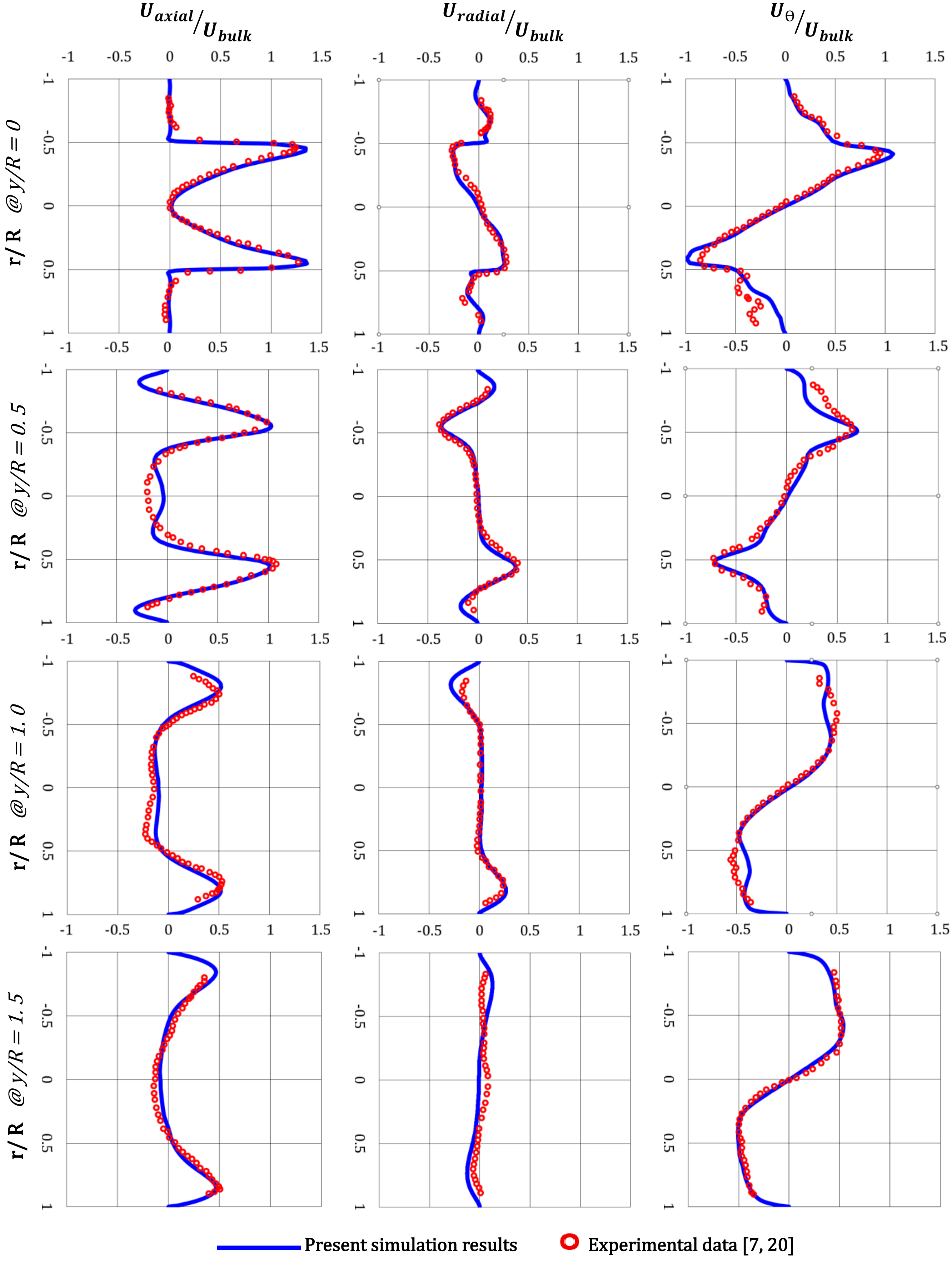

**Figure 2.** Radial plots comparing the experimental data [7, 20] (red circle) with numerical results from the present work (blue line) for the base case. **y/R** is the normalized location from the expansion plane and **R** is the radius before the expansion plane.

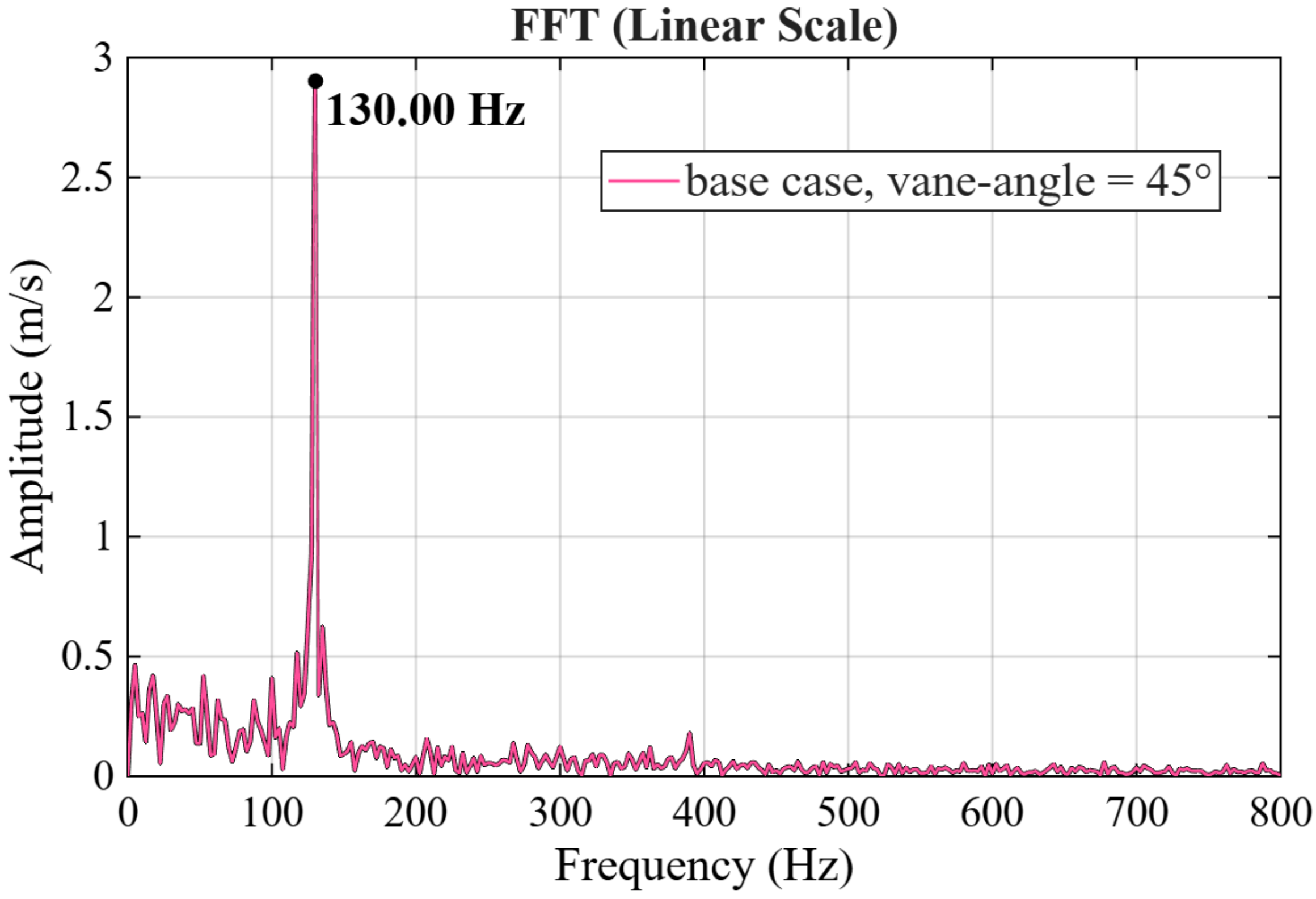

**Figure 3.** Fast Fourier Transform of the x-velocity fluctuations at the geometric center of the cross-stream plane located 0.115 m downstream from the inlet, i.e., 5 mm upstream of the expansion plane.

## 6. Results and Discussions

Apart from the base case that has a vane-angle of $45^0$, used in the validation studies, we simulate five additional cases with vane-angles of $17^0, 25^0, 40^0, 50^0$ and $60^0$ using the validated solver developed in the present work. All flow and other geometric parameters in these cases are identical to those in the base case, which are described in section 2 of the present work.

We assess the onset of VB in the investigated swirl combustor in section 6.1 of the present work to quantify the corresponding critical swirl strength in terms of $SN_g$ [Eq. (3)]. The onset of VB is identified by the disruption in coherency of a VC structure or by the appearance of IRZ. We inspect only the slices of IRZ on a plane in the present work, rather than its full 3-D structure. These slices correspond to regions of non-positive axial-velocity in the core of a flow bounded by iso-contours of zero axial-velocity. These iso-contours are marked as black colour curves to easily locate the slices of IRZ in the present work [57]. IRZ typically start forming at very low swirl strengths but exhibit high spatio-temporal intermittency until a critical swirl strength and so does not appear in the corresponding mean flow fields [58], which is problematic for flame-stabilization. Hence, to identify flows in which VB possesses weak intermittency, stable-VB state [58], it is common to investigate IRZ formation in mean flow fields [58]. Accordingly, in section 6.1 of the present work, we determine the critical swirl strength by locating the first appearance of IRZ in the mean flow field. Next, we examine the degree of generality by which $SN_g$ is more appropriate in representing swirl strength, where we compare the axial variation of $SN_g$ with other prominent $SN$ formulations in the present investigated cases. Further, we address a group of queries on $SN_g$ specified in the *Introduction* section of the present paper by assessing solely the axial variation of $SN_g$ in the investigated cases from different perspectives. In section 6.2 of the present work, we filter out the investigated cases in which the IRZ forms but possess a high degree of spatio-temporal intermittency by investigating instantaneous flow fields.

Next, we resolve in section 6.3 of the present work the VC topologies and their dynamics in the investigated cases. The VC in a swirl combustor is commonly visualized using a suitable valued $Q$-criterion iso-surface that coherently extends from the swirler through regions of elevated streamwise vorticity [20, 32, 38]. $Q$ represents the second invariant of the velocity gradient tensor, Eq. (18) [38, 59]. Its appropriate value is case dependent.

$$Q = \frac{(trace(\nabla\vec{u}))^2 - trace((\nabla\vec{u})^2)}{2} \tag{18}$$

Another commonly used quantity for vortex core visualization is the $\lambda_2$- criterion, where $\lambda_2$ is the second largest eigen value of the tensor $S^2 + \Omega^2$ with $S$ and $\Omega$ denoting the strain-rate and rotation-rate tensors [20]. This criterion has been reported to provide comparable or slightly improved accuracy relative to the $Q$-criterion [38]. However, the $Q$-criterion remains more popular for swirling flows and in general [32, 59]. To strengthen our choice, both criteria were applied to a randomly selected instant in each investigated case, and the resulting iso-surfaces were found to be nearly identical for suitably chosen values. Hence, further use of the $\lambda_2$-criterion was deemed unnecessary for the present study. Here, we first visualize the VC structures at a few instances in our present work using the $Q$-criterion. However, the structures may decay or evolve over time, and therefore we examine the footprints left by the coherent dynamics of the VC over several timesteps to gain a preliminary assessment of the prevailing VC topologies. Finally, spectral methods are employed to identify the dominant VC topologies and the temporal evolution of their dynamics in greater detail.

## Spectral approach

We first perform a spatial cross-spectral analysis of axial-velocity fluctuations recorded over more than three flow-through times at two diametrically opposite locations on the axial-velocity footprints of VC's dynamics, as in [58, 60]. Let these two simultaneous time-series of axial-velocity fluctuations be denoted as $X(t)$ and $Y(t)$. The two point cross-spectral analysis isolates frequencies at which oscillations are phase-coherent between the two time-series and quantifies their correlated energy [58, 60]. The data in each time-series is recorded at the sampling frequency ($f_s$) of $1600$ Hz in the present work. Each time-series for the two-point cross-spectral analysis is split into $9$ overlapping ensembles, $N_e$, of $128$ samples each, $N_{seg}$, with $50\%$ overlap between consecutive ensembles and a Hann window is applied on each ensemble to suppress spectral leakage around the peaks [60]. Accordingly, the frequency bin, discrete frequency slot, spacing, $\Delta f = \frac{f_s}{N_{seg}}$, in the investigated cases is $12.5$ Hz. We did not increase $N_{seg}$ further to sharpen the peaks in the present spectral results due to resource constraints. However, we ensured lower variance in the present results with a reasonable number of ensembles [60]. The Hann window in discrete time is $w[n] = \frac{1}{2}\left(1 - cos\left(\frac{2\pi n}{(N_{seg}-1)}\right)\right)$, $0 \leq n \leq N_{seg} - 1$. It reduces sidelobes, small leakage lobes next to a main peak, at the expense of a modest widening of the main lobe. Hence, peaks in some of the spectral figures in section 6.3 of the present work appear slightly wider but with much lower leakage into neighboring bins. The cross-spectrum, $S_{XY}(f)$, is evaluated by Fourier transforms (DFTs computed using FFT algorithm) of $X(t)$ and $Y(t)$ on each overlapped ensemble, forming the ensemble-level cross-periodogram, $X_r(f)Y_r^*(f)$, and averaging these over all the ensembles $r$, see Eq. (19).

$$S_{XY}(f) = \frac{1}{N_e}\sum_{r=1}^{N_e} X_r(f)\, Y_r^*(f) \tag{19}$$

here $Y_r^*(f)$ is the complex-conjugate of $Y_r(f)$. The magnitude-squared coherence, $\gamma_{XY}^2(f)$, in Eq. (20) is a normalized measure of narrow-band phase consistency between the two time-series at any frequency $f$.

$$\gamma_{XY}^2(f) = \frac{|S_{XY}(f)|^2}{S_{XX}(f)S_{YY}(f)} \tag{20}$$

here $|S_{XY}(f)|$ is the magnitude of $S_{XY}(f)$. It reflects the corresponding correlated energy. A threshold of $\gamma^2$ is widely used to sort out the coherent oscillations from the noise. Manoharan *et al.* [58] used it for the purpose in swirling jets where it performed well. Further, we quantify the azimuthal mode number, $m$, [58] associated with the resolved coherent oscillations to identify their azimuthal symmetry or the shape of the associated flow structure. The structure associated with these coherent oscillations in our setting would ideally be the VC since the probes of $X(t)$ and $Y(t)$ are positioned in the footprints of VC's coherent dynamics. Note that $m = 0$ corresponds to axisymmetric oscillations, $|m| = 1$ to single-helical oscillations and $m = 2$ to double-helical oscillations. The azimuthal mode index ($m$) can be estimated from the unwrapped cross-phase, $\Phi_{XY}(f) = arg\{S_{XY}\}$, via $m = \frac{\Phi_{12}(f)}{\Delta\theta}$, as in [58]. Here $\Delta\theta$ is the azimuthal separation between the probes and $\Delta\theta = \pi$ radian in the present case. Note that any two-point phase cannot distinguish $m = 0$ from $m = 2$ [58] and since VC structures up to $m = 2$ exist in swirling flows [9], we adopt the four-probe azimuthal FFT analysis described next [61].

Azimuthal symmetry of coherent oscillations is resolved using the four simultaneous time-series, $X_k(t)$ for $k = 1$ to $4$, collected at the four probes, each located at a different azimuthal angle $\theta_k \in \{0, \frac{\pi}{2}, \pi, \frac{3}{2}\pi\}$ radian but at the same radius. Two of these four locations are the same as in the two-point cross-spectral analysis. We use the same framing as in the two-point cross spectral analysis; 9 ensembles, $N_{seg} = 128$, 50% overlap and Hann window; for splitting each time-series. We form short-time spectra, i.e., a windowed FFT on each overlapping ensemble for every probe. The FFT coefficient for each pair of frequency bin, $f$, and ensemble index, $r$, are stacked into a column vector $z_r(f)$ shown in Eq. (21).

$$z_r(f) = [Z_1(f,r), Z_2(f,r), Z_3(f,r), Z_4(f,r)]^{\mathrm{T}} \tag{21}$$

Then, a $4 \times 4$ cross-spectral matrix, $S(f)$, is obtained by averaging the ensemble-level outer product of $z_r(f)$ with itself, given in Eq. (22). $S(f)$ distributes energy across azimuthal pattern at the frequency $f$ through quadratic form in Eq. (22).

$$S(f) =< z_r(f) z_r(f)^{\dagger} >_r \tag{22}$$

where $(.)^{\dagger}$ denotes conjugate transpose and $<.>_r$ denotes averaging over ensemble $r$. The projection of $S(f)$ onto the discrete azimuthal basis, $\{v_m\}$, gives the modal energy distribution, $E(m,f) = Re\{{v_m}^{\dagger} S(f) v_m\}$. Here $v_m$ is a complex column vector, given in Eq. (23), that represents ideal $m$-lobed patterns sampled at the four probe angles. Accordingly, the modal energy fraction, $F(m,f)$, is as given in Eq. (24).

$$v_m = \frac{1}{4}[e^{-jm\theta_1}, e^{-jm\theta_2}, e^{-jm\theta_3}, e^{-jm\theta_4}]^{\mathrm{T}} \tag{23}$$

$$F(m,f) = \frac{E(m,f)}{\sum_{m \in \{0,\pm 1,2\}} E(m,f)} \tag{24}$$

$F(m,f)$ identifies the energy distribution across modes at any frequency and so can be used to sort out the symmetry or flow structure that dominates at the frequencies where coherence, $\gamma^2$, is beyond a threshold value, coherent frequencies. We use $\gamma_{XY}^2(f) \geq 0.7$ to identify coherent frequencies and $F(m, f)$ at those frequencies indicates the dominant symmetry.

Next, we examine the temporal evolution of the coherent modes in the time domain for a selected pair $(m_0, f_0)$. We first form the modal time-series, $c_m(t) = Re\{{v_m}^{\dagger} x(t)\}$, where $x(t) = [X_1(t), X_2(t), X_3(t), X_4(t)]^{\mathrm{T}}$ and then use the Hilbert approach without segmenting the time-series. Specifically, we isolate the oscillation about the selected $f_0$ by applying a narrow-zero-phase band-pass $f_0 \pm 10.0$ Hz, a small band is a contiguous frequency interval, to get $\widetilde{c_m}(t)$.

$$\widetilde{c_m}(t) = \mathcal{B}_{f_0,\,10.0}\{c_m(t)\} \overset{\text{def}}{=} \mathcal{F}^{-1}\{\mathcal{B}_{f_0,10.0}(f)\, C_m(f)\} \tag{25}$$

where $C_m(f) = \mathcal{F}\{c_m(t)\}$ with passband $[f_0 - 10.0\ \mathrm{Hz}, f_0 + 10.0\ \mathrm{Hz}]$. In practice we realize $\mathcal{B}_{f_0,\Delta f}$ as a 4th order Butterworth band-pass in SOS, second-order sections, form with forward–backward filtering, zero phase. The analytic signal is then $a_m(t) = \widetilde{c_m}(t) + j\,\mathcal{H}\{\widetilde{c_m}(t)\}$ where $\mathcal{H}\{.\}$ is the Hilbert transform. The magnitude of $a_m(t)$ is the modal narrowband amplitude (envelope), $|a_m(t)|$. The band-passed modal signal $\widetilde{c_m}(t)$, oscillatory trace, and the envelope $\pm|a_m(t)|$ are plotted for the chosen azimuthal mode at the chosen frequency. The envelope $|a_m(t)|$ shows the time-varying amplitude of $(m_0, f_0)$ at the native sampling resolution without ensemble averaging, while $\widetilde{c_m}(t)$ depicts the oscillation itself at the frequency. The band-passed modal signal is sensitive to end effects caused by filtering. To mitigate this issue, we apply reflection padding to the modal scalar $c_m(t)$ and a Tukey taper to the padded $c_m(t)$ prior to filtering, then compute and apply a frequency-aware internal validity mask to the band-passed analytic signal $a_m(t)$, and finally trim 30 ms from each end of $\tilde{c}_m(t)$ and $|a_m(t)|$ for presentation.

Furthermore, coherent oscillations with distinct $m$ may be simultaneously present and interact. To assess this behavior, we compute the squared cross-bicoherence $XX \rightarrow Y$ as in [62] on triads $(f_a, f_b, f_c)$ with $f_c = f_a \pm f_b$ using the same two-point time-series in cross-spectral analysis [58, 60]. The purpose here is to test whether two frequency bands in $X(t)$ combine through quadratic coupling to populate a third frequency band observed in $Y(t)$. This helps decide whether apparently separate spectral lines should be interpreted as mutually dependent rather than independent. We report only the squared cross-bicoherence corresponding to the permutation $X_{f_a}$and $X_{f_2}$ influencing $Y_{f_c}(t)$ in the frequency-domain $b^2(X_{f_a}X_{f_b} \rightarrow Y_{f_c})$, since the squared cross-bicoherence for the reverse permutation $b^2(Y_{f_a}Y_{f_b} \rightarrow X_{f_c})$ showed similar values. The $b^2(X_{f_a}, \sigma X_{f_b} \rightarrow Y_{f_a+f_b})$ is defined as

$$b^2(X_{f_a}, \sigma X_{f_b} \rightarrow Y_{f_a+f_b}) = \frac{|\langle X(f_a)X(\sigma f_b)Y^*(f_a+\sigma f_b)\rangle|^2}{\langle |X(f_a)X(\sigma f_b)|^2\rangle \langle |Y(f_a+\sigma f_b)|^2\rangle} \quad (26)$$

here $\sigma$ can be $+1$ and $-1$. Two frequency bands taken from $X(t)$ act as predictors and their effect is measured on the third band in $Y(t)$. We evaluate the strength of quadratic interaction on both sum and difference triads and display them together on a signed frequency layout. The $f_2$-axis is drawn positive for sum triads ($f_a$ , $f_b$ , $f_a + f_b$ ) and negative for difference triads ($f_a$, $f_b$ → $|f_a - f_b|$).

Before proceeding to sections 6.1–6.3, we inspect the contours of the resolved TKE fraction in figure 4 for all investigated cases on their respective baseline meshes (M1). The results confirm more than $80\%$ resolution in the regions of interest, excluding the immediate swirler vicinity and near-wall zones. Such resolution is considered appropriate in LES [63, 64] and has been shown to yield accurate results [63, 64]. Given these precedents and the validation of our numerical model in section 5 of the present work, we expect the simulations to adequately resolve the mean flow as well as the topology and dynamics of coherent structures in the investigated cases.

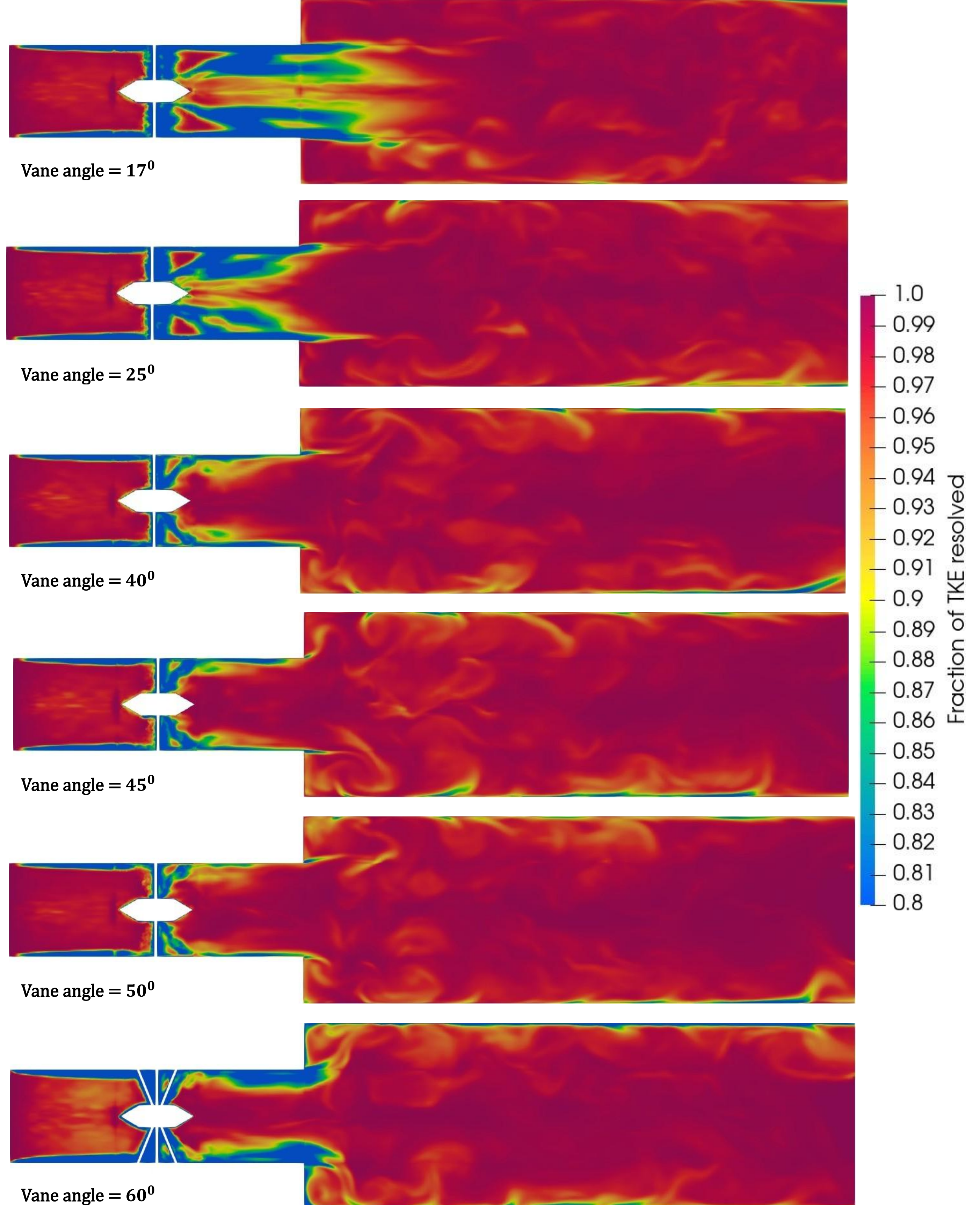

**Figure 4.** Contours of the fraction of resolved turbulent kinetic energy (TKE) on the streamwise plane through the centreline of the computational domain for the cases with different swirler vane-angles. The resolved fraction is mostly greater than 0.8 in the core of the flow.

## 6.1. Onset of VB and most-suitable swirl number formulations

Figure 5 presents the contours of mean axial-velocity for the investigated cases with black color curves in the core of the domain marking the boundary(ies) of the IRZ. The presence of IRZ in a mean flow indicates the occurrence of a stable-VB [57, 58]. It is evident from figure 5 that a stable-VB occurs in the investigated swirl combustor when the vane-angle equals or exceeds $25^0$. Accordingly, the onset of stable-VB in the investigated swirl combustor seems to occur for a vane-angle between $17^0$ and $25^0$. The IRZ spans across the larger-diameter section of the combustor for the vane-angle of $45^0$ and the entire combustor length downstream of the swirler for the vane-angle of $60^0$.

Next, we analyze the streamwise variations of three swirl number formulations, $SN_g$, $SN_c$ and $SN_s$ $[\text{Eqs.} (3), (5) \& (7)]$ for the investigated cases shown in figure 6. The aim is to assess which formulation best represents swirl strength. The reference pressure, $p_{ref}$, required to evaluate $SN_g$, $\text{Eq.} (3)$, is selected based on the guidelines in [29]. Recall that an appropriate $SN$ formulation should remain constant when viscous dissipation is negligible. Otherwise, it should exhibit a net decay along the flow, with weak variation rates, in a region of fixed cross-section away from flow transitions. Figure 6 shows that $SN_g$ is nearly flat in the inlet tube for all the investigated cases. $SN_g$ decays monotonically in the region downstream of the expansion plane; the decay rate is small sufficiently far downstream of the expansion and seems to become even smaller if the combustor length downstream of the expansion plane is increased. In contrast, the values of $SN_c$ and $SN_s$ vary significantly within each region of constant cross-section away from the expansion plane, with $SN_c$ even showing non-monotonic variations. Their variation rates do not appear to reduce appreciably with distance from the expansion plane.

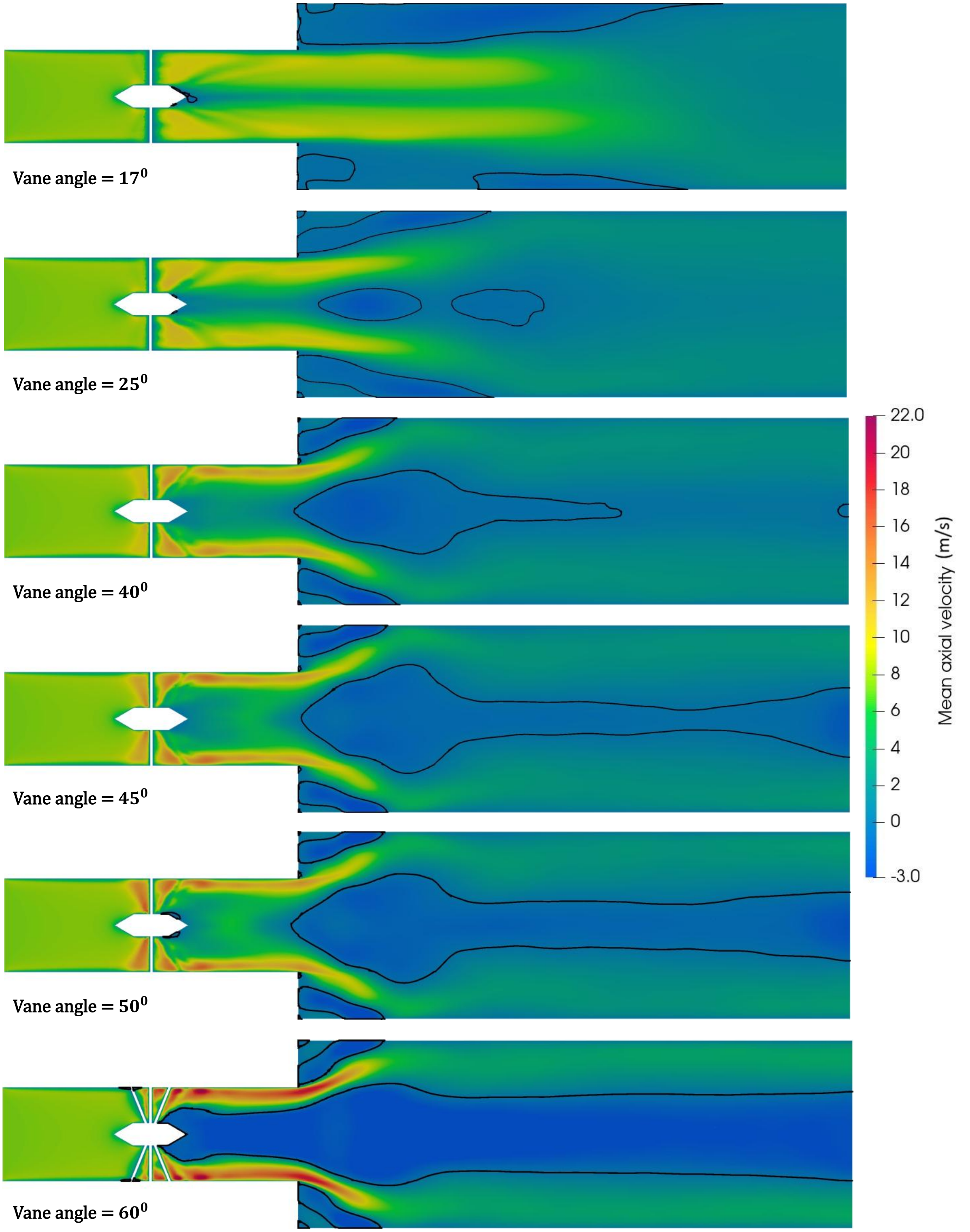

**Figure 5.** Contours of mean axial velocity (m/s) on the streamwise plane through the centreline of the computational domain for the cases with different swirler vane-angles. Black colour curves are isocurves of zero axial velocity.

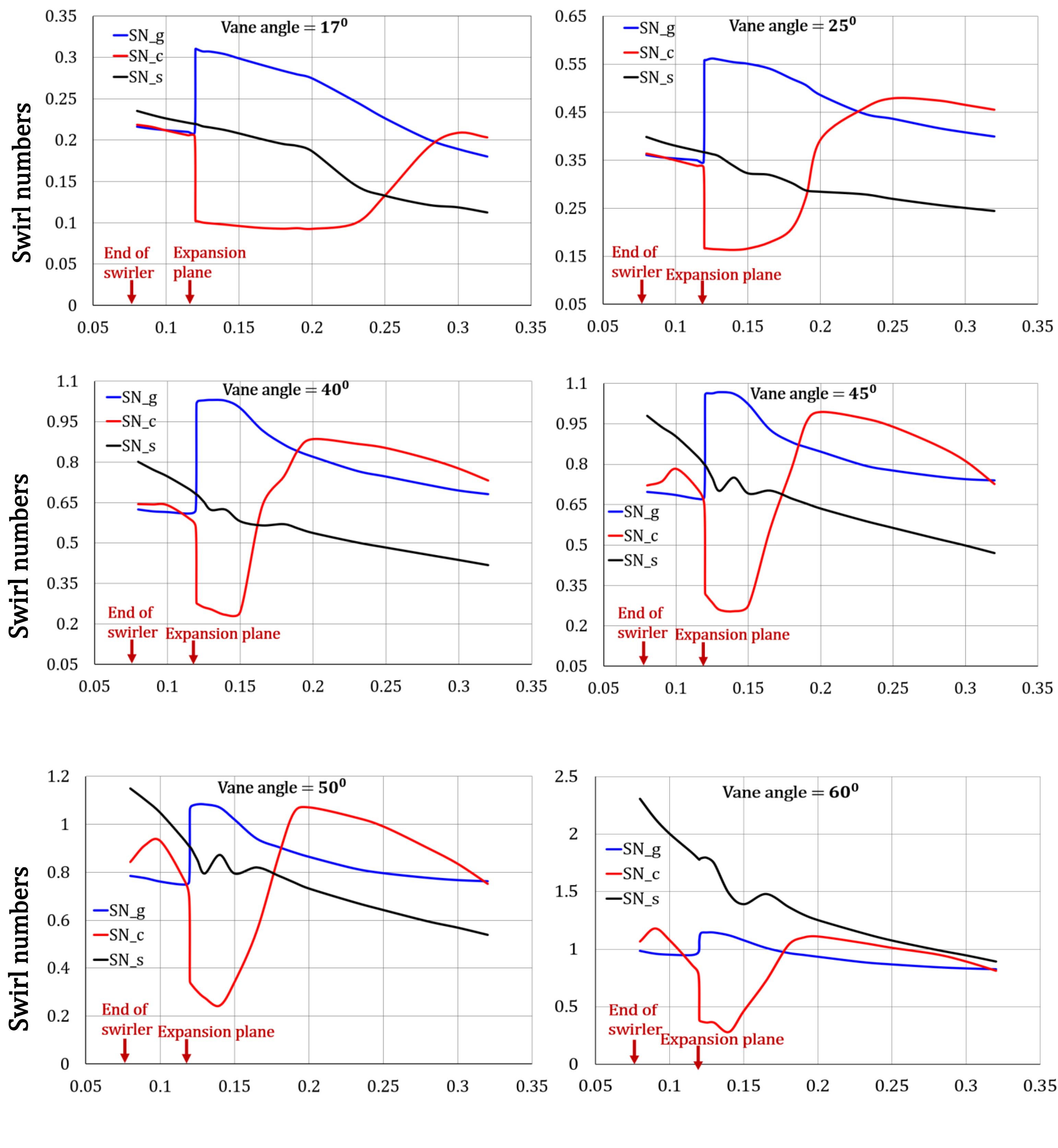

**Figure 6.** Variations in the values of different swirl number formulations along the axial direction for the investigated cases with different swirler vane-angles.

Hence, $SN_g$ satisfies the criteria of an appropriate $SN$ formulation for the investigated cases and therefore, its suitability in representing swirl strength in isothermal flows appears broader than that established for the single case in [29]. The inconsistent behaviour of $SN_c$ and $SN_s$ arises from the simplifications adopted in their respective formulations, as discussed partly in the *Introduction* and elaborated in [29]. A meaningful $SN$ based comparison across combustor flows requires an $SN$ formulation that varies only weakly along the flow, at least or within the shared combustor sections used to quantify swirl strength. The weak variation rates in the values of $SN_g$ in the inlet tube and at locations far downstream from expansion plane provides flexibility in the location where the measured value of $SN_g$ would be able to appropriately distinguish flows with reasonably different swirl strength. Vignat *et al.* [29] did not specify how far the second region should be from the expansion plane. Figure 6 in the present work shows that the distance of the second region increases as the vane-angle decreases. We do not pursue a detailed analysis on this aspect here because measuring $SN_g$ in the inlet tube provides a more robust and practical quantification of swirl strength, as presented in the next paragraph.

Here, we compare the $SN_g$ profiles for the vane-angle of $60^0$ shown in figure 6 and the case studied in [29]. It is apparent that the values of $SN_g$ in the inlet tube of these two cases are approximately $1$, while its values in the downstream sufficiently away from the expansion differ considerably between the two cases. This implies a dependence of the $SN_g$ values in the downstream region on $ER$ and $Re_{inlet}$, which are significantly different between the two cases. Accordingly, using the downstream value of $SN_g$ for comparing different combustor flows can obscure the differences caused by other key parameters like $ER$ and $Re_{inlet}$. Therefore, $SN_g$ measured in the inlet tube of the combustor, i.e., in the region upstream of the expansion plane should be used for quantifying the swirl strength of a flow.

The value of $SN_g$ changes only marginally within the inlet tube in all the cases shown in figure 6. This behavior is attributed to the short length of the inlet tube [29]. If the inlet tube had been longer, measuring $SN_g$ at widely spaced axial positions across investigated cases with different vane-angles may lead to incorrect conclusions. Figure 6 indicates that measuring $SN_g$ within a short region, no more than $40$ mm (it may depend on $\frac{L}{D}$ of inlet tube and weakly on $Re_{inlet}$), just downstream of the swirler would correctly produce different values for cases with reasonably different swirl strength. Consistency in the measurement location, even with this degree of flexibility, appears sufficient to ensure meaningful comparisons across cases. Consequently, the values of $SN_g$ measured within $40$ mm downstream of the swirler in the inlet tube are $0.21, 0.35, 0.61, 0.69, 0.77$ and $0.98$ for the swirl combustor cases with vane-angles of $17^0, 25^0, 40^0, 45^0, 50^0$ and $60^0$, respectively. Referring to the earlier discussions, the critical value of $SN_g$ for the onset of stable-VB in the investigated swirl combustor lies between $0.21$ and $0.35$.

A numerical study of the dependence of this critical value of $SN_g$ on $ER$ and $Re_{inlet}$ would require simulations over a wide range of these parameters. However, LES at higher inlet Reynolds numbers are computationally expensive and beyond our available resources. Hallets and Towes [40], instead of numerical simulations, adopted an experimental approach to examine the dependence of the critical value of $SN_s$, measured in the inlet tube, on $ER$ and $Re_{inlet}$ for $Re_{inlet} \leq 1.6 \times 10^5$. They reported only a weak dependence for $ER \geq 1.5$ and $Re_{inlet} \geq 10^4$. Given our computational limitations, and in the absence of an experimental setup, we rely on reasoning and trends observed in our available simulations to infer the expected behavior of $SN_g$ under similar conditions.

Figure 6 shows the profile of $SN_s$ for the case with a vane-angle of $25^0$, in which the stable-VB state first appears in the swirl combustor. This profile in the inlet tube is nearly flat. Suppose its slope is relatively insensitive to $ER$ and $Re_{inlet}$ in the range $ER \geq 1.5$ and $10^4 \leq Re_{inlet} \leq 1.6 \times 10^5$. Then a weak sensitivity of critical $SN_s$ reported in [40] would suggest that the onset of VB in the present configuration occurs at roughly the same swirler angle. Subsequently, if the slope of $SN_g$ curve in the inlet tube also varies weakly with $Re_{inlet}$ and $ER$ within their ranges given above, then the critical value of $SN_g$ would also exhibit only a weak dependence on these parameters. The latter insensitivity is expected, as the slope of $SN_g$ curve in the inlet tube of the present cases closely matches that observed in [29], despite significant differences in $ER$ and $Re_{inlet}$ between the two configurations.

In conclusion, $SN_g$, the generic swirl number formulation for isothermal flows, exhibits streamwise variation consistent with the theoretical expectations. It remains nearly constant in the short inlet-tube of swirl combustors and is ideally suited for quantifying swirl strength through its values measured just downstream of the swirler. When such values of $SN_g$ are used for comparing swirl strength across different flows, the likelihood of physically incorrect conclusions is minimized. The critical value of $SN_g$ for the onset of stable-VB in the investigated swirl combustor lies between $0.21$ and $0.35$, and is expected to exhibit only weak sensitivity to $ER$ and $Re_{inlet}$ for $ER \geq 1.5$ and $10^4 \leq Re_{inlet} \leq 1.6 \times 10^5$. Next, we investigate the VC structure and its dynamics in the investigated cases.

## 6.2. Vortex breakdown states

Figure 7 presents axial-velocity fields at two randomly selected instances after ten flow-through times for the investigated cases. Black curves along the core of a flow represent the boundary(ies) of the slices of instantaneous IRZ. The IRZ in the smallest vane-angle case in figure 7 appears as a small pebble and displays high spatio-temporal intermittency across the two shown instances. Because of the high level of intermittency, no IRZ appears in the corresponding mean flow field in figure 5. The observed flow corresponds to the near-VB state of [58]. The slices of IRZ in the other cases display an overall weak intermittency across the instances shown in figure 7. Further, the recirculation bubble emanating from outlet appears strongly interacting with the recirculation bubble near the expansion in cases with vane-angles equal to or greater than $45^0$. A separation bubble appears attached to the downstream portion of the CB of swirler across the instances shown in figure 7. However, a high intermittency in the location of separation bubble in the cases with $40^0$ and $45^0$ vane-angle leads to its absence in the corresponding mean fields in figure 5. The recirculation bubbles mentioned above forming near the expansion plane and the outlet merge smoothly with each other and the separation bubble in the largest vane-angle case. It results in an IRZ expanding over the length of combustor downstream of swirler. These swirl combustor flows in figure 7 except the smallest vane-angle flow correspond to the steady-VB state.

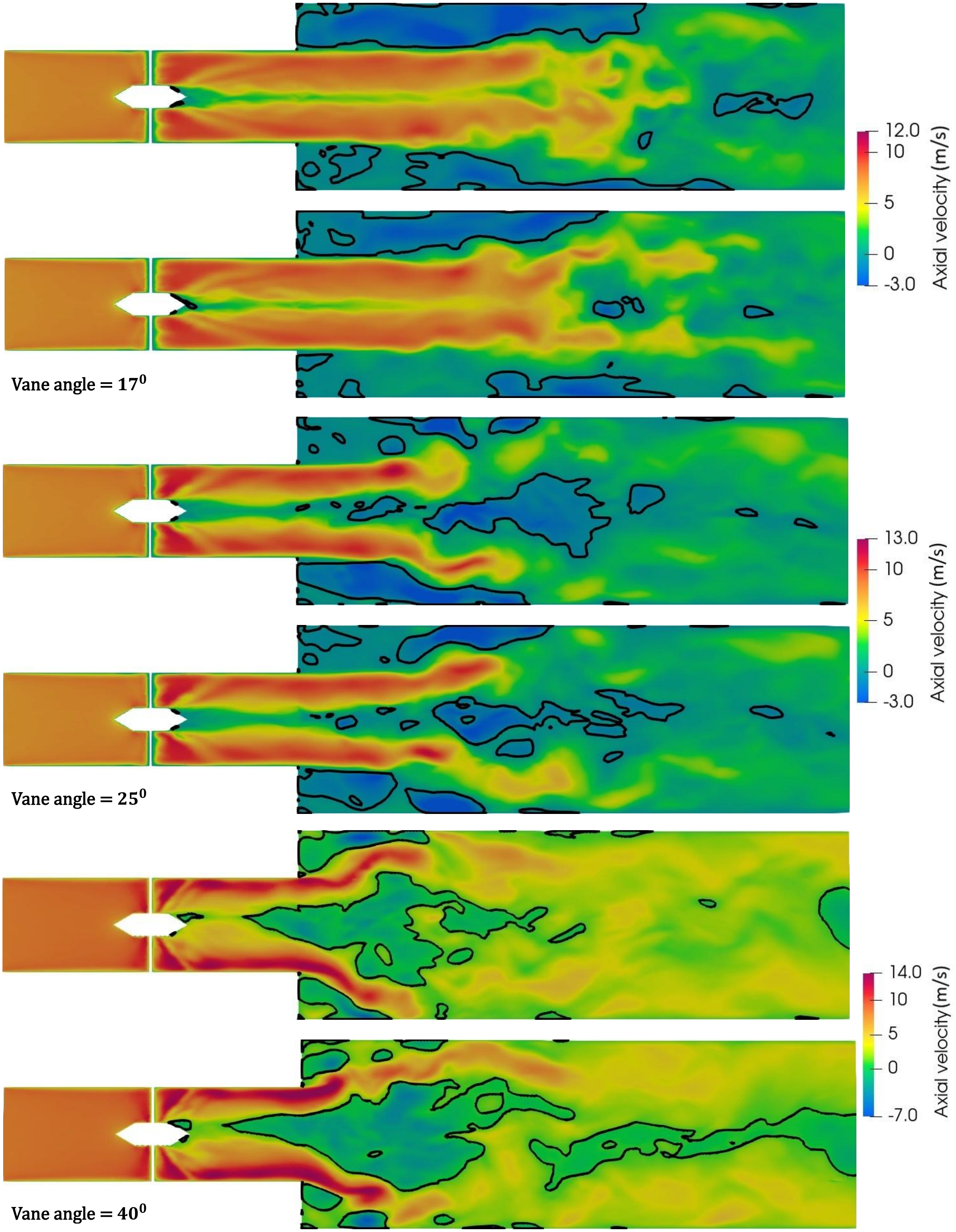

**Figure 7a.** Contours of axial velocity (m/s) at two instances on the streamwise plane through the centreline of the computational domain for the cases with $17^0$, $25^0$ and $40^0$ vane-angles. Black colour curves are isocurves of zero axial velocity.

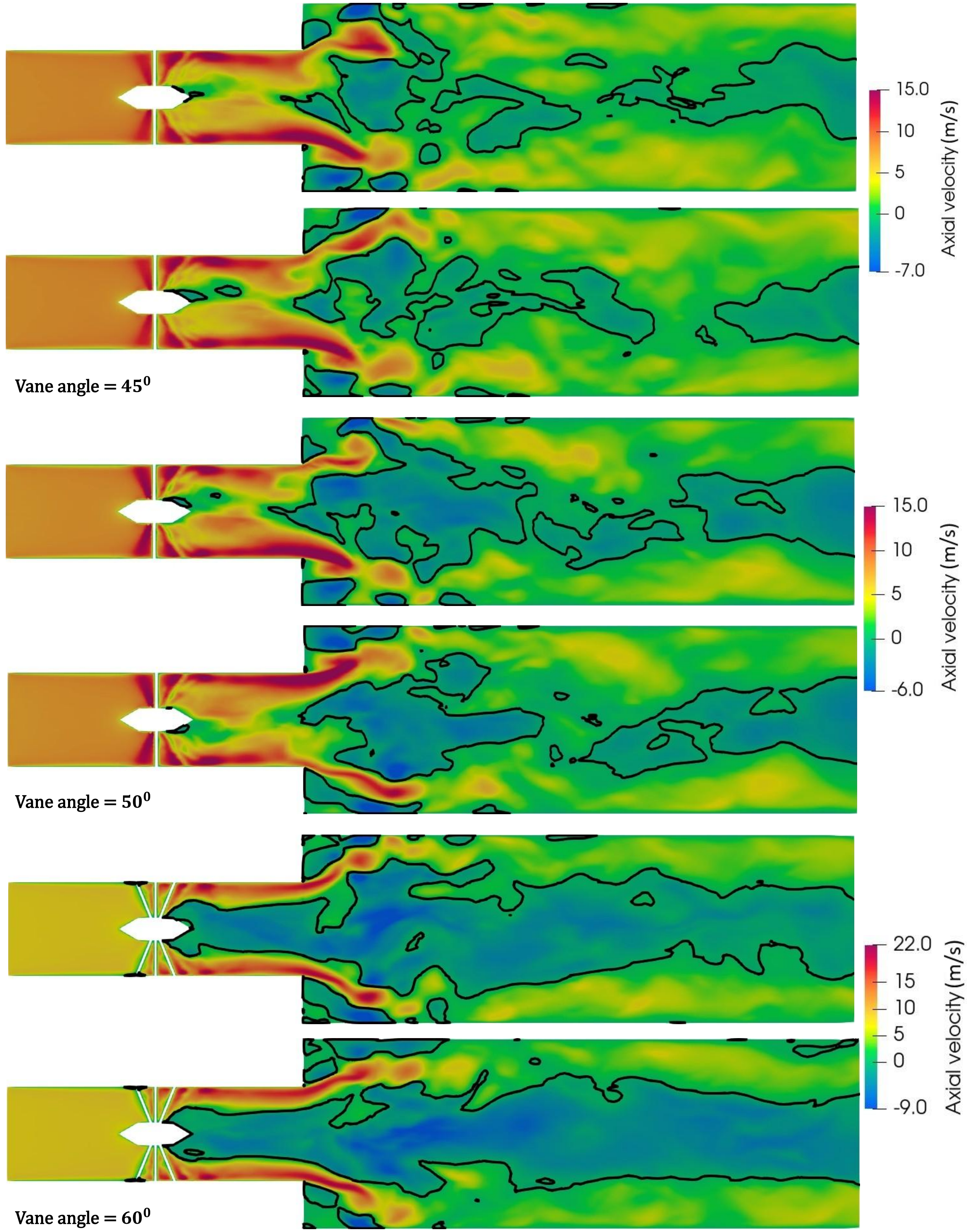

**Figure 7b.** Contours of axial velocity (m/s) at two instances on the streamwise plane through the centreline of the computational domain for the cases with $45^0$, $50^0$ and $60^0$ vane-angles. Black colour curves are isocurves of zero axial velocity.

## 6.3. Topology of vortex core and its dynamics

As stated earlier, in this section, we first visualize the VC structures at few instances and then evaluate their temporal evolution over several timesteps to assess their sustainability or transformations. Further, we locate and identify the response of inner shear layer and outer shear layer to IRZ dynamics, when need arises.

### A. Topology of Vortex core

Figure 8 shows iso-surfaces of the $Q$-criterion for the investigated cases, except $50^0$ vane-angle case that is not shown for brevity. Here the $Q$-value, which differs between cases, isolates an ensemble of flow structures of which a coherent structure is expected to correspond to the VC. The suitable $Q$-value is identified using trial-and-error. Lowering the value by ten percent produced a proliferation of small iso-surfaces apart from coherent structures, whereas increasing it by ten percent led to fragmentation of the expected coherent VC structure. Recall that the VC in a swirl combustor corresponds to a suitable $Q$-valued iso-surface that coherently extends from the swirler through regions of elevated streamwise vorticity [20, 32, 38].

A coherent $Q$ iso-surface with two helical strands twisted in the same sense appears to emanate from the swirler in each of the flow fields in figure 8. More specifically, the two helical strands appear to emanate from the CB of the swirler in the cases with vane-angle < $60^0$. These strands twist in the direction opposite to the rotation of the flow, as evident from comparing their twists with the inclination of swirler-vanes in figure 8. The strands in the cases with vane-angle $< 60^0$ appear to have a near circular cross-section whereas the strands in the largest ($60^0$) vane-angle case appear to have elliptically deformed cross-sections. The two strands in the cases with vane-angle of $17^0$ and $25^0$ in figure 8 first intertwine up to some downstream distance from the swirler and then separate out. One of the strands kinks and coherently spirals out into a larger helix after separation in both the cases. The other strand disappears immediately after separation in the former case, while it persists for a short distance after separation in the latter case but with a poor coherency. This aspect is marked in figure 8 and is shown later in this section with better clarity.

The two strands in the other investigated cases shown in figure 8 separate out from or just after the swirler and then interlace. One of the two strands appears only weakly coherent in these cases except for the largest vane-angle case. A dashed red-color circle highlights the separation region of the two strands in all cases except the smallest ($17^0$) vane-angle case, where circling would obscure this separation region that already clouded by proliferation of smaller iso-surfaces. Note that the coherent structure shown in figure 8 expected to represent VC breaks down along the axial direction even in the smallest vane-angle case, where the flow is in the near-VB state and not in the stable-VB state, as discussed earlier.

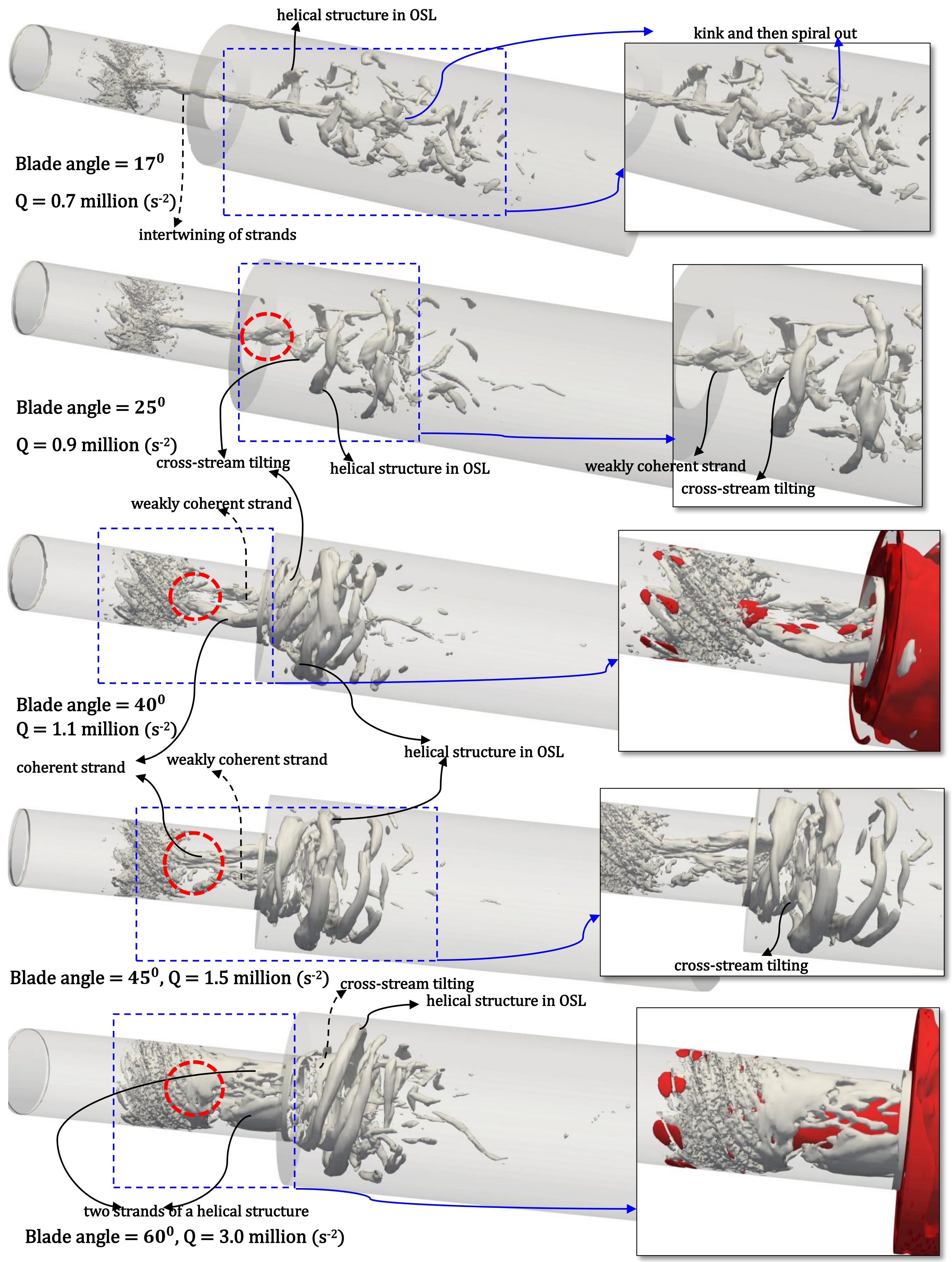

**Figure 8.** Iso-surfaces of Q-criterion (**grey**) and zero axial velocity (**red**, shown in two cases) for cases with different vane-angles.

Vortical structures after the expansion plane in the region outside the core of the flow appear tilted towards the cross-stream. They are coherent in the investigated cases except for the two smallest vane-angle cases. The coherent strand of the double-helix structure, in the core of investigated flows, also begins to tilt towards the cross-stream just after the expansion plane except for the smallest vane-angle case. On the other hand, the weakly-coherent strand does not tilt and dissipates within a short distance from the beginning of this tilting of the coherent strand. The tilting of the coherent strand in the smallest vane-angle case occurs further downstream in figure 8. This cross-stream tilting arises from interaction of VC with the IRZ dynamics in the shear layer surrounding IRZ and the formation mechanism is discussed later as formation of helical rollups around IRZ. Note that both strands appear coherent in the largest vane-angle case in figure 8. They both tilt towards the cross-stream after expansion, unlike the other investigated cases. Although we have isolated the cross-stream tilting of the coherent strand of the double-helix structure by markings in figure 8, observing the disappearance of the weakly-coherent strand in figure 8 is difficult for many cases. Later in figure 12 we present both these aspects clearly.

Although it appears that VC has a double-helix structure in each investigated case in figure 8, a closer examination discussed above reveals that only one of double-helix strands is coherent, except in the largest vane-angle case. The observed difference in coherency between the two strands indicates a possible difference in their origin/formation. Only the coherent strand can be expected to represent the VC, since VC is a characteristic of swirling flows and loses coherency downstream of the VB region.

As mentioned at the beginning of this section, the next check for these strands to be a part of the VC is whether they pass through regions of high streamwise vorticity; the converse may not be true. Figures 9 to 11 present the slices of the $Q$-surfaces of figure 8 as white-color curves over the contours of streamwise vorticity on a cross-stream plane at three different instances. The plane is at a distance of $0.115$ m from the inlet for the cases with vane-angles equal to or greater than $40^0$. This plane lies within the inlet tube, where the $Q$-surfaces of figure 8 only include the double-helix structure, and the other corresponding structures are visible only after expansion. Accordingly, the white-color curves on the selected planes in figures 10 and 11 arise only from slices of the double-helix structures of figure 8. The plane in the two smallest vane-angle cases is positioned downstream of the separation of the double-helix strands so that the slices of the two strands are distinguishable, if both strands persist for some distance after separation. Accordingly, the planes in figure 9 are at the distance of $0.200$ m and $0.130$ m from the inlet for the cases with vane-angles of $17^0$ and $25^0$, respectively. Since these planes are located after expansion, the white-color curves on them arise also from the $Q$-surfaces located away from the core of the flow in figure 8 that are not a part of the double-helix coherent structure. The slices in the core of these planes arise from the double-helix coherent structure.

These white-color curves arising from the double-helix structure on the selected planes appear to pass through regions of higher streamwise vorticity at all the shown instances in figures 9 to 11. The double-helix structure in the two smallest vane-angle cases in figure 8 appears close to the centerline before its strands separate. One of the strands disappears, while the other, coherent strand, kinks and spirals outward after the separation. Accordingly, only the coherent strand intersects the selected plane in figure 9, positioned downstream of the separation, at an eccentric location. However, the slices of both the coherent and the weakly coherent $Q$-surface appear on the investigated cross-stream plane in other investigated cases in figures 10 and 11. It is apparent that the eccentricities from geometric center (white cross in figures 9 to 11) of the slices of the two strands of the helix are not the same in these investigated cases except in the largest vane-angle case. The slice of one of the strands of the double-helix structure appears on a smaller circle, marked in blue-color, while that of the other strand appears on a larger circle, marked in pink-color, in figures 10 and 11a. The markings are shown only for one instance to avoid overshadowing of the clarity of other features in figures 10 and 11. Even the slice of one of the two strands has poor coherency. These traits are typical of a single dominant helix VC topology. Note that slices of both the strands are located at similar eccentricities from geometric center (white cross) in the largest vane-angle case.

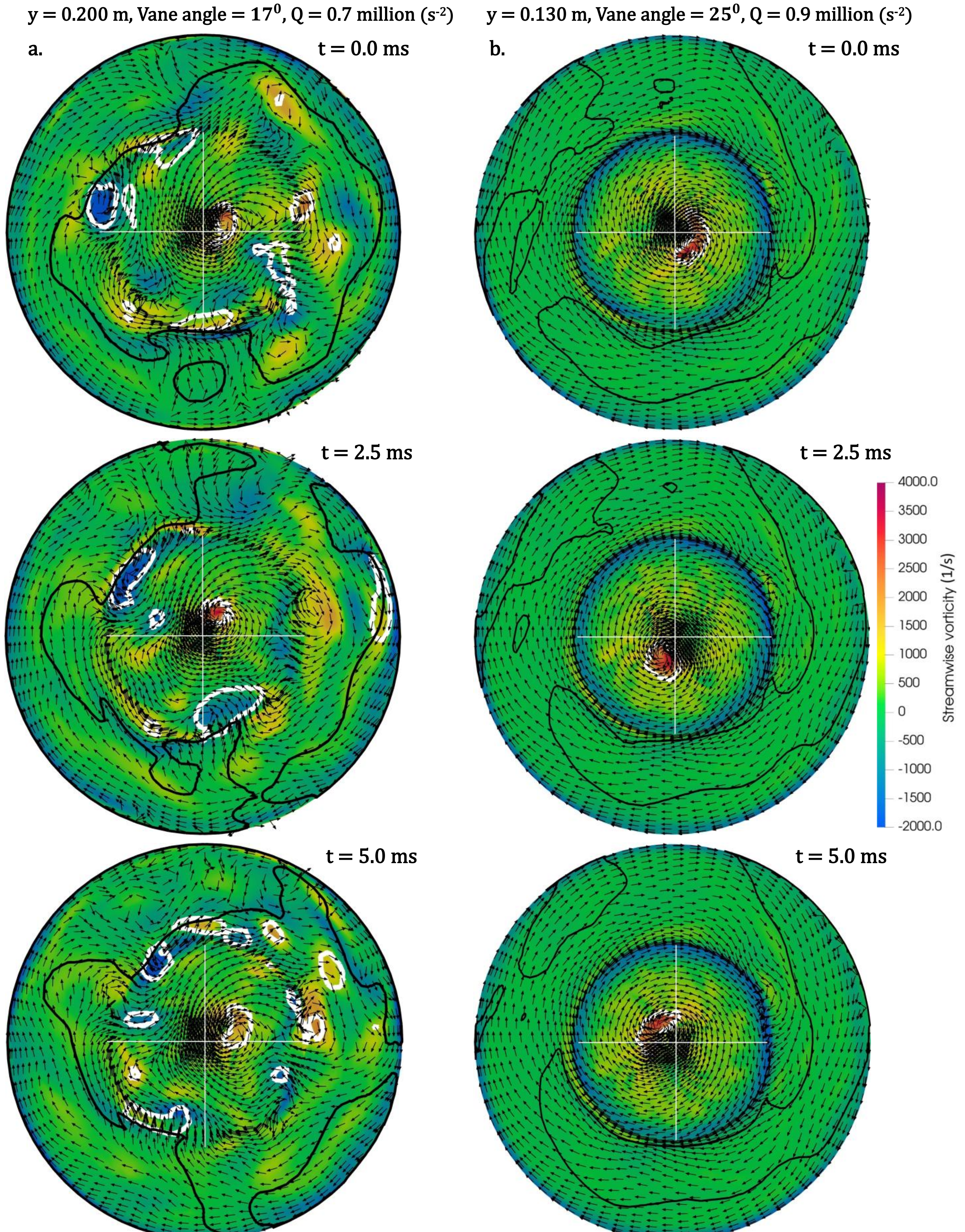

**Figure 9.** Contours of streamwise vorticity, velocity vector, zero-axial-velocity isocurves (**black**) & Q-isolines (**white**) on a cross-stream plane for the cases with $17^0$ and $25^0$ vane-angles at three different instances, showing the relative location of VC, geometric-centre & negative axial-velocity zone. **'y'** is the distance of plane from the inlet.

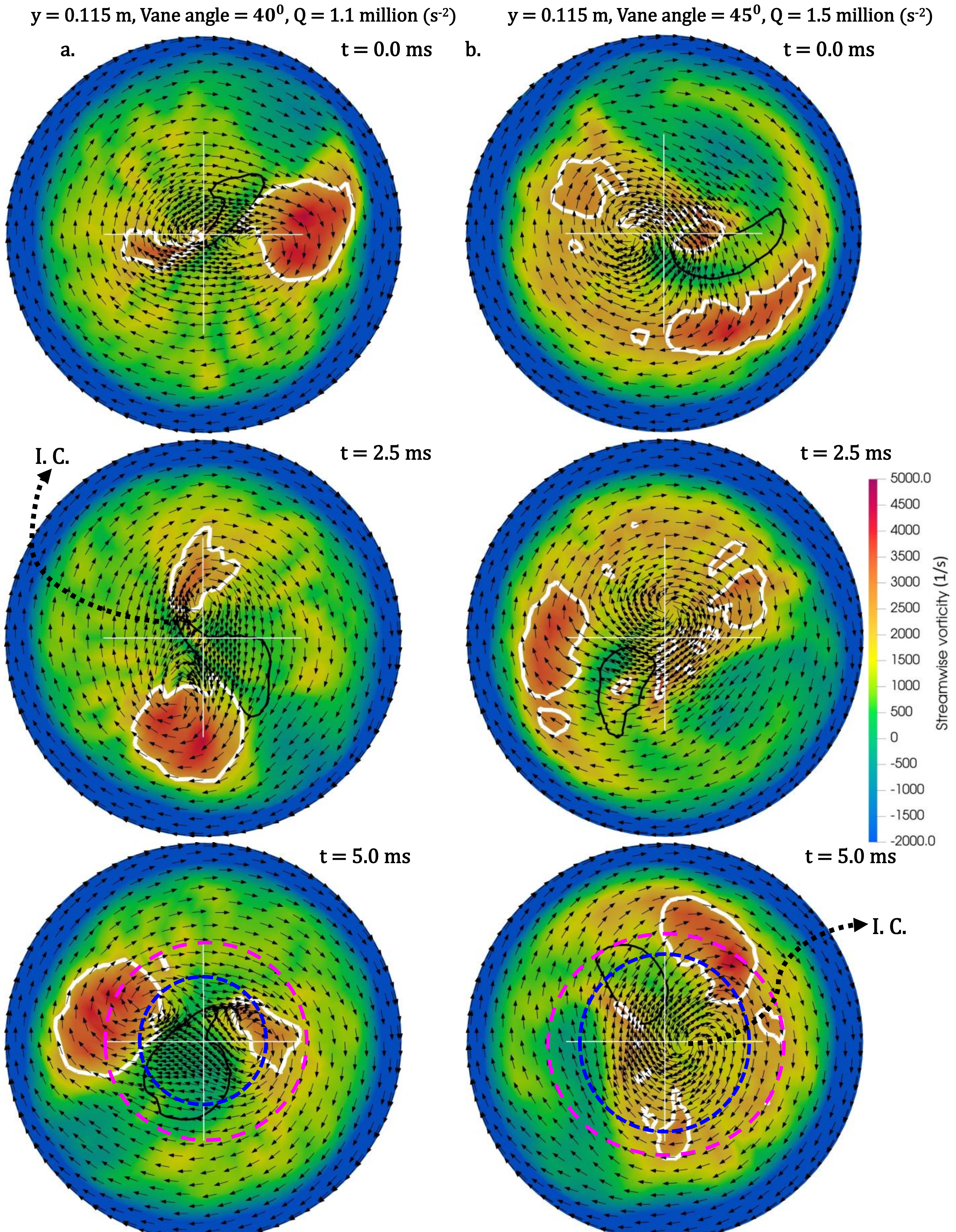

**Figure 10.** Contours of streamwise vorticity, velocity vector, zero-axial-velocity isocurves (**black**) & Q-isolines (**white**) on a cross-stream plane for the cases with $40^0$ and $45^0$ vane angles at three different instances, showing the relative location of VC, geometric-centre & negative axial-velocity zone. **'y'** is the distance of plane from the inlet. The pink and blue dotted circle are the locus of centre of the corresponding VC strand slices

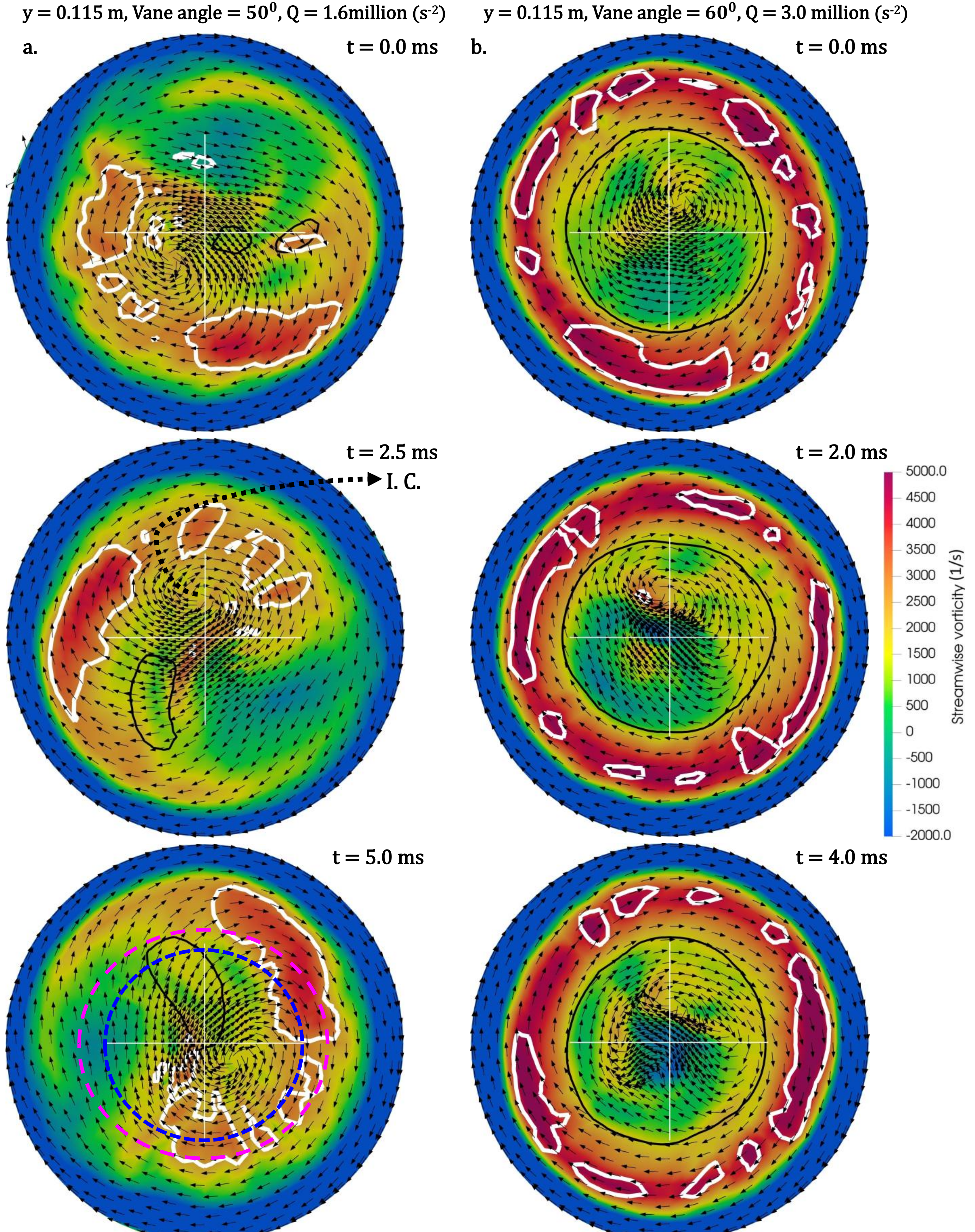

**Figure 11.** Contours of streamwise vorticity, velocity vector, zero-axial-velocity isocurves (**black**) & Q-isolines (**white**) on a cross-stream plane for the cases with $50^0$ and $60^0$ vane angles at three different instances, showing the relative VC, geometric-centre & negative axial-velocity zone. **'y'** is the distance of plane from the inlet. The pink and blue dotted circle are the locus of centre of the corresponding VC strand slices.

## B. Relative position of ISL, OSL and VC

To strengthen the observations in the preceding sections that the weakly-coherent strand is not part of the VC, we investigate the origins of the coherent and weakly coherent strands of the double helix observed in the core of flow fields in figure 8. However, we first present the two strands in a simplified manner in figure 12 to clearly present the flow aspects that could not be presented clearly in figure 8. Figure 12 shows two images for each investigated case. The right-hand images show $Q$-surfaces of figure 8 and black-color curves representing the outlines of recirculation zones over the contours of cross-stream vorticity, while two zig-zag dashed lines appear in place of the $Q$-surfaces in the left-hand images. Each zig-zag line connects the regions where the corresponding $Q$-surface, strand of the double-helix structure, intersects with the plane in view in figure 12. The blue-color zig-zag lines correspond to the weakly-coherent strands, and the white-color zig-zag lines correspond to the coherent strands. The left-hand images show that the weakly-coherent strand in each case disappears within a shorter distance from the swirler than the corresponding coherent strand. The weakly-coherent strands disappear around the location where the corresponding coherent strands begin to bend towards the cross-stream direction. Furthermore, the weakly-coherent strands appear concentrated along the centerline of the domain, while each coherent strand after separating, red-color circle, from the weakly-coherent strand expands out into a larger-diameter helix. This last observation appeared even in figures 9 to 11 but in a rather different manner. The coherent strand kinks and then spirals outward in the two smallest vane-angle cases. The above observations of kinking and cross-stream bending of the coherent strand and disappearance of the weakly-coherent strand in figure 12 could not be clearly resolved from figure 8.

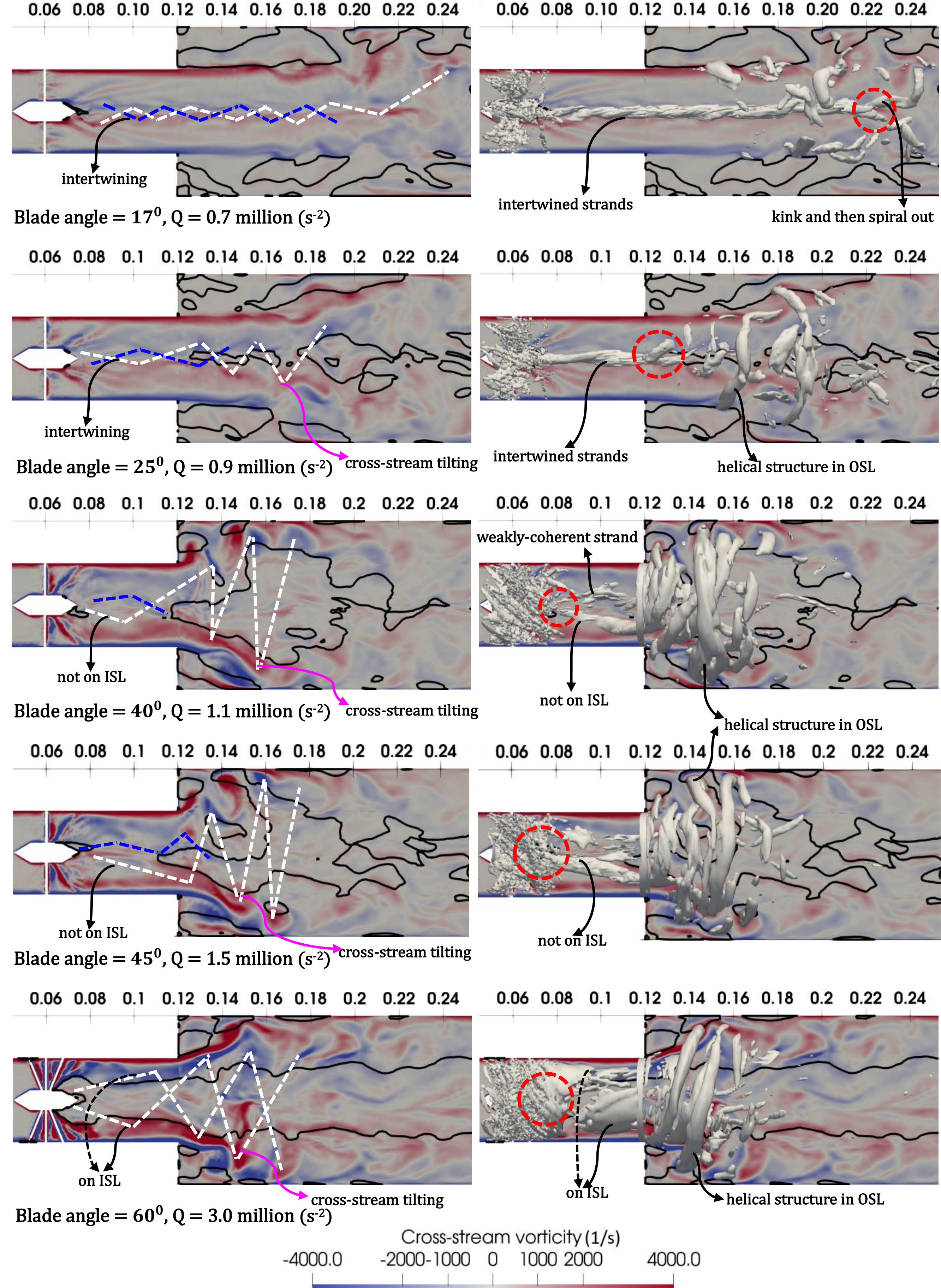

**Figure 12.** Contours of cross-stream vorticity, isolines of zero axial velocity (**black**) and iso-surfaces of Q-criterion (**grey**) on the streamwise plane through the centreline of the domain for the cases with different swirler vane angles.

We now investigate the origin of each strand of the double-helix structure by investigating their positions relative to the IRZ and the shear layers. We begin the discussion by referring to the results of the $40^0$ vane-angle case. The images in figure 12 for the $40^0$ vane-angle case reveal different regions of high cross-stream vorticity that signify different shear layers. We also recall the velocity fields in figure 7 for better understanding of the following discussion.

The boundary-layer in the inlet tube is apparent as the thin region of high cross-stream vorticity adjacent to the wall. High vorticity fluid from this region interacts with the ORZ downstream of expansion, resulting in a high cross-stream vorticity region on the inner periphery of the ORZ in figure 12. This region around the ORZ is referred to as the outer shear layer, OSL. The coherent structure near the ORZ is a depiction of rollup in the OSL. Further, the vanes of the swirler in the $40^0$ vane-angle case appear to direct higher-momentum fluid close to the wall in figure 7. This creates a region of high cross-stream vorticity adjacent to the boundary layer in the inlet tube in figure 12. High vorticity fluid from this region interacts with the IRZ resulting in a high cross-stream vorticity region on the periphery of the IRZ. This entire high-magnitude cross-stream vorticity region after the swirler is referred to as the inner shear layer, ISL, in the present work.

The rotation induced by the swirler creates one or more coherent regions of high streamwise vorticity away from the wall referred to as the vortex core, VC; in the present work this corresponds to the coherent strand of the double-helix structure. This origin of the coherent strand is common for all the investigated cases. The portion of the coherent strand upstream of the IRZ appears tilted away from the centerline but does not appear on the ISL or on any shear layer in the $40^0$ vane-angle case as seen in figure 12, whereas its portion around the IRZ appears on the ISL as a helical rollup. Earlier studies [20, 58] reported that helical rollup of the VC in the ISL arises from the interaction of its portion upstream of IRZ with IRZ precession and a convective instability of the ISL. Flows with a stable IRZ typically admit a global helical eigenmode about the time-averaged mean state, with the wavemaker located on or near the centerline close to the upstream end of the IRZ. This helical mode is mostly marginally stable (growth rate $\approx 0$) but in flows with swirl strength near the onset of a stable IRZ it can be slightly stable (growth rate $\lesssim 0$) yet persist via stochastic forcing from the flow [58, 65]. This eigenmode manifests as a marginally stable (or slightly stable) precessing global hydrodynamic mode on the mean-flow state [58, 65]. The overall flow precesses, with oscillation amplitudes usually largest within the IRZ, while the ISL is relatively static azimuthally [58]. Therefore, IRZ precession seeds azimuthal disturbances in the ISL that typically grow by a convective instability of the layer [67] into the corresponding helical rollups of VC. Thus, the rollup of VC in the ISL is coupled with the formation of the IRZ in the mean flow, and accordingly we observe in figure 12 helical rollup of the coherent strand in the ISL for the cases with vane-angle $\geq 25^0$. We do not infer IRZ precession solely from the formation of the mean IRZ in the present work; instead, we verify it in section 6.3 by assessing the dynamics of the coherent strand via spectral techniques specified earlier in section 6 [20, 58, 66]. Furthermore, the portion of the coherent strand upstream of the mean IRZ appears tilted away from the centerline in the $40^0$ vane-angle case in figure 12, also noted above. However, this tilting of the coherent strand is not a result of shear-layer instabilities because the strand does not lie on any shear layer in the upstream of mean IRZ in figure 12 [20]. This tilting can arise if the marginally stable global helical eigenmode that causes the large-scale precession of the IRZ also has appreciable oscillation amplitude in this upstream region.

Next, moving to the discussion on the weakly-coherent strand recall that it appears concentrated along centerline of the domain, unlike the coherent strand that expands out into a larger-diameter helix after the swirler in the $40^0$ vane-angle case in figure 12. A low-velocity region formed by fluid issuing from the boundary layer of the swirler's CB and surrounded by a nominally high-velocity region appears around the centerline in the $40^0$ vane-angle case in figure 7. The high-gradient zone between the low- and high-velocity regions is commonly referred to as the shear layer in the wake of the swirler's CB. The appearance of this shear layer in figure 12 might not be as clear as the other shear layers owing to its limited radial coverage. Therefore, we present it also using the radial plot of instantaneous axial-velocity at $0.1$ m downstream from inlet in figure 13, where a separate high-velocity-gradient region appears close to the centerline of the domain in the $40^0$ vane-angle case. The weakly-coherent strand is likely a convective-instability growth on this shear layer, seeded by the dynamics of the tilted-coherent-strand. A simple way to check the perturbation source is to compare dynamics of the weakly coherent structure with the dynamics of the presumed source of perturbation. Accordingly, we will evaluate and compare the dynamics of the weakly coherent and the coherent strands of the double-helix structure later in this section of the paper. Hence, the weakly-coherent strand seems not to gain its vorticity strength directly from the rotation generated by the swirler and is therefore not a part of the VC in the $40^0$ vane-angle case.

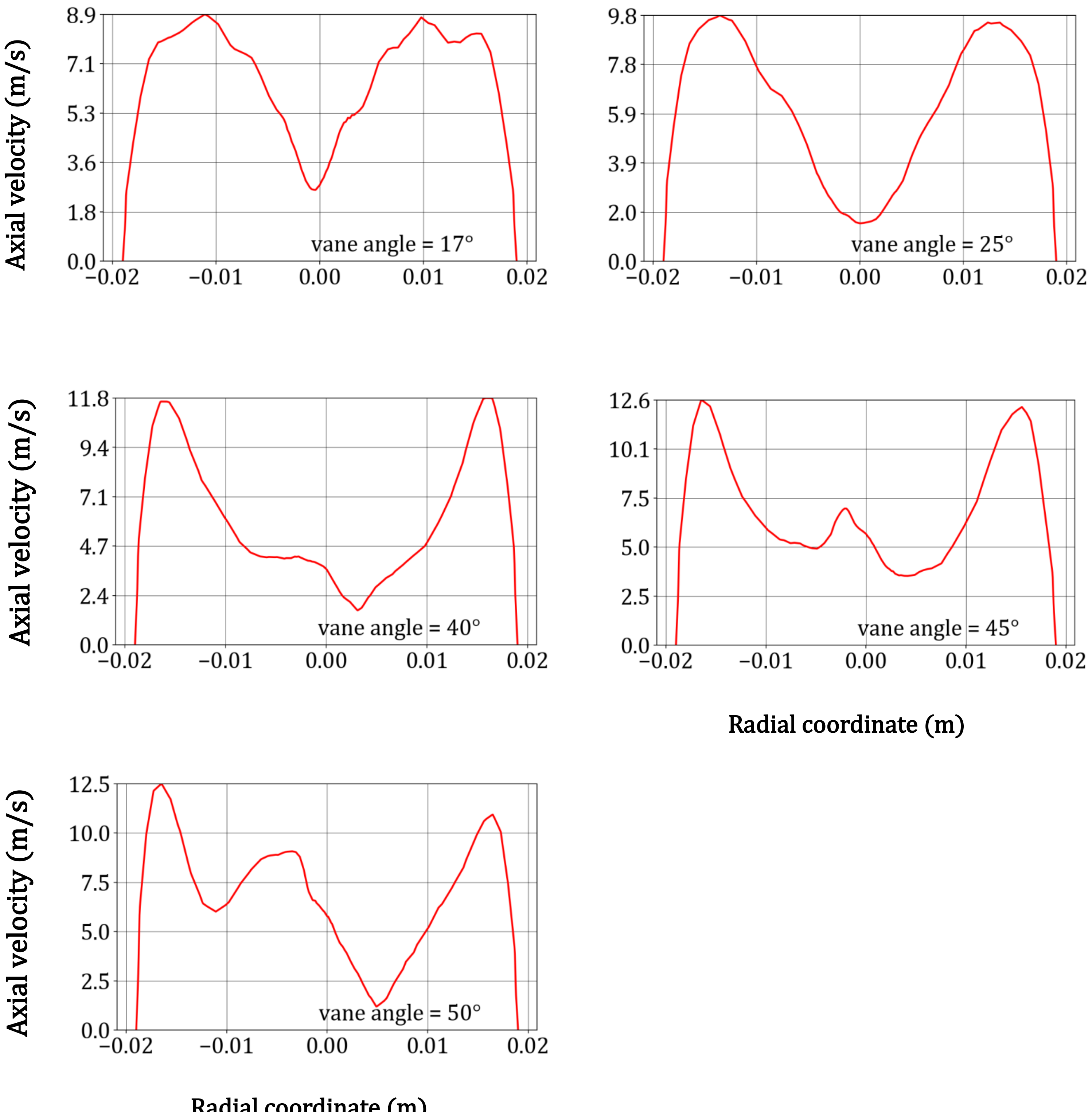

**Figure 13.** Radial variation in axial velocity at the location 0.1 m downstream from the inlet for the cases with different vane angles. An unusual dip in axial velocity appears close to the centreline in the three larger vane angles cases.

Figures 8, 12 and 13 reveal observations similar as above in context to IRZ precession, helical rollup in ISL, the coherent and the weakly-coherent strands in the $45^0$ vane-angle case. The double-helix strands in the $50^0$ vane-angle case have not been shown in figures 8 and 12 for brevity of results. But we found in our investigation that the flow-field behavior in the $50^0$ vane-angle case in context of the present section is similar to that in the $40^0$ vane-angle case. Similarly, the applicability of the mechanisms could even be extended to the $25^0$ vane-angle case. However, the portion of the coherent strand in the upstream of the IRZ appears to have moved towards the centerline in the $25^0$ vane-angle case in figure 12 but its tilt and associated dynamics may still be sufficient for the formation of the weakly-coherent strand. Further, the weakly-coherent strand appears well-formed upstream of the IRZ in the $25^0$ vane-angle case, unlike the $40^0$, $45^0$ and $50^0$ cases. This is consistent with the observation in figures 7 and 13 that the shear layer in the wake of swirler's CB merges with the ISL in the $25^0$ case, likely owing to lower swirl strength, making the wake's shear layer and hence the corresponding helical rollup or strand well-formed. This explains about the weakly coherent structure of strand in the $40^0$, $45^0$ and $50^0$ vane-angle cases. Therefore, we expect the formation mechanisms of the coherent and weakly-coherent strands in the $40^0$ vane-angle case to directly apply also to the $25^0$, $45^0$ and $50^0$ vane-angle cases. In essence, IRZ precession is the factor responsible for the afore-mentioned behavior of the double-helix strands in the $40^0$, $45^0$

and $50^0$vane-angle cases. Before quantifying the precession we visually verify it by revisiting figures 9b, 10 and 11a. The IRZ performs centerline oscillations on a cross-stream plane over the time instances shown in the figures in each of four cases being discussed here, confirming IRZ precession [20]. The slices of the double-helix structure located around or outside the periphery of the IRZ move along with it almost locked in space relative to it. Note that the velocity vectors in figures 9 to 11 are not to scale for the convenience of visualization.

The formation mechanism of the coherent strand in the lowest vane-angle case ($17^0$) is obvious as in the cases discussed above. However, the strand kinks and rolls out into a larger helix in this case well upstream of the IRZ in figure 12, unlike the helical rollup that forms on the ISL around the IRZ in cases with vane-angle $\geq 25^0$. Hence, a reasonable oscillation amplitude of a marginally stable helical mode appears to shape and move the coherent strand directly and not via the IRZ perturbation in the lowest vane-angle case. The precession of the coherent strand of the double-helix after kinking is clearly visible in figure 9a, however, we do not claim that this behavior is persistent. The portion of the coherent strand upstream of the IRZ appears to have moved towards the centerline in the $17^0$ vane-angle case in figure 12, but its tilt and associated dynamics may still be sufficient for the formation of the weakly-coherent strand. Further, the weakly-coherent strand appears well-formed upstream of the IRZ in the $17^0$ vane-angle, like in the $25^0$ vane-angle case, for the same reason as in the $25^0$ vane-angle case discussed above. We reiterate that we will evaluate and compare the dynamics of the weakly-coherent and coherent strands of the double-helix structure later in this section of the paper to resolve if the coherent strand dynamics is the source of perturbation to the formation of the weakly-coherent strand. Overall, the above discussion indicates that the weakly-coherent strand is a secondary structure in the double-helix in cases with vane-angles $\leq 50^0$ and therefore is not a part of the VC in these cases.

At last, we discuss the largest vane-angle ($60^0$) case. We can see from figures 7, 8 and 12 that the IRZ extends over the entire length of swirl combustor downstream of swirler and so there is no portion of the double-helix upstream of the IRZ in the largest vane-angle case. Both strands of the double-helix structure in the $60^0$ case portray similar coherency in figure 8 and their slices are located on the same shear layer, ISL, at similar distances from the center of the domain in figures 11b and 12, appearing as two helical rollups in the ISL. These rollups are likely arising from the interaction of concentrated vorticity generated by swirler with the IRZ precession and a convective instability of the ISL, as in cases discussed above. Hence, both rollups are likely a primary structure and so a double-helix VC forms in the largest vane-angle case of the present work.

The above conclusion is supported upon the presence of IRZ precession. However, the IRZ in the inlet tube of $60^0$ case appears almost axisymmetric, while its portion just after expansion appears asymmetric in figure 12. The movement of the slices of the IRZ on a plane in the inlet tube of $60^0$ case is apparent from figure 11b, and we plot it on a plane just downstream of expansion in figure 14. The slice of the IRZ on the cross-stream plane in figure 11b appears slightly asymmetric over time, whereas it displays a reasonable rotating asymmetry on the cross-stream plane in figure 14. This degree of asymmetry should be sufficient to perturb and cause the helical rollups on ISL in the largest vane-angle case, as rollups are already apparent in figures 8, 11b and 12.

The above discussion indicates that the centerline oscillations of the IRZ are not strongest around its uppermost end across the shown instances in figures 11b and 14, a phenomenon rarely encountered [58, 66]. This uncommon behavior likely arises from the merging of the IRZ with the separation bubble of CB. The wavemaker of the marginally stable helical eigen mode on mean flow state that causes IRZ precession sits upstream of the IRZ [58, 65]. The merger of the IRZ with the separation bubble in a similar combustor [68] was reported to disrupt the wavemaker and stabilize the precessing mode [68]. Precession was still observed in [68], but it was intermittent, driven by stochastic forcing of the stabilized precessing mode by background turbulence. At least an intermittent precessing motion is expected in the present largest vane-angle case since two helical roll-ups appear at the instance shown in figures 8 and 12. Though the intermittency is not evident in the instances shown in earlier figures, it will be resolved via spectral analysis in section 6.3.c of the present work.

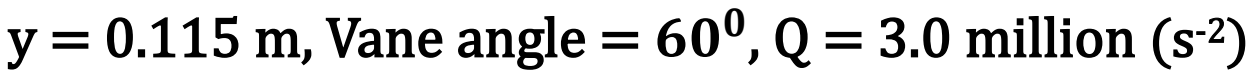

**Figure 14.** Contours of streamwise vorticity, velocity vector, zero-axial-velocity isocurves (**black**) & Q-isolines (**white**) on a cross-stream plane for the $60^0$ vane-angle case at three different instances, showing the relative VC, geometric-centre & negative axial-velocity zone. **'y'** is the distance of plane from the inlet.

## C. Dynamics of the flow

The above comprehensive discussion indicates that the VC has a spiral or a single-helix topology when the vane-angle $\leq 50^0$ and a double-helix topology when vane-angle is $60^0$. Further, a weakly coherent spiral appears in the shear layer behind the swirler's CB when the vane-angle $\leq 50^0$. These conclusions are based on observations at the instant shown in figures 8, 12 and 13. The weakly coherent structure may persist, either continuously or intermittently, or decay, while the VC, though persistent in swirling flows, may change topology over time. By following the work in [65] and [66] we first verify whether the VC and the vortical structure in the CB's shear layer predominate in a helical topology over time. Studies in [65] and [66] imply that the coherent precessing motion of a helical flow structure manifests as amplified azimuthal and radial velocity fluctuations along its path, with strong signatures at a few corresponding locations. Axial-velocity fluctuations are amplified at those locations along the path of helical flow structures that are off-axis or away from the centerline of the computational domain, again with strong signatures at a few corresponding locations. A helical VC rolls up in the ISL around IRZ, which positions the structure significantly away from the flow/ domain centerline. Accordingly, a VC that predominates in a helical topology over time would produce footprints in all the components of velocity fluctuations or turbulent kinetic energy (TKE). However, the second vortical structure of concern in our work appears in the shear layer concentrated about the centerline with no presence in the off-axis location and so its predominance in a helical topology would produce footprints only in the azimuthal and radial components.

Figures 15 and 16 show contours of the azimuthal component of resolved TKE ($TKE_{resolved,tang}$) and the axial component of resolved TKE ($TKE_{resolved,ax}$) for the investigated cases. Here, $TKE_{resolved}$ denotes the resolved TKE computed using time-averaged velocity moments over ten flow-through times, consistent with section 4. We first compare the images in figures 12, 15 and 16 for the $25^0$ vane-angle case. Two maxima in $TKE_{resolved,tang}$ appear in this case in figure 15. The upstream maximum in $TKE_{resolved,tang}$ arises around an intersection of the field with the double-helix strands, while the downstream maximum arises around an intersection of the field with the weakly-coherent strand. The coherent strand or VC intersects the ISL at several locations. A maximum in $TKE_{resolved,ax}$ occurs in the ISL around two of these locations in the plane in figures 16. The value of $TKE_{resolved,tang}$ is significant but not maximal at these ISL locations in the $25^0$ case. Overall similar observations appear in the other cases with vane-angle > $25^0$ upon comparison of corresponding images in figures 12, 15 &16.

A maximum in $TKE_{resolved,tang}$ also appears around the locations in the ISL where $TKE_{resolved,ax}$ is maximum for the cases with vane-angles $\geq 40^0$. Maxima in $TKE_{resolved,tang}$ appear even at off-axis locations in the inlet tube for the cases with vane-angles $\geq 40^0$, because the coherent strand is tilted from the centerline in the inlet tube in these cases. The centerline maxima in $TKE_{resolved,tang}$ for the cases with vane-angles $\geq 40^0$ except the largest vane-angle case appears in figure 15 at locations where the weakly-coherent strand repeatedly intersects. No centerline maximum appears in $TKE_{resolved,tang}$ in the largest vane-angle case as both strands lie along the ISL in this case that appears close to the wall in figure 11. In the smallest vane-angle case, $TKE_{resolved,tang}$ peaks around the location where the coherent strand kinks, while $TKE_{resolved,ax}$ is globally maximum in the vicinity of ORZ likely from the dynamics of rollup in the OSL. The value of $TKE_{resolved,ax}$ is elevated, though not maximal, where the coherent strand intersects the plane in view in figure 16 after kinking. These observations may be consistent with precession of the coherent VC strand in the lowest vane-angle case; however, the present evidence is not conclusive. Note that in figure 16 the contours of $TKE_{resolved,ax}$ are asymmetric in the lowest vane-angle case, this observation has important implications for the discussion that follows. Further, note that the strength of footprints increases with vane-angle in the investigated cases.

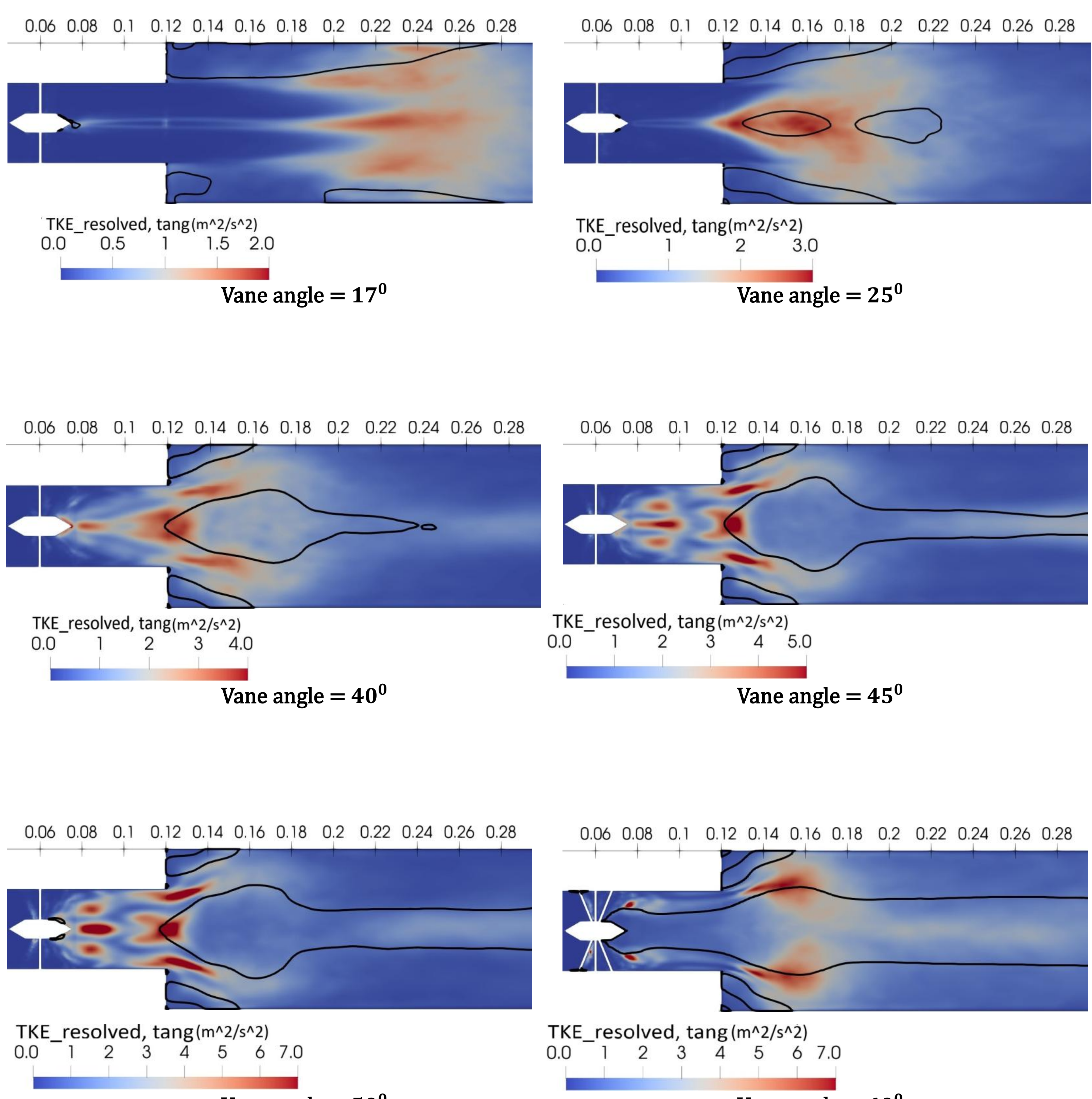

**Figure 15.** Contours of azimuthal component of resolved turbulent kinetic energy $TKE_{resolved,tang}$ ($m^2/s^2$) on the streamwise plane through the centreline of the domain for the cases with different swirler vane angles.

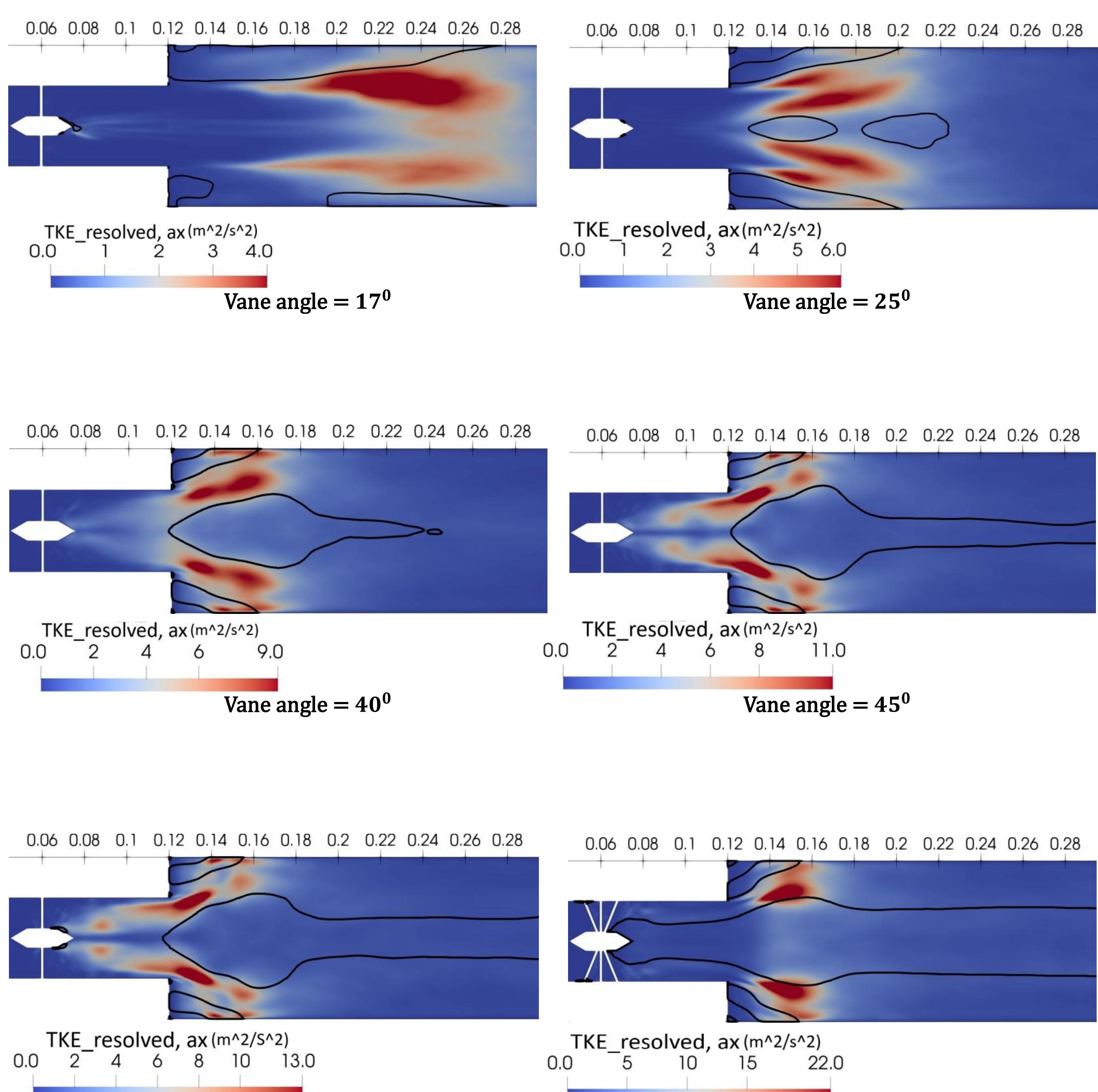

**Figure 16.** Contours of axial component of resolved turbulent kinetic energy $TKE_{resolved,ax}$ ($m^2/s^2$) on the streamwise plane through the centerline of the domain for the cases with different swirler vane angles.

The above discussion indicates that a helical flow structure appears predominantly on the shear layer in the wake of the swirler's CB. The footprints of coherent motion reveal that the VC also predominates in a helical topology over time. However, these footprints do not uniquely determine the VC topology over time. Similar footprints may result from a single-coherent-strand helix, a double-coherent-strand helix, or from an intermittent presence of both [66]. Moreover, the above analysis does not rule out the co-existence of an axisymmetric topology, since radial and axial oscillations of the IRZ can also produce peaks of $TKE_{resolved,ax}$ in the ISL; though not in $TKE_{resolved,tang}$. We now resort to the set of spectral analyses discussed at the beginning of section 6 of the present work to precisely resolve the prevalent VC topologies and to identify the characteristics of their coherent oscillations that give rise to the above observed footprints.

We perform a cross-spectral analysis over the time-series of axial-velocity fluctuations across two diametrically opposite locations lying on the axial-velocity footprints of VC dynamics, i.e., at the locations where $TKE_{resolved,ax}$ peaks in figure 16. The locations of these probe pairs in each investigated case are listed in table 4.

**Table 4.** Coordinates of points for cross-spectral analysis.

| Vane-angle for the investigated cases | Point 1 $(r, z)$ | Point 2 $(r, z)$ |
|---|---|---|
| $17^0$ | (0.0030 m, 0.2460 m) | (−0.0030 m, 0.2460 m) |
| $25^0$ | (0.0140 m, 0.1680 m) | (−0.0140 m, 0.1680 m) |
| $40^0$ | (0.0170 m, 0.1330 m) | (−0.0170 m, 0.1330 m) |
| $45^0$ | (0.0181 m, 0.1330 m) | (-0.0181 m, 0.1330 m) |
| $50^0$ | (0.0186 m, 0.1330 m) | (−0.0186 m, 0.1330 m) |
| $60^0$ | (0.0220 m, 0.1450 m) | (−0.0220 m, 0.1450 m) |

Figure 17 shows the variation of the magnitude of the cross-spectrum, $|S_{XY}(f)|$, for the investigated cases, where $X(t)$ and $Y(t)$ correspond to the time-series of axial-velocity-fluctuations at two diametrically opposite points. Further, Figure 18 shows the variation of the magnitude-squared coherence, $\gamma_{XY}^2(f)$. The peaks in $\gamma_{XY}^2(f)$ in figure 18 and in some scenarios in $|S_{XY}(f)|$ in figure 17 are not razor-sharp since the ensemble length, $N_{seg}$, is modest and the frequency-bin, discrete slot, spacing is 12.5 Hz, as noted in the spectral approach in section 6 of the present work. We use a coherence threshold of $\gamma^2 = 0.70$ to isolate coherent oscillations, i.e., phase-consistent oscillations, because as evident from figure 18 most of the other peaks lie well below this value and qualify as noise except in the largest vane-angle case. Figure 19 shows the unwrapped cross-phase, $\Phi_{XY}(f)$, between the two time-series, which, using $m \approx \frac{\Phi_{XY}(f),}{\pi}$ gives a preliminary estimate of the azimuthal symmetry (mode number) of the coherent oscillations. We reiterate that two-point phase cannot distinguish $m = 0$ from $m = 2$ and since VC structures up to $m = 2$ have been reported in swirling flows [9], we adopt the four-probe azimuthal FFT analysis to identify the correct azimuthal mode [61].

Figures 20 to 25 present the results of an azimuthal FFT applied to the time-series of axial-velocity fluctuations from four probes located at the same radius separated azimuthally by $\frac{\pi}{2}$ radian. Panels ***a*** in figures 20 to 25 show the modal energy distribution, $E(m, f)$, where $m$ is the azimuthal mode number. Panels ***b*** show the modal energy fraction, $F(m, f)$ $[\text{Eq.}\,(24)]$. $F(m, f)$ is distinguishably high at the coherent frequency for a particular mode number, indicating the dominant azimuthal symmetry of the oscillation and thus of the associated flow structure. The threshold value of $F(m, f)$ for "distinguishably high" is case specific, as is apparent in panels ***b***, since it is interpreted only at frequencies that first pass the coherence threshold in figure 18. Panels ***d*** in Figures 20 to 25 show the time trace of the band-passed modal signal $\widetilde{c_m}(t)$ together with its envelope $\pm|a_m(t)|$ for the selected $m$ and $f$. Here $\widetilde{c_m}(t)$ is the oscillatory narrow-band signal and $|a_m(t)|$ is the modal narrow-band amplitude, the slow envelope, for the selected $m$ and $f$ at the native sampling resolution. The native sampling resolution in the present work is the 250$^{\text{th}}$ simulation time-step.

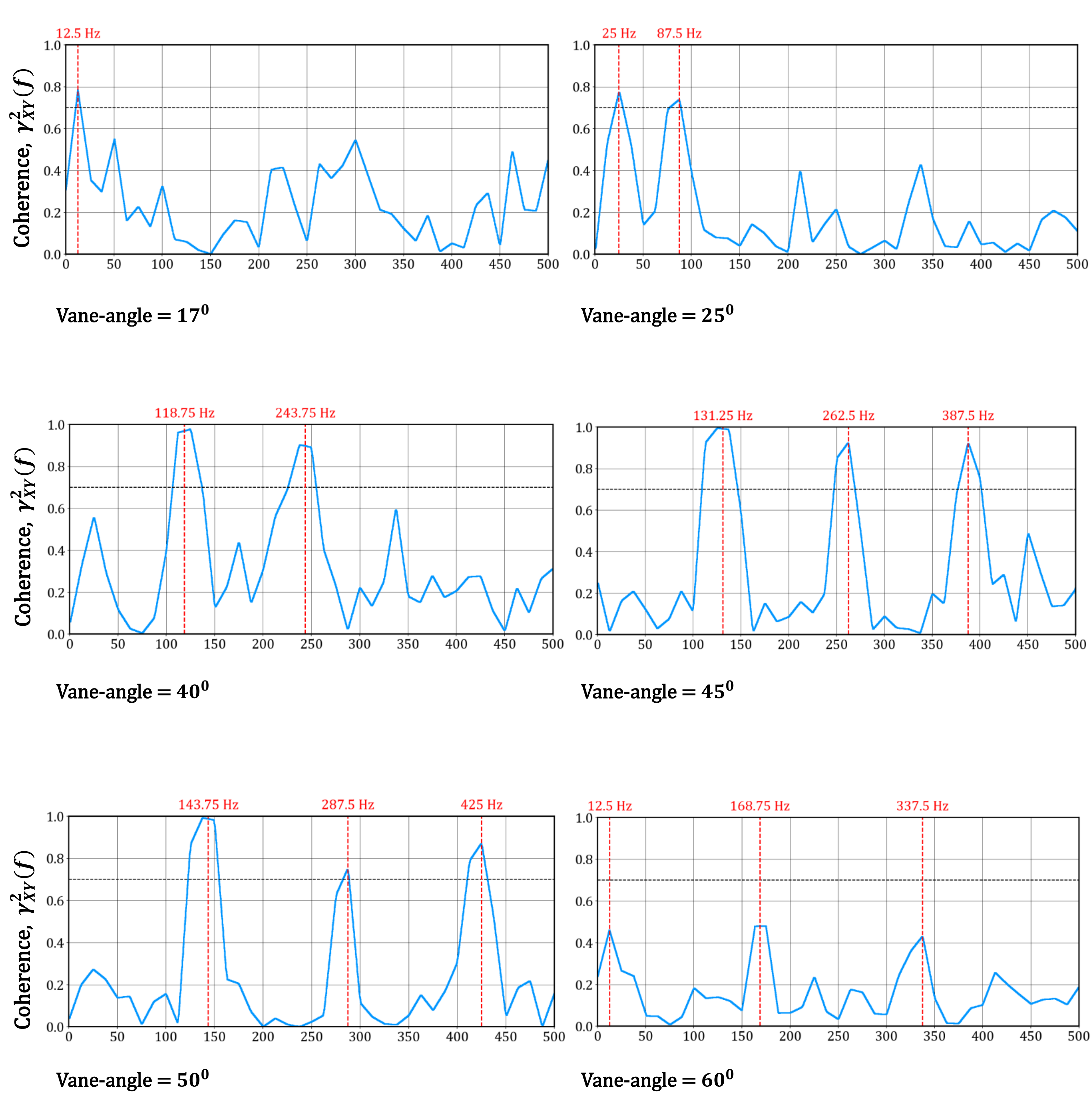

**Figure 17.** Variations of coherence, $\gamma_{XY}^2(f)$, for the cases with different swirler vane angles. The dotted black line marks the threshold value of coherence, which is 0.7. The dotted red lines mark the peaks above this threshold except in the last case.

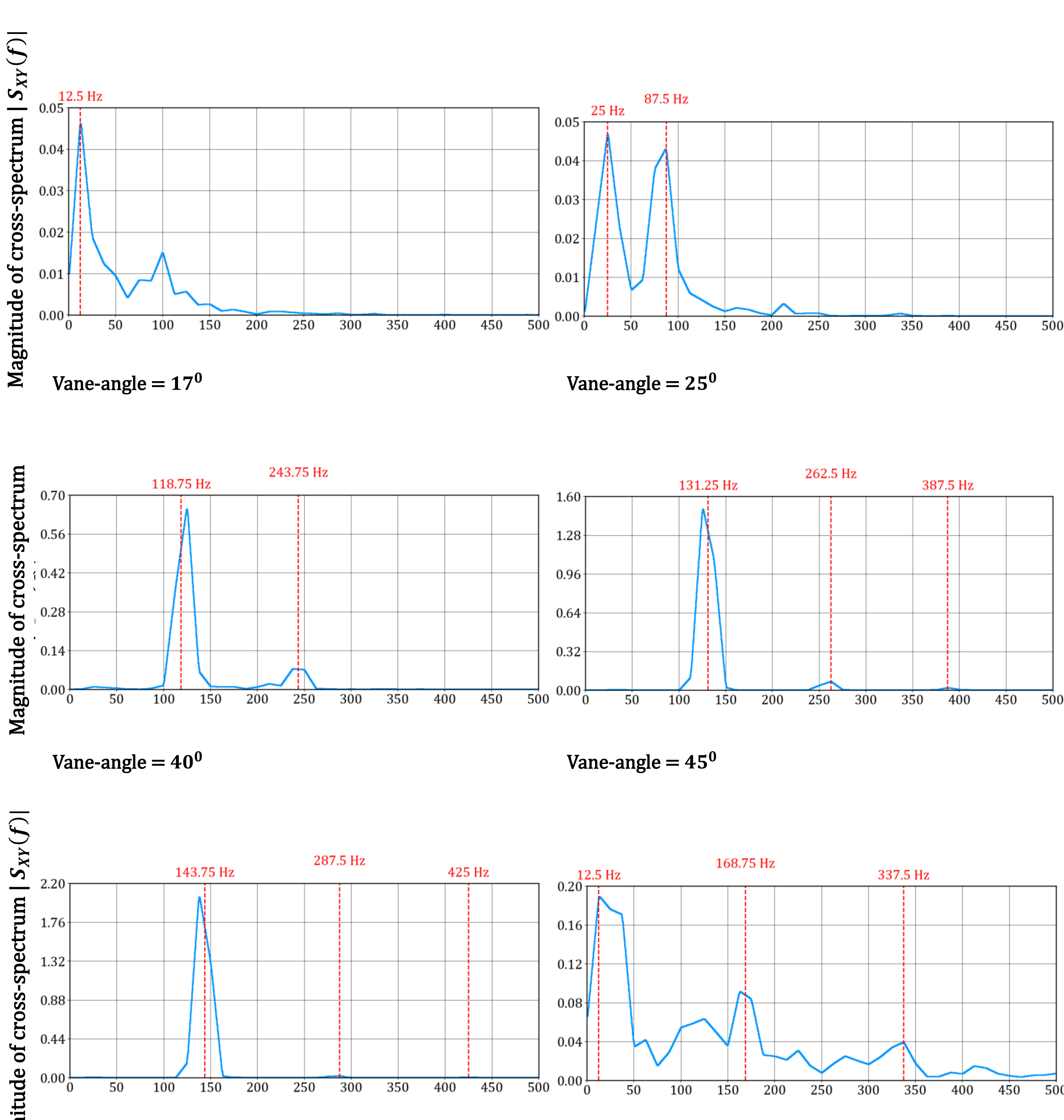

**Figure 18.** Variations Magnitude of cross-spectrum, | $S_{XY}(f)$|, **(m²/s²)** for the cases with different swirler vane angles. The dotted red lines correspond to the frequencies marked in figure 17.

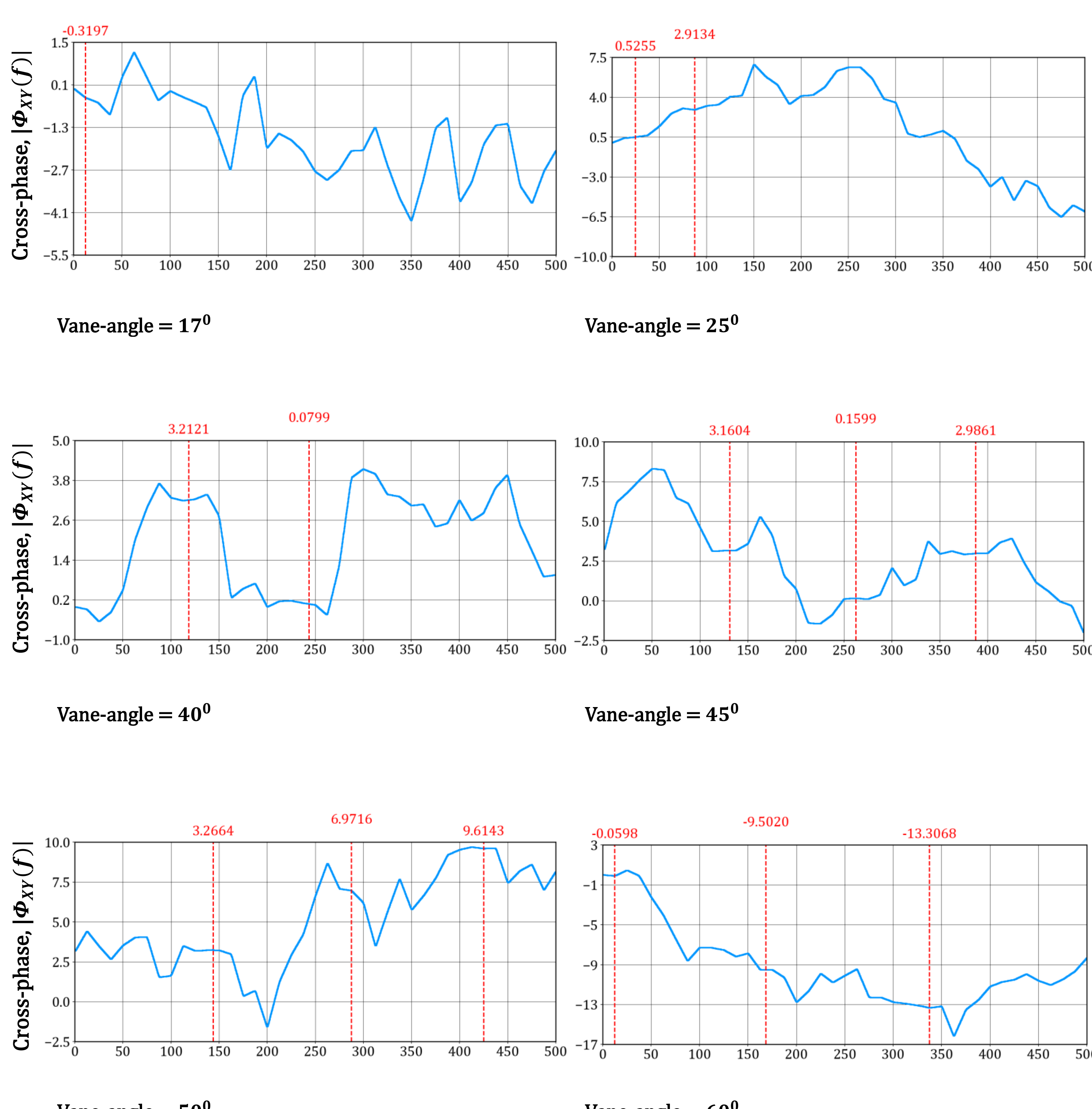

**Figure 19.** Variations cross-phase, $|\Phi_{XY}(f)|$, **(radian)** for the cases with different swirler vane angles. The dotted red lines correspond to the frequencies marked in figure 17.

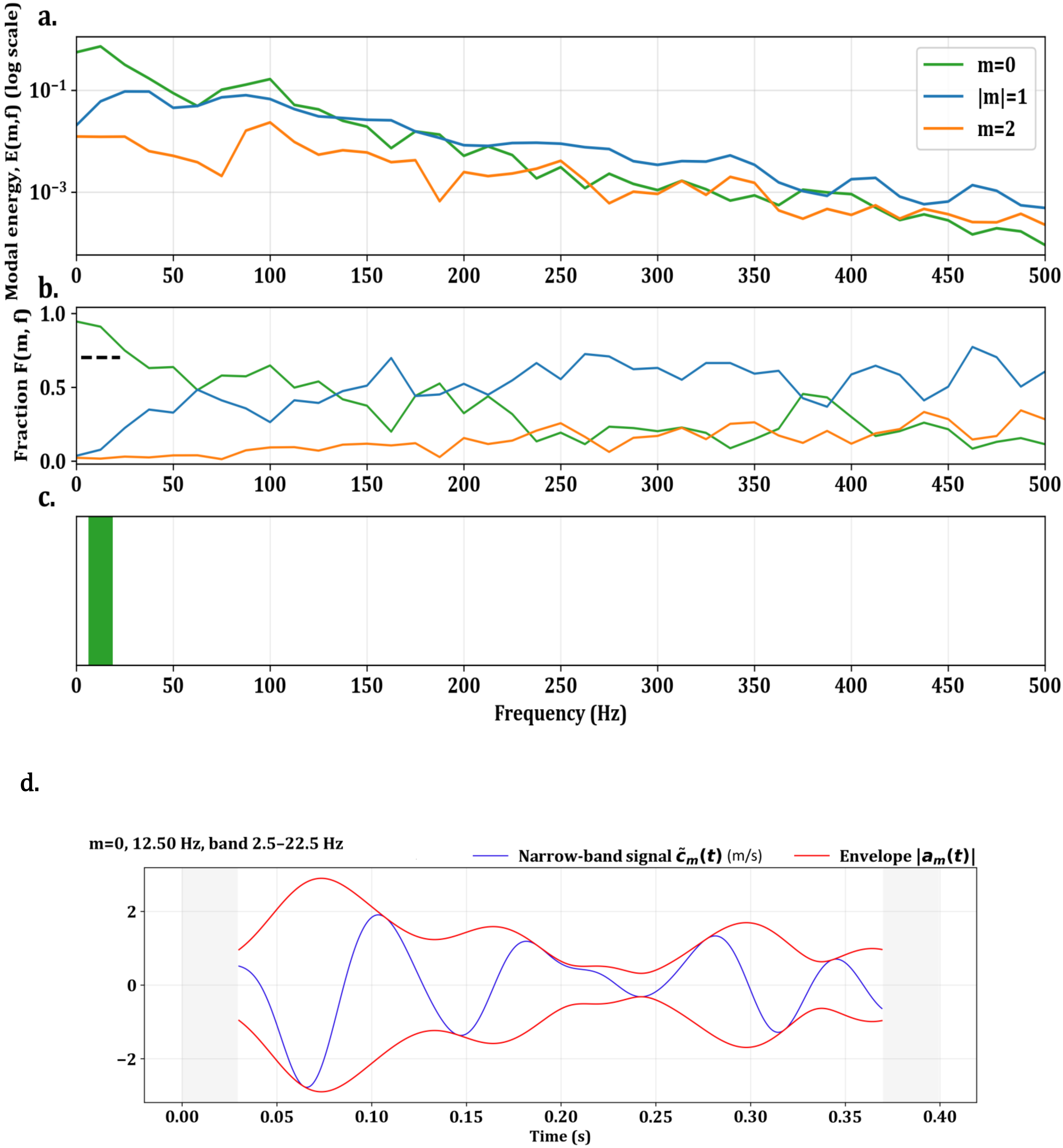

**Figure 20.** **(a)** Variation of modal energy **($m^2/s^2$)**, **(b)** variation of the fraction of modal energy, **(c)** dominant modes at the coherent oscillations and **(d)** temporal variation of the signal, $\widetilde{c_m}(t)$, **(m/s)** and its envelope, $|a_m(t)|$, **(m/s)** for the coherent oscillations in the $17^0$ vane-angle case. The black dashed lines in **(b)** around coherent frequencies indicate the benchmark of 0.7.

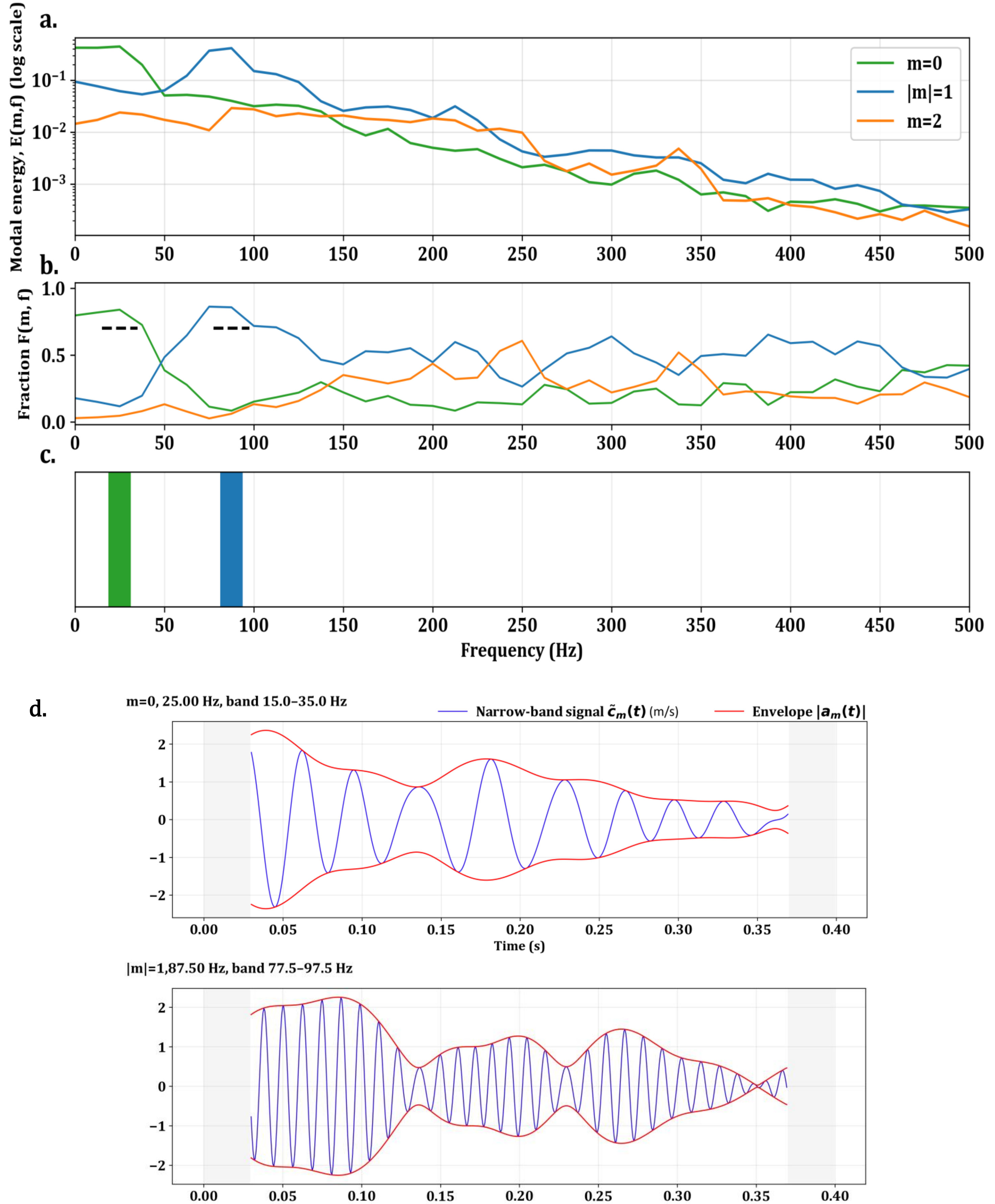

**Figure 21.** **(a)** Variation of modal energy **($m^2/s^2$)**, **(b)** variation of the fraction of modal energy, **(c)** dominant modes at the coherent oscillations and **(d)** temporal variation of the signal, $\widetilde{c_m}(t)$, **(m/s)** and its envelope, $|a_m(t)|$, **(m/s)** for the coherent oscillations in the $25^0$ vane-angle case. The black dashed lines in **(b)** around coherent frequencies indicate the benchmark of 0.7.

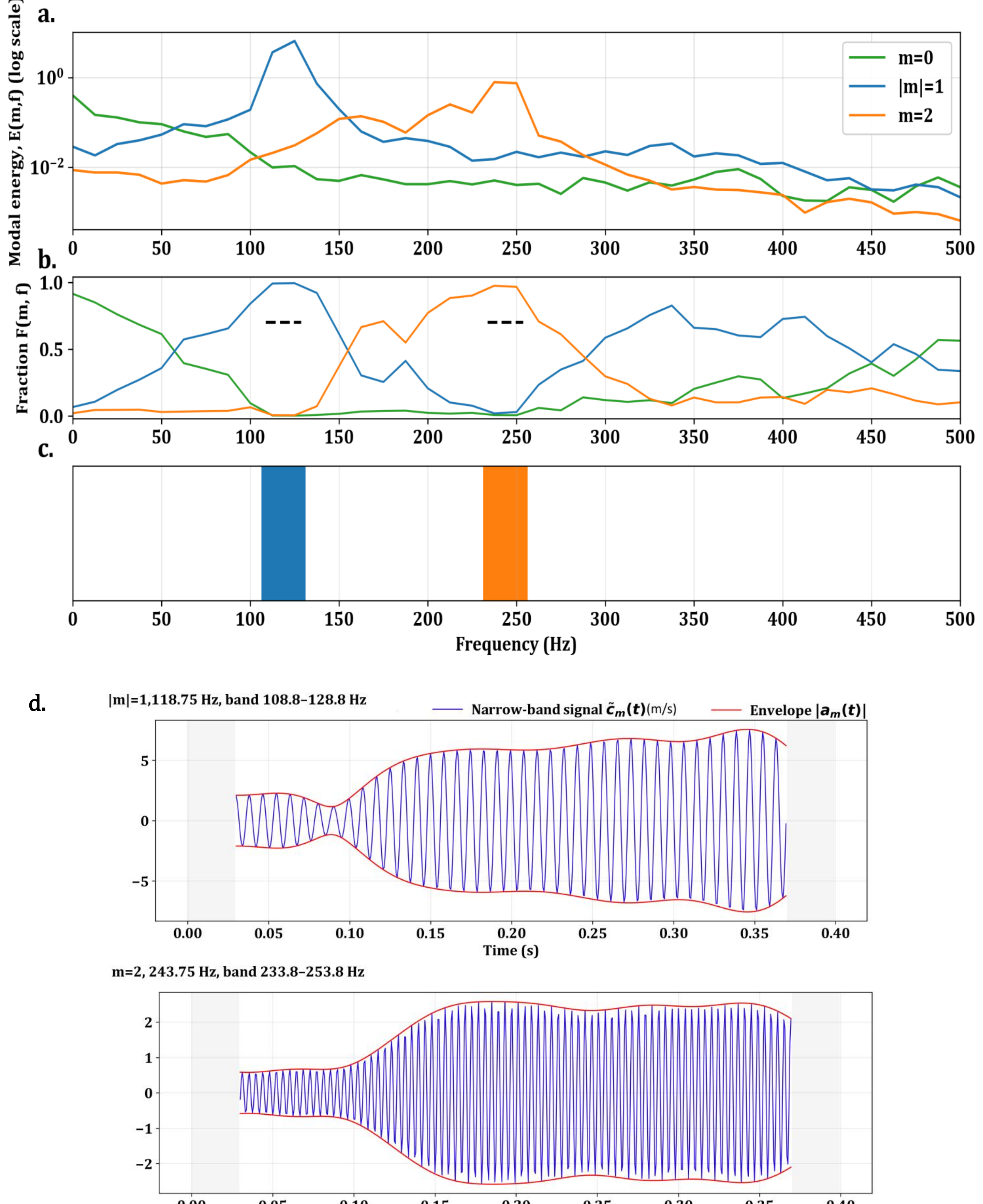

**Figure 22. (a)** Variation of modal energy **(m²/s²)**, **(b)** variation of the fraction of modal energy, **(c)** dominant modes at the coherent oscillations and **(d)** temporal variation of the signal, $\widetilde{c_m}(t)$, **(m/s)** and its envelope, $|a_m(t)|$, **(m/s)** for the coherent oscillations in the $40^0$ vane-angle case. The black dashed lines in **(b)** around coherent frequencies indicate the benchmark of 0.7.

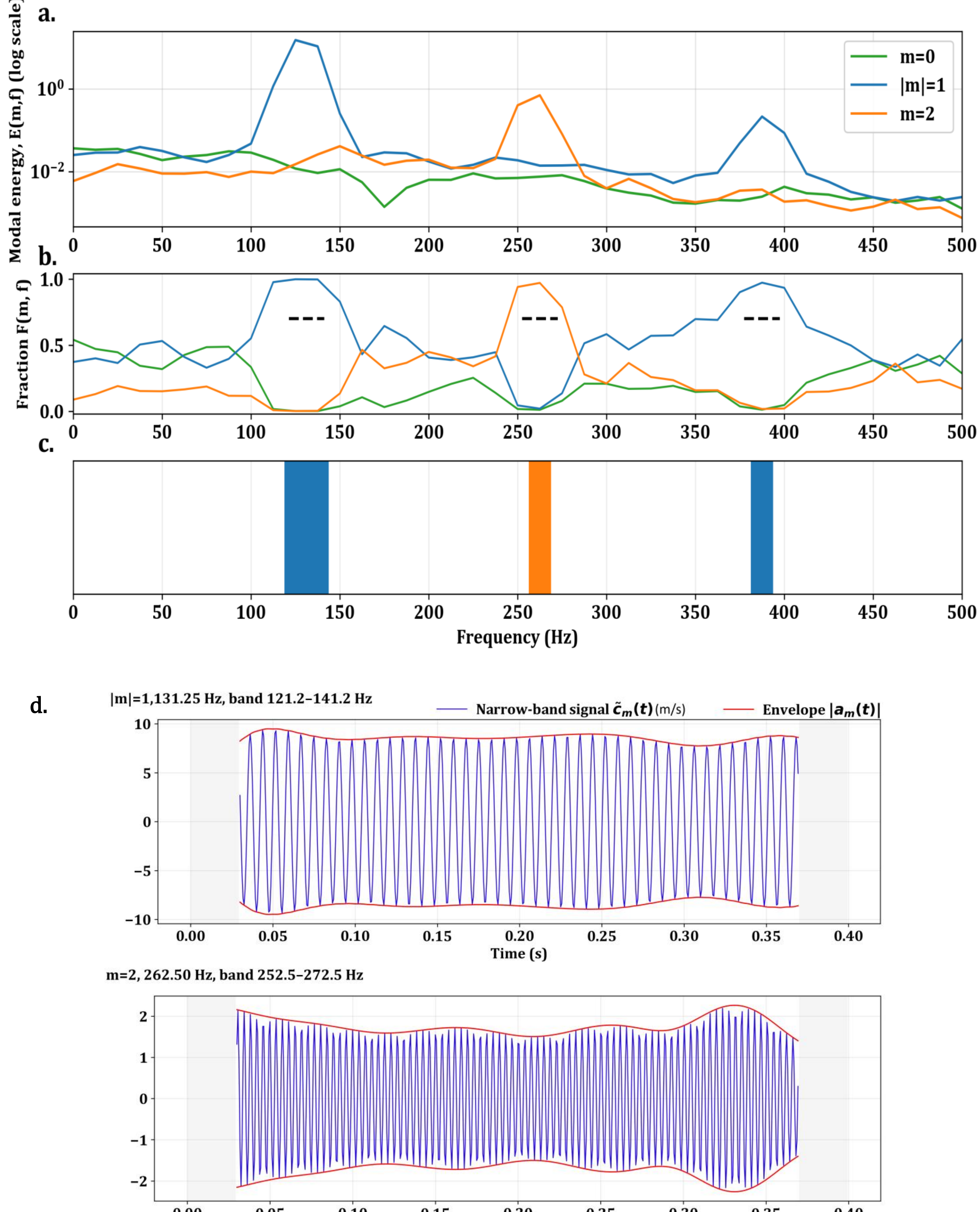

**Figure 23. (a)** Variation of modal energy **(m²/s²)**, **(b)** variation of the fraction of modal energy, **(c)** dominant modes at the coherent oscillations and **(d)** temporal variation of the signal, $\widetilde{c_m}(t)$, **(m/s)** and its envelope, $|a_m(t)|$, **(m/s)** for the coherent oscillations in the 45$^0$ vane-angle case. The black dashed lines in **(b)** around coherent frequencies indicate the benchmark of 0.7.

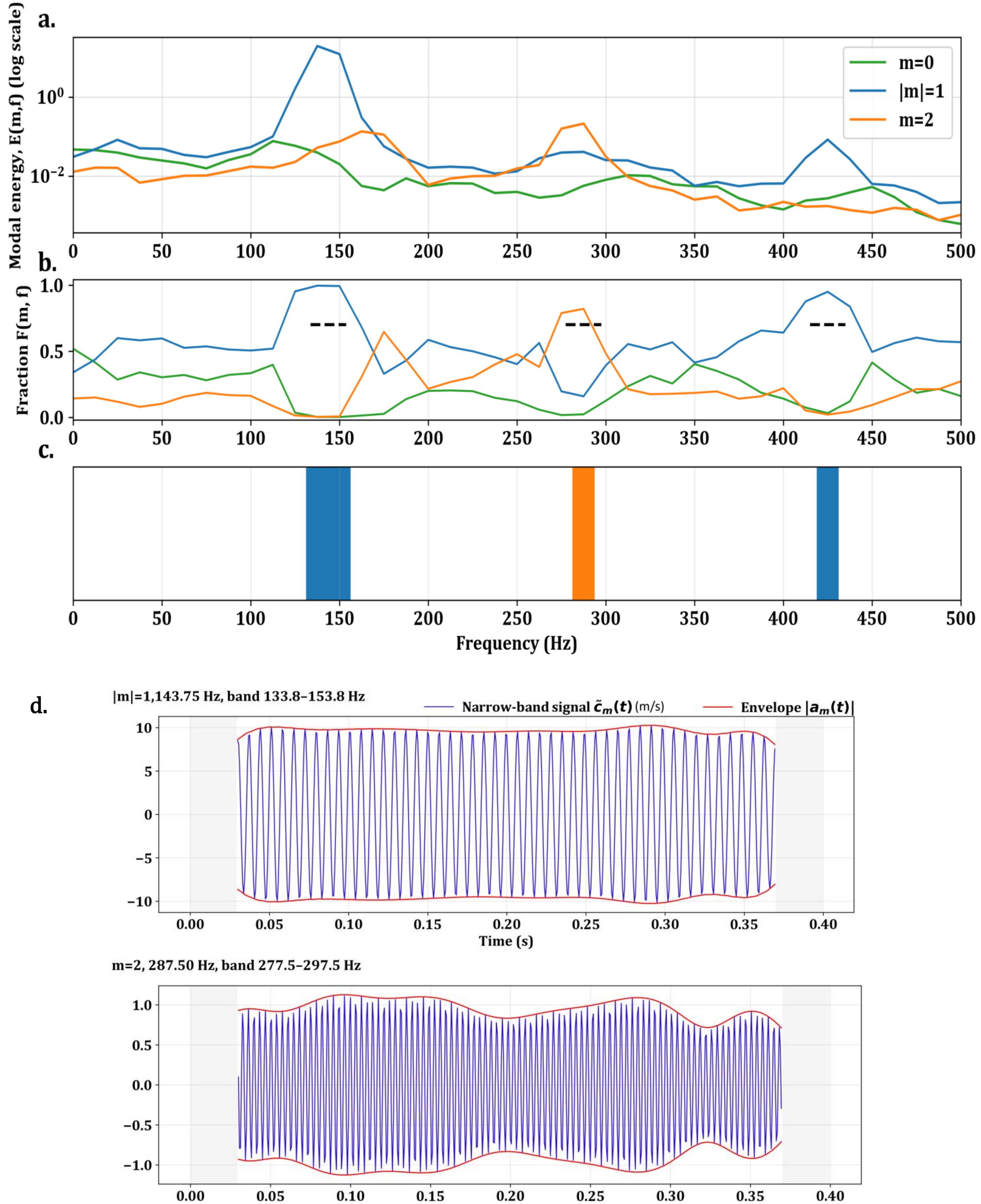

**Figure 24.** **(a)** Variation of modal energy **($m^2/s^2$)**, **(b)** variation of the fraction of modal energy, **(c)** dominant modes at the coherent oscillations and **(d)** temporal variation of the signal, $\widetilde{c_m}(t)$, **(m/s)** and its envelope, $|a_m(t)|$, **(m/s)** for the coherent oscillations in the $50^0$ vane-angle case. The black dashed lines in **(b)** around coherent frequencies indicate the benchmark of 0.7.

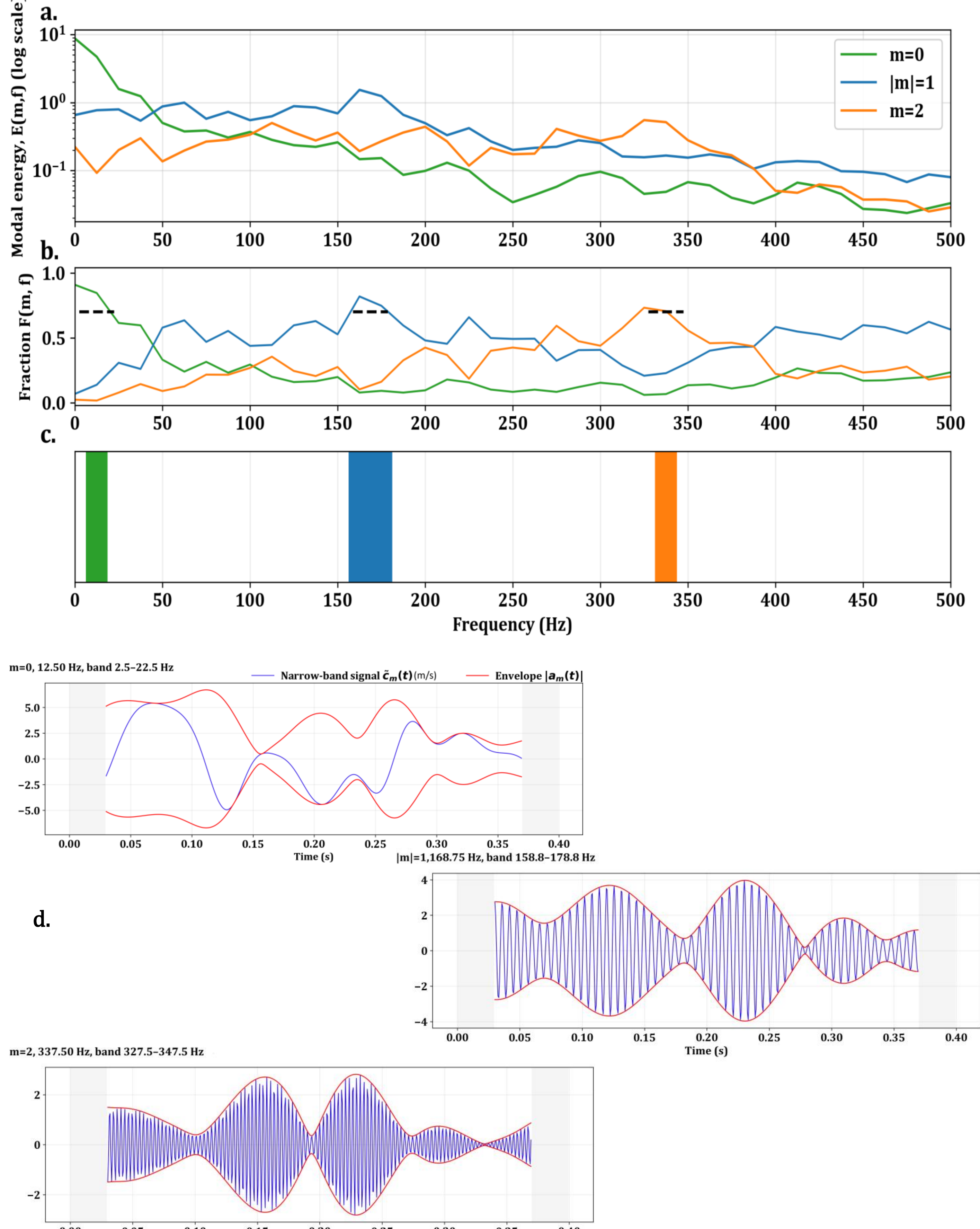

**Figure 25. (a)** Variation of modal energy **($m^2/s^2$)**, **(b)** variation of the fraction of modal energy, **(c)** dominant modes at the coherent oscillations and **(d)** temporal variation of the signal, $\widetilde{c_m}(t)$, **(m/s)** and its envelope, $|a_m(t)|$, **(m/s)** for the coherent oscillations in the $60^0$ vane-angle case.

Figure 18 shows red dashed lines that correspond to the frequencies where the magnitude-squared coherence, $\gamma_{XY}^2(f)$, peaks and is above the set threshold except in the largest vane-angle case. This first marking appears at $f_1 = 118.75, 131.25$ and $143.75$ Hz for the cases with vane-angle of $40^0$, $45^0$ and $50^0$, respectively. The $|S_{XY}(f)|$in figure 17 also peaks around these frequencies, indicating that the dominant correlated energy resides in the associated dynamics. Figures 22a to 24a, $E(m, f)$, confirm that among the coherent oscillations, the oscillations at $f_1$ carry the largest energy in the three cases and they also show that this frequency corresponds to $|m| = 1$. Figure 19 shows that $\Phi_{XY}(f) \approx \pi$ radian or $m \approx 1$ at $f_1$ in the three cases. The frequencies at the other markings in figure 18, corresponding to coherence beyond threshold, are almost integral multiples of $f_1$, i.e., $f_2 \approx 2f_1$ and $f_3 \approx 3f_1$; coherence at $f_3$ is low in the $40^0$ vane-angle case. A high value of $\gamma_{XY}^2(f)$ shows that the two time-series remain consistently phase-correlated at these frequencies. However, $|S_{XY}|$ is relatively small at $f_2$ and $f_3$ in the three cases, implying small correlated energy in the corresponding oscillations, as also reflected in figures 22a to 24a. $|S_{XY}|$ is specifically quite small at $f_3$ and is therefore not discussed further. Figure 19 shows that $\Phi_{XY}(f) \approx 2\pi$ radian or $m \approx 2$ at $f_2$ in the $50^0$ vane-angle case and $\Phi_{XY}(f) \approx 0$ radian or $m \approx 0$ at $f_2$ in $40^0$ and $45^0$ vane-angle cases. However, recall that the two-point phase cannot distinguish between 0 and $2\pi$ radians and so between $m = 0$ and $m = 2$ [58]. This is also evident upon comparing figure 19 with figures 22b to 24b. Figures 22b to 24b, arising from the four probes azimuthal FFT, show that $m = 2$ at $f_2$ in each of the three cases of interest. Hence, a VC exhibiting strong single-helix signatures, $|m| = 1$, and weaker double-helix signatures, $m = 2$ , is expected in the $40^0$, $45^0$ and $50^0$ vane-angle cases .

The band-passed modal signal $\widetilde{c_m}(t)$ and its envelope $|a_m(t)|$ for $|m| = 1$ and $m = 2$ at their characteristic frequencies $f_1$ and $f_2$, respectively, in figures 22d to 24d show that the oscillation amplitudes are higher for the ($|m| = 1, f_1$) pair or tone. Further, it is apparent that the amplitudes for ($|m| = 1$, $f_1$) and ($m = 2$, $f_2$) in the $45^0$ and $50^0$ vane-angle cases are nearly constant over time, indicating stable limit-cycle oscillations for these tones. This behavior is consistent with precession in the $45^0$ and $50^0$ vane-angle driven by a marginally stable hydrodynamic mode on the mean flow state. The oscillation amplitude for ($|m| = 1$, $f_1$) and ($m = 2$, $f_2$) in the $40^0$ vane-angle case remains nearly constant for a while and then grows to a higher, nearly constant level. This behavior is consistent with slow drift of the mean flow (stochastic variations) that alters the characteristics of the marginally stable helical mode driving the precession.

We now test whether ($m = 2$, $f_2 \approx 2f_1$) arises from quadratic self-coupling of ($|m| = 1$ , $f_1$) or is an azimuthal coherent oscillation independent of it, by investigating the corresponding squared cross-bicoherence, $b^2(X_{f_1}, X_{f_1} \rightarrow Y_{2f_1})$. A high $b^2$ for a triad ($f_a$,$f_b$,$f_c$) indicates reproducible second-order phase-locking, a phase-sum/-difference relation, i.e., $\Phi_{XY}(f_c) \approx \Phi_{XY}(f_a) \pm \Phi_{XY}(f_b)$, across ensembles, while a low $b^2$ indicates no such stable phase-locking despite any spectral energy. The $b^2(X, X \rightarrow Y)$ maps for the investigated cases are plotted in figure 26. To account for finite bin spacing, $\Delta f = 12.5$ Hz, and window broadening, we draw guide bands so that bin mismatch is not misread as a weak coupling. We place constant-sum anti-diagonals at the measured second peak with its own uncertainty, $f_a + f_b = f_2 \pm \sigma_{f_2}$ and we set $\sigma_{f_2} = 10$ Hz. The white-color solid anti-diagonal marks $f_a + f_b = f_2$ and the white-color dashed anti-diagonals mark the tolerance band $f_a + f_b = f_2 \pm 10$ Hz. We also mark the self-pair $(f_1, f_1)$ with a green-color circle and draw an axis-aligned green-color square $| f_a - f_1 | \leq \frac{10}{\sqrt{2}}$ Hz, $| f_b - f_1 | \leq \frac{10}{\sqrt{2}}$ Hz around it, which is RMS uncertainty. Within the overlap of this square and the constant-sum band, the observed $b^2(X_{f_a}, X_{f_b} \rightarrow Y_{f_a+f_b})$ is very high in all three cases in figure 26, although $\gamma_{XY}^2(f_2 \approx 2f_1)$ is high. Therefore, $m = 2$ oscillation at $f_2 \approx 2f_1$ is predominantly the first harmonic of $|m| = 1$ oscillation at $f_1$ in the investigated swirl combustor cases with vane-angles of $40^0$, $45^0$ and $50^0$. In other words, $m = 2$ oscillation at $f_2 \approx 2f_1$ most-likely arises from quadratic self-interaction of $|m| = 1$ oscillation at $f_1$ in these cases. Its implication on VC is that the double-helix signatures apparent on it from ($m = 2$, $f_2 \approx 2f_1$) arise from self-coupling of single-helix signatures from ($|m| = 1$ , $f_1$) and therefore the two signatures oscillate in a phase-locked manner.

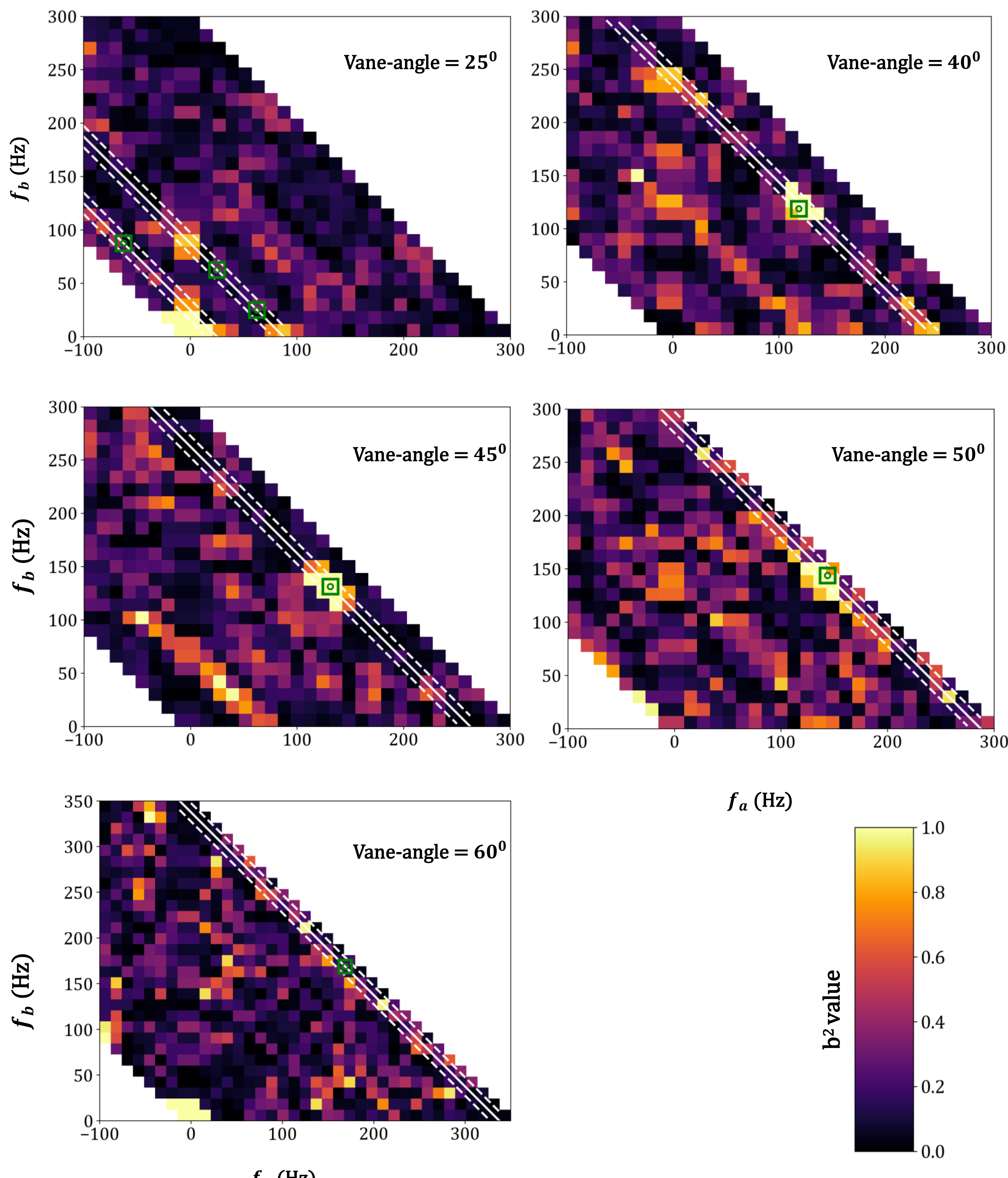

**Figure 26.** Squared cross-bicoherence, $b^2$, map for the investigated cases with different vane angles.

Figure 18 shows two frequencies, 25.0 and 87.5 Hz, where the magnitude-squared coherence, $\gamma_{XY}^2(f)$, peaks and is above the set threshold of 0.7 in the case with $25^0$ vane-angle. Note that the $\gamma_{XY}^2(f)$ value at each peak in the present case is just above the threshold value of 0.7, whereas it is very close to 1.0 at corresponding peaks in the $40^0$, $45^0$ and $50^0$ vane-angle cases. $|S_{XY}(f)|$ and $E(m,f)$ in figures 17 and 21a are relatively high and comparable at these frequencies, indicating two dominant coherent oscillations in the flow. $F(m,f)$ in figure 21b shows the dominance of $m = 0$ around 25.0 Hz and $| m |= 1$ around 87.5 Hz. Next, we investigate an indirect interaction between these two coherent modes via the triad 25.0, 62.5 → 87.5 Hz since its permutations are frequency consistent. However, figure 17 shows that the magnitude-squared coherence, $\gamma_{XY}^2(f)$, around 62.5 Hz is well below the threshold value for a coherent motion. Figures 18 and 21 show that $|S_{XY}(f)|$ is also very small around 62.5 Hz relative to its value at 31.5 and 87.5 Hz and the $E(m,f)$ values are similar for the three investigated modes. However, a relatively large fraction of modal energy, $F(m,f)$, is concentrated in $|m| = 1$ at 62.5 Hz in figure 21. Motivated by this last observation, we investigate the extent of interaction among the triads from the relevant $b^2$ values. The values of $b^2$ are low for 25.0, 62.5 → 87.5 Hz, for 62.5, 25.0 → 87.5 Hz and for 87.5, 62.5 → 25.0 Hz in figure 26. Accordingly, the axisymmetric oscillations ($m = 0$, 25.0 Hz) and the helical oscillations ($|m| = 1$, 87.5 Hz) are quadratically independent. The band-passed modal signal $\widetilde{c_m}(t)$ and the envelope $|a_m(t)|$ for $m = 0$ and $|m| = 1$ at their characteristic frequencies, 25.0 Hz and 87.5 Hz, respectively, are shown in figure 21d. These signals have comparable amplitudes, with dips to small values and subsequent recovery. The $|m| = 1$ signal appears more dynamic. The amplitudes for ($|m| = 1$, $87.5$ Hz) wax and wane with large variations, approach near zero at times and then re-grow toward a limit-cycle-like oscillation. This evolution suggests that precession in the $25^0$ vane-angle case is driven by stochastic forcing from background turbulence acting on a slightly stable global helical (hydrodynamic) mode. Taken together, a single-helix VC coexists with an axisymmetric oscillation (likely from axisymmetric IRZ motion) in the $25^0$ vane-angle case, with the two tones quadratically independent and sustained by stochastic forcing on slightly stable axisymmetric and helical modes.

Figure 18 shows a single above-threshold coherence peak at 12.5 Hz in the lowest vane-angle case, $17^0$. $|S_{XY}(f)|$ and $E(m, f)$ are also the highest near the same frequency in figures 17 and 20a. $F(m, f)$ for $m = 0$ dominates around 12.5 Hz in figure 20b. The corresponding band-passed modal signal $\widetilde{c_m}(t)$ and the envelope $|a_m(t)|$ in figure 20d exhibit pronounced waxing and waning of the signal's amplitude. Accordingly, we attribute this coherent motion to the axial, axisymmetric, oscillation arising from the spatio-temporal intermittency of the IRZ in the present case, which has already been discussed in sections 6.1 and 6.2 of the present work. No sustained helical coherence is detected in the present case. It could result from the helical oscillations being either largely absent or incoherent, or from the asymmetry of the flow evident in figure 16. The asymmetry biases the VC toward one side of the domain so that the four probes do not receive a balanced helical signature. Spectral arrays respond best when the coherent structure affects all sensors similarly. Further, visual checks using the $Q$-criterion over about $640$ instances show sporadic helical structure, supporting the interpretation that the absence of helical coherence is either a numerical artifact due to flow asymmetry or from genuinely incoherent helical motion of the VC.

Finally, we investigate the largest vane-angle case, $60^0$, where the magnitude-squared coherence, $\gamma_{XY}^2(f)$, does not exceed the set threshold of 0.7 at any investigated frequency. However, $\gamma_{XY}^2(f)$ shows distinguishable peaks, above the value 0.4, at $12.5, 168.75$, and $337.5$ Hz relative to other frequencies in figure 18. Thus, the precession flow dynamics in the largest vane-angle case would appear weakly coherent. The values of $|S_{XY}(f)|$ and $E(m,f)$ in figures 17 and 25a are highest around $12.5$ Hz and reasonable at the other two characteristic frequencies. $F(m,f)$, from azimuthal decomposition, in figure 25b shows that $m=0$ dominates around $12.5$ Hz, $|\ m\ |=1$ dominates around $168.75$ Hz and $m=2$ dominates around $337.5$ Hz. Thus, this dynamic system exhibits all key aspects of helical oscillations prevalent in swirling flows [9]. Interestingly, $b^2$ is low for the $168.75, 168.75 \rightarrow 337.5$ Hz triad in figure 26, indicating that the $m=2$ oscillation at $337.5$ Hz is not the first harmonic of the $|\ m\ |=1$ oscillation at $168.75$ Hz in the largest vane-angle case. This contrasts with the behavior in cases with vane-angles up to and including $50^0$. The temporal signal and its envelope for $m=0$, $|m|=1$ and $m=2$ at their characteristic frequencies, $12.5$ Hz, $168.75$ Hz and $337.5$ Hz, respectively, in figure 25d show largest oscillation amplitudes for $m=0$, smaller for $|m|=1$ and smallest for $m=2$. None of these signals persist at a constant amplitude; their envelopes wax and wane strongly and approach near-zero at times. This behavior is consistent with the characteristic oscillations being sustained by stochastic forcing from background turbulence on slightly stable hydrodynamic modes. The IRZ extends across the combustor downstream of the swirler in the largest vane-angle case and thereby merges with the separation bubble of CB. This merger is known to disrupt the wavemaker of the global precessing mode and stabilize it [68]. Any residual precession then arises from stochastic forcing of this stabilized mode, leading to large amplitude modulations as observed. Conclusively, a VC forms in the largest vane-angle case with strong single-helix characteristics and weaker double-helix characteristics interacting with a stronger axisymmetric oscillation (likely from axisymmetric IRZ motion); these helical and axisymmetric tones are sustained by stochastic excitation of stable hydrodynamic modes. Notably, the double-helix tone does not arise from quadratic self-interaction of the single-helix tone (low $b^2$); instead, the two tones are quadratically independent, indicating a distinct helical hydrodynamic mode associated with the double-helix tone.

Finally, the lowest coherent precession frequencies from figures 17 to 25 for the case with vane-angles of $25°, 40°, 45°, 50°$ and $60°$ are $87.5, 118.75, 131.25, 143.75$ and $168.75$ Hz, respectively. The precession frequency for the smallest vane-angle, if it exists, could not be resolved using the above spectral methods.

To resolve the origin of the weakly-coherent strand discussed earlier for cases with vane-angles of $25°, 40°, 45°$ and $50°$ we compute an FFT of the time-series of x-velocity (tangential-vel) fluctuations at the location where the weakly-coherent strand leaves its strongest tangential footprint, i.e., at the domain centreline where $TKE_{resolved,tang}$ is maximum, see figure 15. A weakly-coherent strand intertwined with the coherent strand of the VC also appears in the lowest vane-angle ($17°$) case. However, its origin could not be argued based on the above logic since the dynamics of VC cannot be resolved by the spectral approaches discussed in the present work, discussed earlier. Nevertheless, the FFT on the time-series of x-velocity (tangential-vel) fluctuations at the location discussed above is shown in figure 27 for the smallest vane-angle case along with other cases for completeness. Figure 27 shows that strongest FFT amplitudes occur at $50.0, 92.5, 122.5, 130.0$ and $142.5$ Hz for the cases with vane-angles of $17°, 25°, 40°, 45°$ and $50°$, respectively. These frequencies are consistent with the lowest precession frequencies mentioned above, supporting the interpretation that the strand originates from VC precession dynamics. Note that the spectra for the two lowest vane-angle cases are very noisy around their global peak in amplitude.

In summary, a VC with strong single-helix characteristics appears to prevail in the swirl combustor with $ER = 2.0$ operating at $Re_{inlet} = 2 \times 10^4$ for all the investigated vane-angles; up to $60^0$ , i.e., $SN_g \leq 0.98$. Manoharan *et al.* [58] likewise observed a single-helix VC to prevail in a swirling jet with $Re_{inlet} \approx 6 \times 10^4$ over a wide range of swirl numbers. Drawing on this background and studies [11-13], recall literature review, VB in any isothermal swirl-combustor flow with $Re_{inlet} \geq 10^4$ and $ER \geq 2.0$ is expected to produce a VC with strong single-helix characteristics. Notably, weak but observable double-helix characteristics appear for vane-angles above $25^0$ in the present work. The double-helix tone is found to result from quadratic self-interaction of the single-helix tone for vane-angles below $60^0$. These two tones are quadratically independent in the $60^0$ vane-angle case, i.e., they are associated with distinct helical hydrodynamic modes. We believe the foregoing discussions address most of the open issues identified in the literature review and therefore we proceed to the conclusion of the present work.

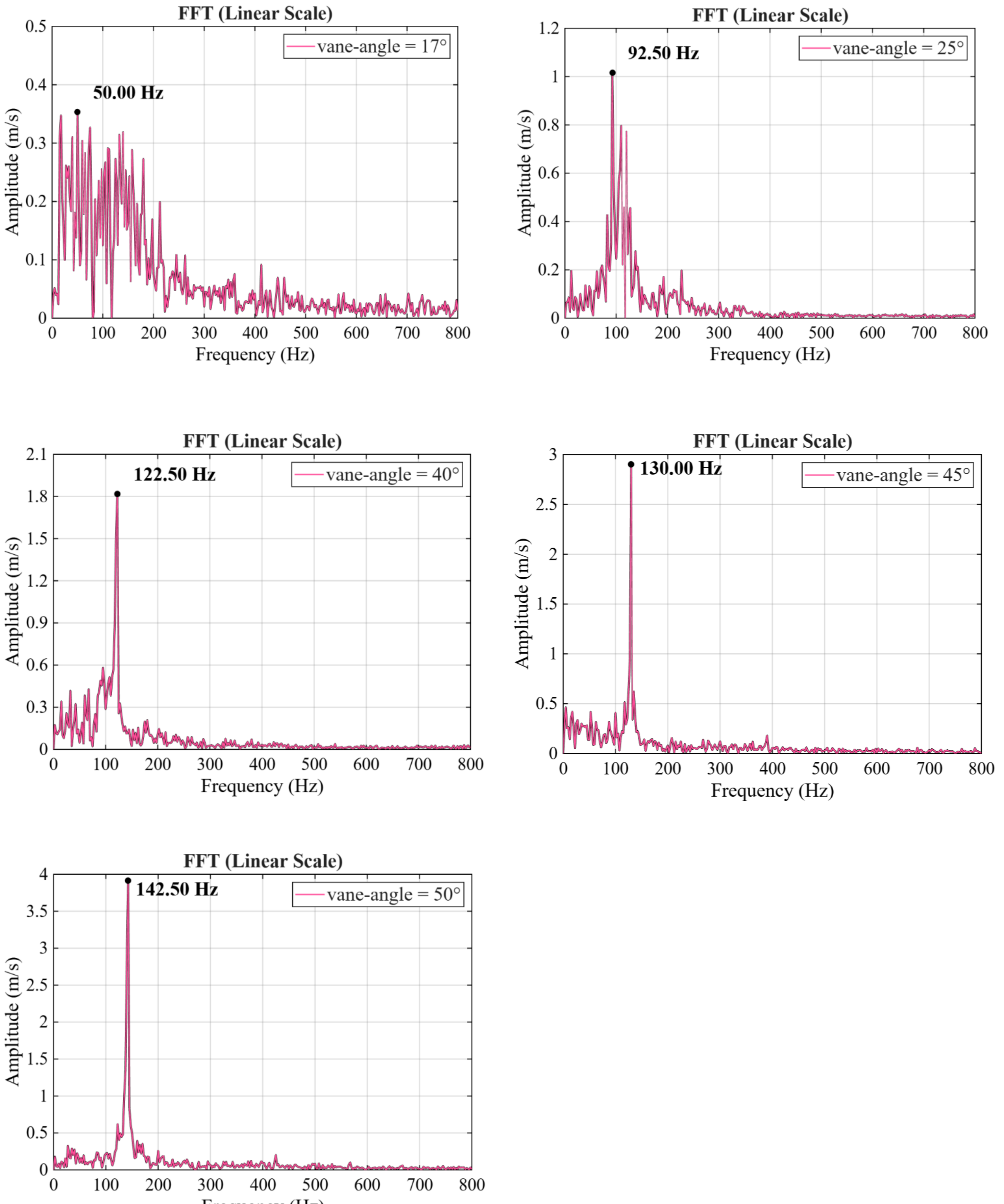

**Figure 27.** Fast Fourier Transform of the fluctuation of x-velocity component at the geometric centre of the cross-stream plane located 0.115 m downstream from the inlet for the investigated cases with different swirler vane angles.

## Conclusions

The present study maps the onset and topology of vortex breakdown with swirl number in an isothermal swirling flow within a well-documented lab-scale combustor facility with expansion ratio, $ER$, of $2.0$ operating at inlet Reynolds number, $Re_{inlet}$, of $2 \times 10^4$. This is done by systematically varying only the swirler vane-angle around the experimental, base-case, vane-angle value and numerically investigating each corresponding flow field. The present study uses a large-eddy-simulation-based solver validated against the experimentally obtained velocity field in the base-case of the lab-scale swirl combustor. The onset of stable-VB is identified by the appearance of the internal recirculation zone (IRZ) in the mean flow. Highly intermittent IRZs may appear at lower swirl strengths in instantaneous flows but are undesirable because they cannot support flame stabilization. A stable VB first appears in the selected swirl combustor flow at the swirler vane-angle of $25^0$ and persists at higher vane-angles. The onset of stable-VB is also quantified in terms of swirl number, to enable comparisons across different flow scenarios, using the generic swirl number formulation, $SN_g$. $SN_g$ appears as the most appropriate formulation for assessing swirl strength in isothermal flows, as its variation within a flow domain (along the flow direction) of fixed cross-section remains comparatively small and monotonically decreasing, in accordance with the expected characteristics of a flow strength measure. Measuring $SN_g$ within $40$ mm downstream of the swirler offers a reliable metric for comparing swirl strength across different flow fields and substantially reduces the likelihood of drawing physically inconsistent conclusions. The critical value of $SN_g$ for the onset of stable-VB lies between $0.21$ and $0.35$ and is argued to exhibit only weak sensitivity to $ER$, and $Re_{inlet}$ for $ER \geq 1.5$ and $10^4 \leq Re_{inlet} \leq 1.6 \times 10^5$.

Three-dimensional vortex core (VC) structures visualized using $Q$-criterion iso-surfaces and their coherent dynamics characterized from velocity time-series sampled at the footprints of the predominant motion reveal that a single-helix VC prevails across all investigated vane-angles; up to $60^0$, i.e., $SN_g = 0.98$. Weaker double-helix characteristics also appear in the investigated cases, arising from quadratic self-interaction of the single-helix tone for vane-angles at or below $50^0$, $SN_g = 0.77$. In contrast, the double-helix tone is quadratically independent of the single-helix tone in the largest vane-angle case, $60^0$, i.e., $SN_g = 0.98$. Thus, the double-helix tone is associated with a helical hydrodynamic mode distinct from the one associated with the single-helix tone. Coherent axisymmetric IRZ oscillations coexist and interact with the helical dynamics of the VC in the $25^0$ and $60^0$(largest vane-angle) cases. The $40^0$, $45^0$ and $50^0$ vane-angle cases exhibit nearly constant-amplitude single-helix precession and associated double-helix precession, consistent with stable limit-cycle oscillations driven by a marginally stable helical mode on the mean state. In contrast, the amplitude of the characteristic helical oscillations in the $25^0$ and $60^0$ vane-angle cases waxes and wanes strongly, approaching zero at times. The $25^0$ vane-angle case corresponds to the onset of the stable IRZ, and in the $60^0$ vane-angle case the IRZ merges with the separation bubble of centerbody; both scenarios are known to stabilize precessing modes. This waxing-waning and stabilizing behavior indicates that the coherent helical oscillations in $25^0$ and $60^0$ vane-angle cases are sustained by stochastic forcing from background turbulence acting on the stabilized helical modes. These observations are argued to be generically applicable to isothermal swirl-combustor flows with $Re_{inlet} \geq 10^4$ and $ER \geq 2.0$.

Besides a coherent VC strand, a weakly-coherent strand also appears to originate from the swirler. The lowest precession frequencies of the coherent strands identified by the spectral analysis used in the present work are consistent with independently evaluated oscillation frequencies of the weakly-coherent strand in the investigated cases, corroborating a VC-precession origin for the latter structure.

Overall, these results provide (i) a suitable criterion for the onset of stable VB, (ii) guidelines on where to evaluate the criterion to minimize spurious variability, and (iii) a VC topology map against swirl strength that can inform reactor design and diagnostics under isothermal conditions. Although the present analysis is restricted to non-reacting flows, it contains the ingredients for extension to reacting flows.

## Acknowledgements

The authors thank IIT (ISM) Dhanbad and the High-Performance Computing facility at IIT Delhi for the computational support.

## Funding statement

The present work is supported by the Faculty Research Scheme (FRS) Grant; FRS (217)/2024-25/FMME of IIT-ISM Dhanbad.

## Competing interests

The authors have no conflicts of interest.

## Data availability statement

The data that supports the findings of this study are available from the corresponding author upon reasonable request.

## Author ORCID

N.K. Sahu, https://orcid.org/0000-0001-9941-215X

A. Dewan, https://orcid.org/0000-0002-1561-713X

M. Kumar, https://orcid.org/0000-0002-4478-8798

## CRediT

**N.K. Sahu**: Conceptualization, Data curation, Formal analysis, Funding acquisition, Investigation, Methodology, Project administration, Resources, Software, Validation, Visualization, Writing – original draft and Writing – review & editing.

**A. Dewan**: Project administration, Writing – review & editing.

**M. Kumar**: Conceptualization, Funding acquisition, Resources, Writing – review & editing.

## Ethical statements

This study involved no human participants or animals and required no ethical approval.

## Use of AI tools

An AI-assisted grammar checker (**ChatGPT**, version current as of **28 September 2025**) was used for language polishing only. All scientific content, analyses and conclusions are the authors' own.